\documentclass{nature}

\usepackage{lineno}
\usepackage{setspace}
\usepackage{enumitem}
\usepackage[margin=0.7in]{geometry}
\usepackage{caption}
\usepackage{amsfonts}
\usepackage{amssymb}
\usepackage{amsmath}

\usepackage{placeins}
\usepackage{graphicx}
\usepackage[T1]{fontenc}
\usepackage{arydshln}

\usepackage[export]{adjustbox}

\usepackage[labelformat=empty]{caption}

\usepackage{xspace}
\usepackage{subfig}
\usepackage{ulem}
\usepackage{longtable}
\usepackage{multirow}
\usepackage{adjustbox}
\usepackage{booktabs}
\usepackage{array}
\usepackage{float}

\newcommand{\apjl}{The Astrophysical Journal Letters}
\newcommand{\apjs}{The Astrophysical Journal Supplement Series}

\newcommand{\aap}{Astronomy and Astrophysics}

\newcommand{\arcsec}{{$^{\prime\prime}$}}

\newcolumntype{C}{>{\footnotesize\raggedright\arraybackslash}c}

\newcolumntype{L}{>{\footnotesize\raggedright\arraybackslash}l}

\newcounter{mybodyfigure}
\newcounter{myedfigure}
\newcounter{mysdfigure}

\newcommand{\beginbodyfigures}{\renewcommand{figure}{\let\caption\NAT@figcaption}{{\themybodyfigure}}}
\newcommand{\beginedfigures}{\renewcommand{figure}{\let\caption\NAT@figcaption}{{\themyedfigure}}}
\newcommand{\beginsdfigures}{\renewcommand{figure}{\let\caption\NAT@figcaption}{{\themysdfigure}}}

\newcommand{\stepbodyfigure}{\refstepcounter{mybodyfigure}}
\newcommand{\stepedfigure}{\refstepcounter{myedfigure}}
\newcommand{\stepsdfigure}{\refstepcounter{mysdfigure}}

\newcommand{\bodyfigurelabel}[1]{\bf{Fig.\,\bodyfigure{#1}:}}
\newcommand{\edfigurelabel}[1]{\bf{Extended Data Fig.\edfigure{#1}:}}
\newcommand{\supfigurelabel}[1]{\bf{Supplementary Fig.\sdfigure{#1}:}}

\DeclareRobustCommand{\bodyfigure}[1]{\stepbodyfigure\label{#1}{\themybodyfigure}}
\DeclareRobustCommand{\edfigure}[1]{\stepedfigure\label{#1}{\themyedfigure}}
\DeclareRobustCommand{\sdfigure}[1]{\stepsdfigure\label{#1}{\themysdfigure}}

\makeatletter
\g@addto@macro\caption@prepareslc{%
  \renewcommand{\stepbodyfigure}{\caption@l@stepcounter{mybodyfigure}}}
\makeatother

\makeatletter
\g@addto@macro\caption@prepareslc{%
  \renewcommand{\stepedfigure}{\caption@l@stepcounter{myedfigure}}}
\makeatother

\makeatletter
\g@addto@macro\caption@prepareslc{%
  \renewcommand{\stepsdfigure}{\caption@l@stepcounter{mysdfigure}}}
\makeatother

\DeclareCaptionLabelFormat{supplementarytablelabel}{\bf{Supplementary Table #2}} % Define the table label format

\DeclareCaptionLabelFormat{tablelabel}{\bf{Extended Data Table #2}} % Define the table label format

\usepackage[super,comma,sort&compress]{natbib}

\usepackage{hyperref}
\usepackage{verbatim}

\usepackage{ulem} %provides various types of underlining that can stretch between words and be broken across lines
\usepackage{rotating} % rotate any object by arbitrary angle
\usepackage{soul}%h y p h e n -a t a b l e l e t t e r s p a c i n g ( s p a c i n g o u t ) , underlining and some derivatives such as overstriking and highlighting

\usepackage{color} %change color of text

\usepackage[colorinlistoftodos]{todonotes}

\makeatletter
\newsavebox\myboxA
\newsavebox\myboxB
\newlength\mylenA

\newcommand*\xoverline[2][0.75]{%
    \sbox{\myboxA}{$\m@th#2$}%
    \setbox\myboxB\null% Phantom box
    \ht\myboxB=\ht\myboxA%
    \dp\myboxB=\dp\myboxA%
    \wd\myboxB=#1\wd\myboxA% Scale phantom
    \sbox\myboxB{$\m@th\overline{\copy\myboxB}$}%  Overlined phantom
    \setlength\mylenA{\the\wd\myboxA}%   calc width diff
    \addtolength\mylenA{-\the\wd\myboxB}%
    \ifdim\wd\myboxB<\wd\myboxA%
       \rlap{\hskip 0.5\mylenA\usebox\myboxB}{\usebox\myboxA}%
    \else
        \hskip -0.5\mylenA\rlap{\usebox\myboxA}{\hskip 0.5\mylenA\usebox\myboxB}%
    \fi}
\makeatother

\usepackage{multibib}

\newcites{main}{\noindent{\bfseries \LARGE References}}
\newcites{methods}{\noindent{\bfseries \LARGE Additional References}}

\title{Signatures of Suppressed Matter Clustering revealed by Fast Radio Bursts}

\author{\large 
Kritti Sharma$^{1,^\ast}$, 
Elisabeth Krause$^{2,3}$, 
Vikram Ravi$^{1,4}$, 
Liam Connor$^{5}$, 
Dhayaa Anbajagane$^{6,7}$,  
Pranjal R. S.$^{2}$
}

\begin{document}

\maketitle

\begin{affiliations}
\item Cahill Center for Astronomy and Astrophysics, MC 249-17 California Institute of Technology, Pasadena, CA 91125, USA.
\item Department of Astronomy/Steward Observatory, University of Arizona, 933 North Cherry Avenue, Tucson, AZ 85721, USA
\item Department of Physics, University of Arizona, 1118 E Fourth Street, Tucson, AZ 85721, USA
\item Owens Valley Radio Observatory, California Institute of Technology, Big Pine, CA 93513, USA.
\item Center for Astrophysics | Harvard $\&$ Smithsonian, Cambridge, MA 02138-1516, USA
\item Department of Astronomy and Astrophysics, University of Chicago, Chicago, IL 60637, USA
\item Kavli Institute for Cosmological Physics, University of Chicago, Chicago, IL 60637, USA
\item[] $^{\ast}$Correspondence email: kritti@caltech.edu
\end{affiliations}

\begin{abstract}

Complex astrophysical processes regulate the growth of galaxies by injecting energy and momentum into their surroundings~\citep{2019OJAp....2E...4C}, redistributing baryons across megaparsec scales. The clustering of matter (baryons and dark matter) on these scales, as measured via weak gravitational lensing and galaxy surveys, encodes critical cosmological information -- including on the dynamical dark energy, the nature of dark matter and the sum of neutrino masses. The suppression of matter clustering due to feedback processes in galaxy formation  limits the interpretation of cosmological measurements~\citep{2025A&A...703A.158W, 2026arXiv260210065D}. Multiple probes of the cosmic baryon distribution, including X-ray~\citep{2022PASJ...74..175A, 2023MNRAS.526.6103K, 2024arXiv241116555P, 2025arXiv250910455S, 2025arXiv251202954S}, kinematic~\citep{2024MNRAS.534..655B, 2025arXiv251202954S} and thermal~\citep{2025arXiv250704476D, 2025arXiv250607432P} Sunyaev-Zel’dovich effect measurements, have attempted to quantify the strength of feedback via measurements of suppression in the matter power spectrum. The dispersion measures (DMs) of fast radio bursts~\citep{2022A&ARv..30....2P} (FRBs) have emerged as a powerful new probe of cosmic baryons, with the advantage over other probes of being unbiased with respect to density and temperature~\citep{2014ApJ...780L..33M}. Analyses of FRB samples have so far been limited to estimates of the diffuse-baryon density~\citep{2020Natur.581..391M}, and the partitioning of baryons between halos and the cosmic web~\citep{2024ApJ...973..151K, 2025NatAs...9.1226C}. Here, we use a sample of 109 FRBs with redshifts and DMs to directly measure the spatial fluctuations in the baryon density field, quantifying the effects of feedback on the matter power spectrum at scales of $k \sim 0.1-3~h$\,Mpc$^{-1}$, and the gas fraction in galaxy groups and clusters ($10^{13}-10^{15}~M_\odot$). We use a state-of-the-art halo-model prescription to conduct inference, and find that FRB data reduces the posterior variance at $k \sim 1~h$\,Mpc$^{-1}$  by a factor of $\sim 8$ relative to the prior. With just 109 FRBs, the statistical precision of inferred FRB constraints is similar to other baryon tracers, while probing a complementary redshift regime ($z \lesssim 0.3$). A comparison with several hydrodynamical simulations excludes extreme large-scale feedback scenarios like Illustris~\citep{2015A&C....13...12N} and OWLS-AGN~\citep{2010MNRAS.406..822M} at $\sim 2\sigma$ confidence. This work establishes FRBs as a sensitive probe of feedback-regulated structure formation. As next-generation experiments deliver orders-of-magnitude larger samples, FRBs are poised to drive the constraints on baryonic physics in the era of precision cosmology.

\end{abstract}

% Importance of breaking cosmology-feedback degeneracy with baryon probes - what can be measured with small scales - how they’re dealt with now - what the ideal future here is.

Large-scale structure surveys -- such as with the Vera Rubin Observatory (LSST)~\citep[][]{2019ApJ...873..111I} and Euclid~\citep{2011arXiv1110.3193L} -- are designed to map the cosmic matter distribution, with percent-level precision on the matter power spectrum, using weak gravitational lensing~\citep{2025A&A...703A.158W, 2026arXiv260210065D}. By extending analyses into the non-linear regime ($k \gtrsim 1\,h\,\mathrm{Mpc}^{-1}$), these Stage IV experiments aim to unlock the full statistical power of small-scale structure, enabling stringent tests of departures from the standard $\Lambda$CDM paradigm and delivering precise constraints on the dark energy equation of state and the sum of neutrino masses. However, the imprints of cosmological parameters, as encoded in the matter power spectrum at small scales $k \gtrsim 0.1\,h\,\mathrm{Mpc}^{-1}$, are strongly degenerate with galaxy formation processes. If unaccounted for, uncertainties in galaxy formation physics can significantly bias cosmological parameter inference, with expected biases in the dark energy equation-of-state parameters and the sum of neutrino masses reaching $5\sigma$ and $20\sigma$, respectively~\citep{2019OJAp....2E...4C}. To mitigate such biases, current weak lensing analyses either adopt conservative scale cuts~\citep{2025A&A...703A.158W, 2026arXiv260210065D}, thereby sacrificing a substantial fraction ($\sim 55\%$ signal-to-noise~\citep{2026arXiv260210065D}) of cosmological information encoded on small scales, or marginalize over data-driven astrophysical feedback models~\citep{2019JCAP...03..020S}, which inevitably weakens their constraining power. Independent insights on feedback-induced modifications to the small-scale matter power spectrum are therefore essential to fully exploit the potential of next-generation weak lensing surveys, while ensuring unbiased cosmological inference.

% Simulations of feedback - why feedback is important in galaxy formation - uncertainties on modes of feedback - why it’s not constrained observationally.

Cosmological simulations that explicitly model the hydrodynamic evolution of baryons alongside dark matter provide a physically motivated framework for studying the impact of galaxy formation processes on small-scale matter clustering. In these simulations, stellar and active galactic nucleus (AGN) feedback redistributes gas within halos and can eject it beyond the virial radius, thereby suppressing matter clustering on scales characteristic of galaxy groups and clusters ($\sim 0.1-1$~Mpc). Since these feedback mechanisms operate on scales far below the simulation resolution, they are implemented through phenomenological models calibrated to selected observables~\citep{2022PASJ...74..175A, 2023MNRAS.526.6103K, 2024arXiv241116555P, 2025arXiv250910455S}. However, the key aspects of baryon coupling efficiency and feedback implementation, including thermal versus kinetic AGN modes, bursty versus continuous energy injection and the treatment of gas ejection and re-accretion, remain uncertain. Consequently, despite being calibrated to reproduce benchmark observables, such as the galaxy stellar mass function, the star-forming main sequence, and cluster gas fractions, different hydrodynamical simulations predict markedly different amplitudes and scale dependence of matter power spectrum suppression over $k \sim 0.1-10\,h\,\mathrm{Mpc}^{-1}$~\citep{2019OJAp....2E...4C}. This large theoretical spread in small-scale clustering predictions underscores the need for direct observational probes of baryons to robustly constrain feedback physics.

% Baryon probes and their uncertainties/sensitivities - Galaxy surveys, X-ray, SZ: underlying assumptions and systematics - Discuss sensitivities in M/z/k - what results exist now - What does observational progress require in each probe.

Joint analyses of weak lensing with complementary baryonic tracers provide a powerful avenue to break the cosmology-feedback degeneracy. Combining lensing measurements with X-ray gas mass fractions in galaxy groups and clusters~\citep{2022PASJ...74..175A, 2023MNRAS.526.6103K, 2024arXiv241116555P, 2025arXiv250910455S, 2025arXiv251202954S}, diffuse X-ray emission from hot gas~\citep{2024PhRvL.133e1001F}, stacked kinematic Sunyaev-Zel’dovich (kSZ) effect profiles of group-scale halos~\citep{2024MNRAS.534..655B, 2025arXiv251202954S}, thermal Sunyaev-Zel’dovich (tSZ) effect-selected clusters~\citep[][]{2025arXiv250704476D} and Compton-$y$ maps~\citep[][]{2025arXiv250607432P} enables direct constraints on the distribution of baryons within halos, thereby disentangling astrophysical feedback effects from cosmological parameters. Rapid progress in both X-ray and microwave surveys positions joint lensing-baryon analyses as an increasingly mature and promising strategy for anchoring the small-scale matter power spectrum. Forthcoming higher-resolution, higher-sensitivity all-sky maps from eROSITA will improve the characterization of X-ray gas properties~\citep{2024A&A...682A..34M}. Similarly, next-generation multi-frequency CMB experiments, such as the Simons Observatory, in combination with spectroscopic galaxy surveys, will enhance signal-to-noise in SZ measurements and enable increasingly precise velocity-field reconstructions~\citep{2025JCAP...08..034A}. Each probe, however, is subject to distinct systematic uncertainties. X-ray gas fraction measurements are sensitive to modeling assumptions~\citep{2025arXiv251204203E} and sample selection effects~\citep{2022PASJ...74..175A, 2023MNRAS.526.6103K, 2024arXiv241116555P, 2025arXiv250910455S}; in particular, X-ray luminosity-selected clusters potentially exhibit higher hot gas fractions than optically-selected systems. Measurements of diffuse X-ray emission are additionally contaminated by unresolved AGN sources~\citep{2026arXiv260202484M}. Current kSZ constraints are limited by contamination from primary CMB anisotropies~\citep{2019PhRvD.100j3532M}, uncertainties from halo mis-centering, unknown satellite galaxy fractions, and finite beam resolution~\citep{2025MNRAS.536..572D}. Likewise, tSZ measurements are affected by cosmic infrared background (CIB) contamination~\citep{2025MNRAS.540.1055E}. Together, these considerations underscore both the promise and the complexity of joint lensing-baryon analyses, where continued progress hinges on incorporating new, statistically powerful baryon tracers with largely independent systematics.

% FRB DM-z relation as a baryon probe - method, including your previous papers - robustness to systematics, including DM-host evolution - results to date.

Fast Radio Bursts (FRBs), millisecond-duration transients of extragalactic origin~\citep{2022A&ARv..30....2P}, probe the distribution of baryons along cosmological sightlines. The dispersion measure (DM) imprinted on the FRB spectrum encodes the frequency-dependent dispersive delay induced by intervening plasma and directly traces the integrated column density of free electrons along the line of sight~\citep{2020Natur.581..391M}. The extragalactic contribution DM$_\mathrm{exgal}$, can be decomposed into a cosmic component, $\mathrm{DM}_\mathrm{cosmic}$, arising from the intergalactic medium (IGM) and intervening halos, together with the host galaxy contribution, $\mathrm{DM}_\mathrm{host}/(1+z_\mathrm{s})$, where DM$_\mathrm{host}$ is defined in the rest-frame of the source at redshift $z_\mathrm{s}$. Hydrodynamical simulations suggest that the rest-frame DM$_\mathrm{host}$ is well described by a log-normal distribution~\citep{2020ApJ...900..170Z}. While plausible redshift evolution in DM$_\mathrm{host}$, expected if FRBs trace star-formation~\citep{2024Natur.635...61S}, does not impact current measurements, it may become an important systematic for next-generation experiments (see Extended Data Fig.~\ref{fig:HMcode_cornerplot} and Methods). Likewise, selection effects arising from the FRB luminosity function, source redshift evolution, and instrument DM selection functions are not expected to bias present constraints, but will require careful modeling as FRB catalogs grow~\citep{2026ApJ...999..202S}. After accounting for the host contributions, the remaining sightline-to-sightline variance in $\mathrm{DM}_\mathrm{cosmic}$ arises from fluctuations in the ionized gas distribution within intervening collapsed structures~\citep{2014ApJ...780L..33M}. This variance, denoted as $\sigma[\mathrm{DM}_\mathrm{cosmic}(z_\mathrm{s})]$, is directly governed by the feedback-dependent electron power spectrum $P_\mathrm{ee}(k,z)$ (see Equation~\ref{eqn:variance} in Methods) with tight correlation to the suppression in matter power spectrum upto scales of $k \lesssim 10~h$~Mpc$^{-1}$, thus providing a sensitive probe of small-scale baryonic physics~\citep{2025ApJ...989...81S}.

% FRB sensitivity in comparison with other probes - complementary nature, what can be probed - observational prospects for future surveys.

We illustrate the complementary nature of FRBs as a novel probe of baryonic structure in Fig.~\ref{fig:sensitivity_analysis}, where we compare the halo mass ($M_{200}$), redshift ($z$) and scale ($k$) sensitivity of the FRB observable, $\sigma[\mathrm{DM}_\mathrm{cosmic}(z)]$, with that of conventional baryon probes (see Methods for details of the sensitivity calculations). Current cosmic shear ($\xi_\pm$) measurements, such as those from the Dark Energy Survey (DES-Y3)~\citep[][]{2022PhRvD.105b3514A} require precise knowledge of baryon distribution in $M_\mathrm{200} \sim 10^{13}-10^{15}\,M_\odot$ halos up to redshifts $z \sim 0.5$, necessitating stringent scale cuts and restricting reliable interpretation to $k \lesssim 1~h$\,Mpc$^{-1}$. This effectively discards the rich small-scale information at $k \sim 1-10~h$\,Mpc$^{-1}$, where baryonic feedback effects dominate. Recent studies have lifted this limitation by combining weak lensing with complementary probes, including cross-correlations with the ACT-Planck Compton $y$-parameter ($\xi^{\gamma y}$)~\citep[][]{2025arXiv250607432P}, the ACT galaxy cluster tSZ $Y-M$ relation~\citep{2025arXiv250704476D}, X-ray observations of eROSITA eRASS1 galaxy clusters~\citep{2025arXiv251202954S}, and ACT kSZ stacks on the DESI bright galaxy (BGS) and luminous red galaxy (LRG) samples~\citep{2024MNRAS.534..655B, 2025arXiv251202954S}. These probes collectively cover the halo mass and redshift ranges relevant for precise weak-lensing interpretation. FRBs offer a complementary view: the sensitivity of $\sigma[\mathrm{DM}_\mathrm{cosmic}(z)]$ peaks at low redshifts, with the top two-thirds of sensitivity extending to $z \sim 0.3$, and probes electron distributions on scales $0.1-3~h$\,Mpc$^{-1}$. Their halo mass sensitivity ($M_\mathrm{200} \sim 10^{13.5}-10^{15}\,M_\odot$) coincides with the dominant contributors to matter power spectrum suppression. With Stage-IV weak lensing experiments~\citep[][]{2019ApJ...873..111I} extending sensitivity down to $M_\mathrm{200} \sim 10^{12}\,M_\odot$ at $z \lesssim 1$, FRBs will provide a complementary probe across the full halo mass and redshift range (see Extended Data Fig.~\ref{fig:sensitivity_forecast}), enabled by next-generation facilities such as the Deep Synoptic Array (DSA)~\citep[][]{2019BAAS...51g.255H}, the Square Kilometer Array (SKA)~\citep[][]{2004NewAR..48..979C}, and the Canadian Hydrogen Observatory and Radio-Transient Detector (CHORD)~\citep[][]{2019clrp.2020...28V}.

% Halo model inference with the current FRB sample - E.g., start paragraph with “We use a halo model framework to perform inference on the existing sample of FRBs with secure host galaxies...” - The introduce the sample in a few sentences, and then get into the halo model.

We perform inference on the existing sample of FRBs with secure host associations and spectroscopic redshifts, leveraging a halo model framework. Our FRB compilation (summarized in Extended Data Table~\ref{table:FRBsample}) includes 87 sub-arcsecond to arcsecond-scale localizations from ASKAP, CHIME, DSA-110, FAST, MeerKAT, Parkes and VLA-RealFast, and 22 sub-arcminute to arcminute-scale localizations from CHIME at redshifts $z \lesssim 0.3$. We show the DM$_\mathrm{exgal}-z$ distribution of this sample in Fig.~\ref{fig:dmz}. For Bayesian inference, we model the feedback-dependent power spectrum using \textsc{BCEmu7} halo model, where gas profiles are analytically modeled as a cored radial profile truncated at ejection radius, $\theta_\mathrm{ej}$. The mass-dependent slope of the profile is governed by $\log M_c$, the characteristic mass below which the profile becomes shallower than the NFW form, and by $\mu_\beta$, which controls the mass scaling of this slope. The parameters $\delta$ and $\gamma$ regulate the slope beyond truncation. The total gas fraction is tied to the total stellar fraction, with $\eta$ and $A$ denoting the slope and normalization of the stellar-to-halo mass (SHM) relation. The stellar mass locked in the central galaxy is modulated by $\eta_\delta$. The efficacy of these profiles in reproducing gas power spectra, matter power spectrum suppression and gas fractions in hydrodynamical simulations is well established~\citep{2021JCAP...12..046G}. We summarize the free parameters in our model in Extended Data Table~\ref{table:parameters} and key equations in the Methods.

% Best-fit P(DM$|$z) - Compare DM$_\mathrm{host}$ results with other works - What can then be inferred (focus on gas fractions and power-spectrum suppression).

A covariant principal components analysis of the inferred \textsc{BCEmu7} posterior distribution quantifies the effective dimensionality constrained by FRB data, $N_{\mathrm{eff}} \simeq 1.2$, with a $224\%$ reduction in the variance of the dominant eigenmode relative to prior (see Extended Data Fig.~\ref{fig:BCEmu_cornerplot}, \ref{fig:spk_BCEmu7_CPC} and Supplementary Fig.~\ref{fig:drop_one_out_validation}). The leading Karhunen-Lo\`eve mode projects primarily onto $\log M_c$, accounting for $\sim 55\%$ of its variance, while $\eta$, $\gamma$, and $\eta_\delta$ project almost entirely onto unconstrained modes and remain prior dominated. Motivated by these results, we introduce reduced-complexity models: \textsc{BCEmu1}, in which only $\log M_c$ is allowed to vary, and \textsc{BCEmu4}, where $\log M_c$, $\theta_\mathrm{ej}$, $\mu_\beta$, and $\delta$ are varied while the remaining parameters are fixed to fiducial values. Since the assumed $\eta$ value impacts the inferred power spectrum suppression constraints (see Extended Data Fig.~\ref{fig:spk_vary_feedback_cosmology}), we adopt a physically informed prior on $\eta$ using SHM measurements of galaxy clusters from literature~\citep{2018AstL...44....8K}, defining the baseline \textsc{BCEmu5} model for our results (see Extended Data Fig.~\ref{fig:eta_prior}). We have validated the ability of our analysis pipeline to discriminate between extreme feedback scenarios using synthetic data (see Supplementary Fig.~\ref{fig:extreme_scenario_test}). We show the best-fit $p(\mathrm{DM}_\mathrm{exgal}|z_{\mathrm{s}})$ distribution in Fig.~\ref{fig:dmz} and summarize parameter constraints in Extended Data Table~\ref{table:HMcode_BCEmu_constraints}. The measured mean $\langle \mathrm{DM}_\mathrm{host} \rangle = 128.60_{-17.50}^{+18.33}$~pc\,cm$^{-3}$ and variance $\sigma[\mathrm{DM}_\mathrm{host}] = 150.25_{-25.66}^{+28.88}$~pc\,cm$^{-3}$ of the rest-frame host galaxy contribution are consistent with both observational constraints \citep{2024ApJ...973..151K, 2025NatAs...9.1226C} and our theoretical expectations for FRBs originating in star-forming galaxies~\citep{2024Natur.635...61S}. From MCMC posterior distributions of halo gas profile parameters, we derive two physical quantities: (i) gas density profiles and gas mass fractions in groups and clusters, thus partitioning baryons between IGM and halos (Fig.~\ref{fig:fgas_rhogas}), and 
(ii) the resulting baryon-induced suppression of the matter power spectrum (Fig.~\ref{fig:spk_constraints}). Given the current constraining power of baryon probes, both these metrics are robust and unbiased probes of feedback strength in the Universe (see Methods).

% Gas fractions - inference - comparison with other work - discuss pre/post eROSITA observational uncertainty.

The gas mass fractions in galaxy groups and clusters inferred from FRB observations favor more moderate feedback scenarios than those implied by hot gas fraction measurements from eROSITA X-ray stacks of clusters~\citep{2024arXiv241116555P, 2025arXiv250910455S}, with $1.9\sigma$ confidence for $10^{14}~M_\odot$ halos. 
Gas fractions derived from HSC-XXL~\citep{2022PASJ...74..175A, 2023MNRAS.526.6103K} X-ray-selected clusters are likewise higher, potentially reflecting modeling assumptions~\citep{2025arXiv251204203E} and selection effects~\citep{2024arXiv241116555P} that bias the inference of X-ray observations. Importantly, FRBs probe the total ionized gas content, whereas X-ray observations are primarily sensitive to the hot, dense phase of the intragroup medium (IGrM) and intracluster medium (ICM), due to Bremsstrahlung emission, which scales as $L_X \propto n_e^2 T^{0.5}$, with electron density $n_e$ and temperature $T$. The higher total gas fractions inferred from FRBs relative to eROSITA measurements are not surprising, as FRBs are sensitive to diffuse, lower-density gas that contributes negligibly to X-ray emission. This interpretation is consistent with stacked tSZ measurements of thermal pressure profiles in the circumgalactic medium (CGM) of group-scale halos, which show no compelling evidence for flattened pressure profiles due to extreme heating, nor for a significant baryon reservoir displaced far beyond $R_{200}$~\citep{2025ApJ...991..205D}. The $\sim 0.6-0.8\sigma$ preference for systematically larger gas densities within $R_{200}$ relative to comparable halo model-based analyses of stacked kSZ measurements from ACT DR5~\citep{2024MNRAS.534..655B} and Compton-$y$–cosmic shear cross-correlations~\citep{2025arXiv250607432P} may indicate more extreme feedback operating at higher redshifts probed by the kSZ and tSZ observations. Together, these results suggest a broadly consistent picture in which a substantial fraction of baryons reside in a cooler, diffuse phase, implying that halos may retain much of their baryonic content in the local universe, with feedback processes likely becoming progressively more efficient at higher redshifts. Future FRB samples will enable more fine-grained tests of this picture, allowing us to probe baryonic processes in halos across cosmic time.

% Power spectrum - comparison with other work - comparison with simulations, and what can be inferred on feedback / suppression methods (axions)

The level of matter power spectrum suppression implied by FRB observations disfavors scenarios with extreme large-scale feedback operating in the local universe at $\sim 2\sigma$ confidence. The FRB data contributes $\sigma^2_\mathrm{prior}/\sigma^2_\mathrm{posterior} \sim 8$ reduction in the variance of posterior relative prior at scale $k \sim 1~h\,\mathrm{Mpc}^{-1}$, underscoring their constraining power on small scales. We also exclude a zero-feedback scenario with $1.6\sigma$ significance at $k \sim 1~h\,\mathrm{Mpc}^{-1}$, finding no evidence for feedback strengths within the power spectrum suppression parameter space previously ruled out using independent techniques~\citep{2025ApJ...991L..25L, 2025arXiv250717742R}. Our measurement is consistent with constraints from cosmic shear~\citep{2025arXiv250903582A, 2025A&A...703A.158W, 2025arXiv251025596X} and X-ray observations~\citep{2025arXiv251202954S} at $<0.3\sigma$-level, while favoring more moderate feedback (at $\sim 0.7\sigma$ confidence level) than constraints derived from tSZ~\citep{2025arXiv250704476D, 2025arXiv250607432P} and kSZ~\citep{2024MNRAS.534..655B, 2025arXiv251202954S} measurements. This difference may again reflect the distinct redshift sensitivities of these probes: FRB and X-ray observations primarily constrain lower-redshift halos, whereas SZ measurements preferentially access higher-redshift regimes (Fig.~\ref{fig:sensitivity_analysis}), supporting progressively inefficient feedback beyond $\sim 10^{13}\,M_\odot$ halos at lower redshifts. These results motivate the development of halo model prescriptions that explicitly allow for redshift-dependent feedback beyond the implicit evolution encoded through the halo mass function, enabling unbiased multi-probe joint analyses in the Stage-IV era. When compared with hydrodynamical simulations, our inferred suppression at $k\sim1~h$\,Mpc$^{-1}$ is consistent (at $< 1\sigma$ level) with BAHAMAS, SIMBA, and FLAMINGO (both, fiducial and $f_\mathrm{gas}$-$8\sigma$ variant), while it is discrepant with Illustris~\citep{2015A&C....13...12N} ($2.6\sigma$), OWLS-AGN~\citep{2010MNRAS.406..822M} ($1.8\sigma$) and IllustrisTNG~\citep{2019ComAC...6....2N} ($1.4\sigma$). Interpreting these results in the context of AGN feedback requires careful consideration of other physical mechanisms capable of producing similar signatures in the matter power spectrum. For instance, cosmic rays can drive extended outflows that suppress gas accretion without excessive CGM heating, yielding reduced central gas densities, flatter gas profiles, and diminished small-scale clustering~\citep{2025OJAp....8E..66Q}. Similarly, the nature of dark matter, such as ultra-light axions, can suppress structure formation below the Jeans wavenumber ($k \sim 8~h\,\mathrm{Mpc}^{-1}$)~\citep{2025MNRAS.542.2698P}. Precise measurements of matter power spectrum suppression therefore provide a powerful avenue for disentangling baryonic feedback, cosmic-ray-driven processes, and exotic dark matter physics.

By probing the total ionized gas content, FRBs offer an independent census of baryons across halo masses. Our inferred baryon partition shows agreement with previous FRB-based measurements, revealing a coherent picture of baryon distribution across clusters, groups, and lower-mass halos, emerging from FRB analyses. For cluster-scale halos ($M \geq 10^{14}\,M_\odot$), we obtain an ICM baryon fraction of $f_\mathrm{ICM} = 0.035^{+0.005}_{-0.007}$ (the uncertainties denote 68\% confidence interval), consistent with $f_\mathrm{ICM} = 0.0375^{+0.005}_{-0.005}$, derived from cluster mass functions combined with Chandra halo baryon fraction relations~\citep{2025NatAs...9.1226C}. At group scales ($5\times10^{12} \leq M < 10^{14}\,M_\odot$), we infer $f_\mathrm{IGrM} = 0.048^{+0.019}_{-0.024}$, in agreement with Chandra-derived gas mass fraction estimate $f_\mathrm{IGrM} = 0.054^{+0.010}_{-0.010}$~\citep{2025NatAs...9.1226C}. For lower-mass halos ($10^9 \leq M < 5\times10^{12}\,M_\odot$), we measure a CGM fraction of $f_\mathrm{CGM} = 0.013^{+0.031}_{-0.010}$, consistent with FRB inference calibrated against hydrodynamical simulations~\citep{2025NatAs...9.1226C}. We further find that the combined fraction $f_\mathrm{CGM}+f_\mathrm{IGrM} = 0.06^{+0.03}_{-0.04}$ is statistically consistent with measurements from density field reconstruction in the foreground of FRBs~\citep{2024ApJ...973..151K}. 
Collectively, these comparisons reinforce the robustness and coherence of baryon inventory derived from diverse independent FRB-based methodologies (see Methods).

% Forecasts for larger FRB samples - What can be done with more DM/z measurements, and what systematics need to be addressed (including in modeling). Show a figure with different scenarios, based on your previous papers. - How does this compare with next-gen X-ray/galaxy/SZ probes, including in their systematics (say what you think here!) - and there’s more! importance of cross-correlation studies with even larger samples, prospects for joint inference with WL

Looking ahead, the scientific return of FRB cosmology is poised to scale rapidly with the orders-of-magnitude growth in well-localized bursts anticipated from next-generation facilities such as DSA~\citep{2019BAAS...51g.255H}, SKA~\citep{2004NewAR..48..979C} and CHORD~\citep{2019clrp.2020...28V}. Precise DM-redshift (including photometric redshift) measurements will be possible in synergy with upcoming galaxy surveys, including LSST~\citep[][]{2019ApJ...873..111I} and Euclid~\citep{2011arXiv1110.3193L}. Realizing the immense potential of this dataset requires careful control of systematics in DM$-z$ analyses~\citep{2026ApJ...999..202S}. The redshift evolution of DM$_\mathrm{host}$ and source population, instrument selection effects, and propagation-induced biases must be modeled or empirically calibrated using correlations with host galaxy properties~\citep{2025A&A...696A..81B} and FRB-level observables~\citep{2026ApJ...998....1M}. However, these systematics can be effectively mitigated by performing statistical cross-correlations with large-scale structure, where FRBs act as backlights that trace intervening matter~\citep{2026ApJ...998..252C}. In parallel, emerging model-agnostic techniques provide a promising route to circumvent uncertainties associated with non-linear clustering modeling~\citep{2025arXiv250919514L}. The expected $\sim 10^4$ localized FRBs with extended redshift baseline ($z \lesssim 1$) will enable stacked measurements of gas profiles down to galaxy-scale halos ($M_{200} \gtrsim 10^{11}~M_\odot$), and cross-correlations with large-scale structure surveys (see Extended Data Fig.~\ref{fig:sensitivity_forecast})~\citep{2019PhRvD.100j3532M, 2026ApJ...998..109S}. Tomographic stacking of FRB sightlines on galaxy samples will directly chart the redshift evolution of galaxy formation processes, while joint analyses with galaxy clustering, galaxy-galaxy lensing and cosmic shear will enable stringent tests of extensions to $\Lambda$CDM cosmological model in a unified framework (see Extended Data Fig.~\ref{fig:fisher_forecast}), pinning down the sum of neutrino masses ($\Delta~\Sigma m_\nu \sim 0.12$~eV) and dynamical dark energy ($\Delta w_0 \sim 0.08, \Delta w_a \sim 0.4$). Moreover, the sensitivity of FRBs to matter clustering positions them as an avenue for addressing the longstanding $S_8$ ($\Delta \sigma_8 \sim 0.006$) tension~\citep{2025PDU....4901965D}, bridging the gap between cosmology and astrophysics. In concert with tSZ and kSZ measurements from next generation CMB experiments such as the Simons Observatory, FRBs will enable a comprehensive census of baryons across the full thermodynamic spectrum of cosmic gas. Beyond serving as an independent probe, FRBs can also calibrate key systematic uncertainties in other observables -- for example, breaking the optical-depth degeneracy that limits the accuracy of large-scale velocity reconstruction from kSZ effect~\citep{2025arXiv250621657H}, enhancing the constraining power of upcoming CMB experiments on primordial non-Gaussianity, cosmic growth rate and amplitude of density fluctuations by an order-of-magnitude~\citep{2019PhRvD.100j3532M}. FRBs will become an essential component of multi-probe efforts over the next decade, unlocking a uniquely precise and decisive window onto new physics.

\newpage

\begin{figure}[ht!]
    \includegraphics[width=\textwidth]{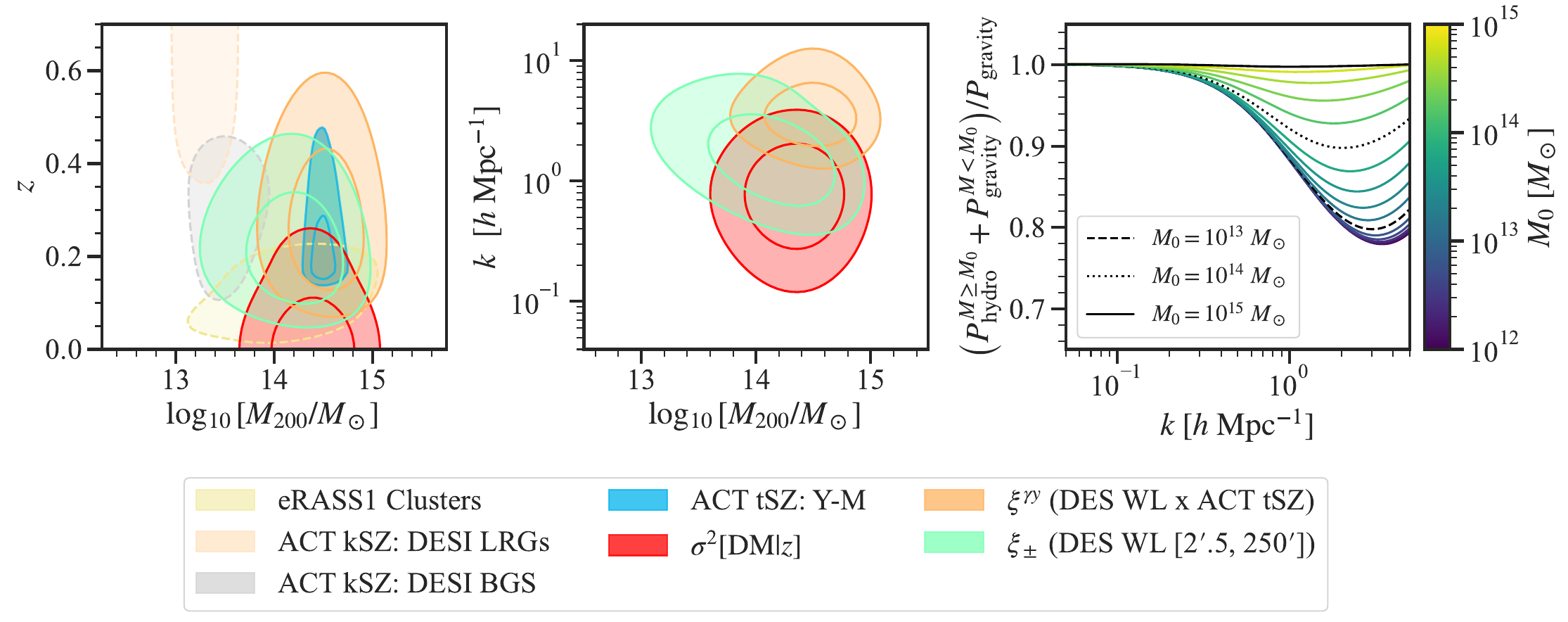}
    \caption{{\bodyfigurelabel{fig:sensitivity_analysis}} \textbf{Investigation of the distinct feedback regimes probed by baryon tracers and their role in unveiling the small-scale matter power spectrum suppression.} The halo mass ($M_{\mathrm{200}}$), redshift ($z$) and scale ($k$) sensitivity of the imprint of feedback on various baryon tracers are shown in panels \textbf{a} and \textbf{b}. The sensitivity of cosmic shear ($\xi_\pm$)~\citep{2022PhRvD.105b3514A}, Compton $y$-parameter--cosmic shear cross-correlations ($\xi^{\gamma y}$)~\citep{2025arXiv250607432P} and tSZ Y-M relation~\citep{2025arXiv250704476D} are computed using their analysis likelihoods and covariance matrices, with contribution from the one-halo term to the observable. The contours correspond to top one-third and two-thirds of the sensitivity. For X-ray observations and kSZ effect, we show the 86th percentile contours for $M_{\mathrm{200}}$ and $z$ distributions of eROSITA eRASS1 clusters~\citep{2025arXiv250910455S} and kSZ stacks on DESI BGS and LRG samples~\citep{2025arXiv250910455S}. All probes are sensitive to halo mass range of $M_\mathrm{200} \sim 10^{13}-10^{15}\,M_\odot$, with complementary redshift and scale sensitivity. The halo mass sensitivity of matter power spectrum suppression at various scales is illustrated in panel \textbf{c} through the ratio of modified to gravity-only power spectra, where in the modified power spectrum construction, we assume that only halos above $M_0$ are impacted by baryonic feedback. The $M_\mathrm{200} > 10^{13}\,M_\odot$ halo mass regime, probed by most baryon tracers, contributes dominantly to the suppression of matter power spectrum at scales $k \sim 0.1-5~h$\,Mpc$^{-1}$.}
\end{figure}

\newpage
\clearpage
\newpage
\newpage

\begin{figure}[h]
    \centering
    \includegraphics[width=0.6\textwidth]{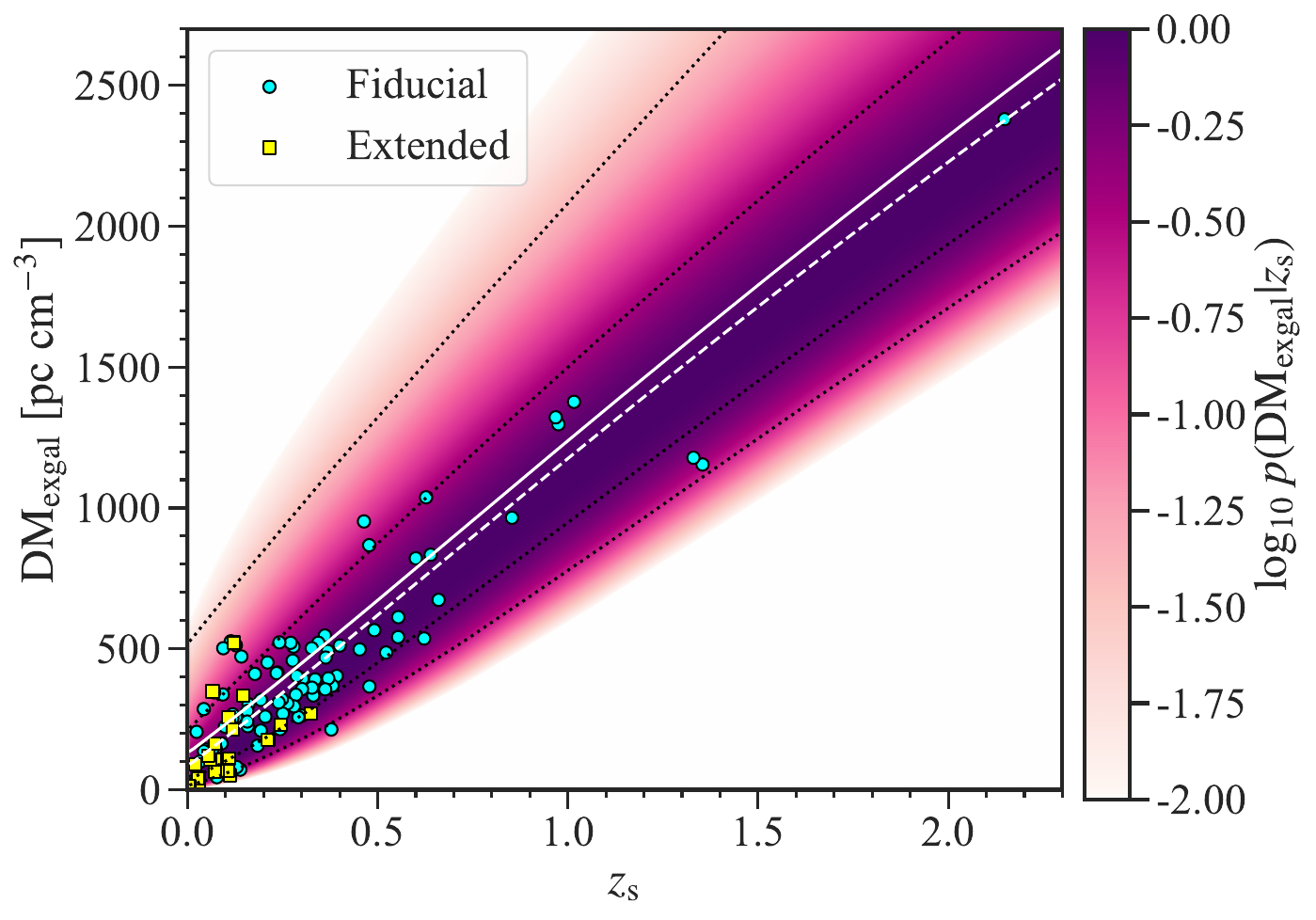}
    \caption{{\bodyfigurelabel{fig:dmz}} \textbf{The observed sightline-to-sightline variance in FRB dispersion measures as a probe of baryon clustering.} The heatmap corresponds to the best-fit $p(\mathrm{DM}_\mathrm{exgal}|z_\mathrm{s})$ distribution, as inferred from the halo model prescription with flexible analytical gas profiles. Published FRBs with sub-arcsecond to arcsecond-scale localizations, confident host associations ($P_\mathrm{host} \geq 0.9$) and spectroscopic redshifts are shown as blue points (see Extended Data Table~\ref{table:FRBsample}). These include FRBs from various instruments: ASKAP, CHIME, DSA-110, FAST, MeerKAT, Parkes and VLA-RealFast. In our extended sample, we also include sub-arcminute to arcminute-scale localizations by CHIME at redshifts $z \lesssim 0.3$ with confident host associations, which are shown as yellow points. The mean (solid line), median (dashed line) and 68/95th percentiles (dotted lines) are shown for reference. The variance (i.e. spread in DM for fixed $z$), quantified by the degree of clustering of baryons, arises from intervening halos and is sensitive to the strength of feedback~\citep{2025ApJ...989...81S}.}
\end{figure}

\newpage
\clearpage
\newpage
\newpage

\begin{figure}[ht!]
    \includegraphics[width=\textwidth]{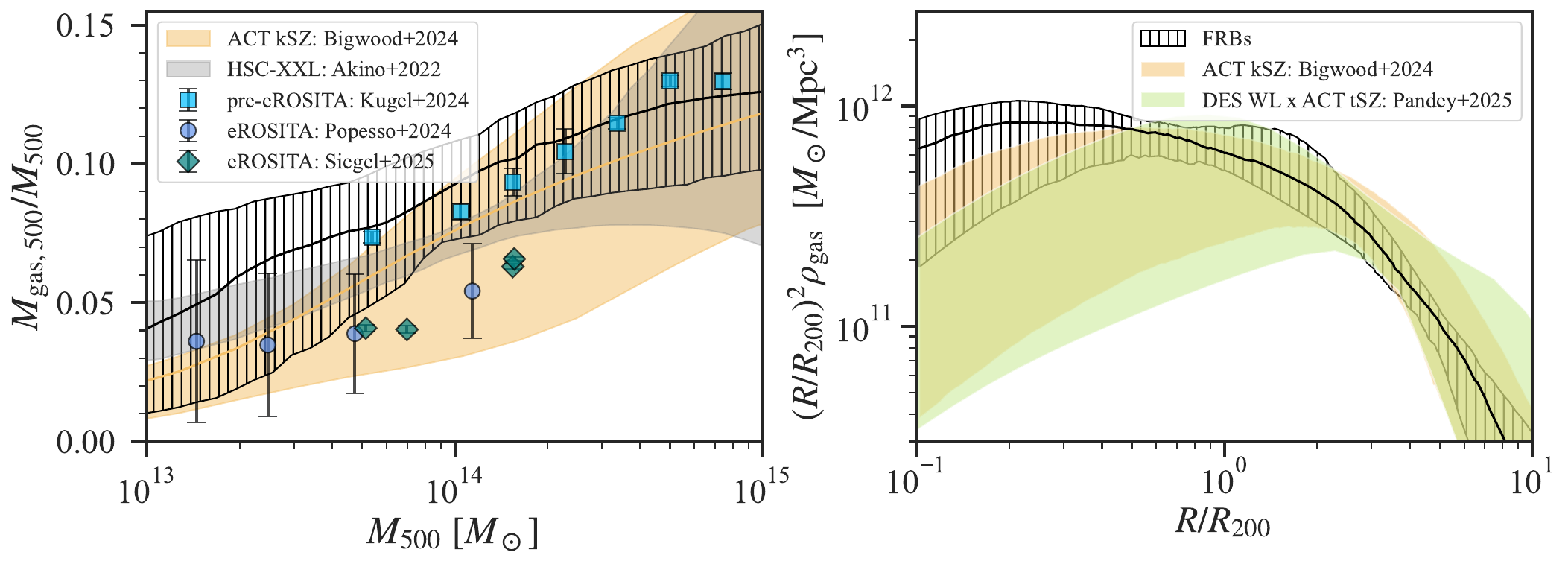}
    \caption{{\bodyfigurelabel{fig:fgas_rhogas}} \textbf{Inferred gas mass fractions and halo gas profiles.} The constraint on the fraction of mass in gas in galaxy groups and clusters, $M_\mathrm{gas,500}/M_{500}$, as a function of halo mass $M_{500}$, obtained from the posterior distribution of analytical gas profile parameters inferred with FRBs, is shown in panel \textbf{a} (68\% confidence interval and median). The measurements in literature from other probes of baryons, including X-ray constraints from HSC-XXL~\citep{2022PASJ...74..175A}, pre-eROSITA~\citep{2023MNRAS.526.6103K} and eROSITA~\citep{2024arXiv241116555P, 2025arXiv250910455S} stacks, and kSZ constraints from ACT DR5 stacks~\citep{2024MNRAS.534..655B}, are shown for comparison. The FRB observations indicate gas fractions that are $\sim 1.9\sigma$ larger compared to recent stacked measurements from eROSITA in $10^{14}\,M_\odot$ halos~\citep{2024arXiv241116555P, 2025arXiv250910455S}. This suggests a larger cool gas content within halos that is not strongly X-ray luminous. The constraint on the gas profile of $M_\mathrm{200}=10^{14}\,M_\odot$ halo is compared with constraints from similar halo model-based inferences conducted using kSZ effect ACT DR5 stacks~\citep{2024MNRAS.534..655B} and Compton $y$-parameter--cosmic shear cross-correlations~\citep{2025arXiv250607432P} in panel~\textbf{b} (68\% confidence interval). The $\sim 0.6-0.8\sigma$ preference for higher gas mass density within $R_{200}$ compared to recent kSZ~\citep{2024MNRAS.534..655B} and tSZ~\citep{2025arXiv250607432P} constraints, points toward relatively weaker feedback in the local Universe, as probed by FRBs.}
\end{figure}

\newpage
\clearpage
\newpage
\newpage

\begin{figure}[ht!]
    \includegraphics[width=\textwidth]{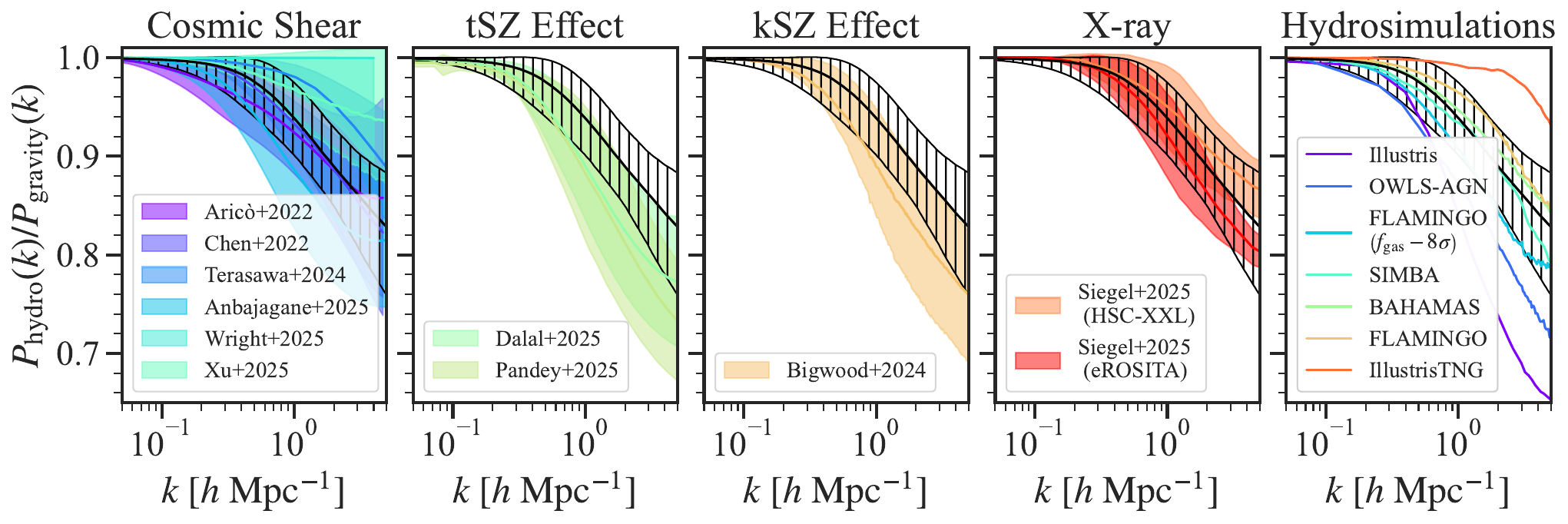}
    \caption{{\bodyfigurelabel{fig:spk_constraints}} \textbf{Comparison of the inferred matter power spectrum suppression with constraints from other baryon probes and hydrodynamical simulations.} We show our constraints as the median and 68\% confidence interval from the flexible analytical gas profiles halo model (vertical hatch). The entropy of the posterior relative to the prior distribution implies that the FRB data contribute an information gain of $\sigma^2_\mathrm{prior}/\sigma^2_\mathrm{posterior} \sim 8$ at scale $k \sim 1~h\,\mathrm{Mpc}^{-1}$.  Overall, the agreement with cosmic shear~\citep{2025arXiv250903582A, 2025A&A...703A.158W, 2025arXiv251025596X} and X-ray observations~\citep{2025arXiv251202954S} (at $<0.3\sigma$ level), with a $\sim 0.7\sigma$ preference for more moderate feedback than inferred from recent tSZ~\citep{2025arXiv250704476D, 2025arXiv250607432P} and kSZ effect~\citep{2024MNRAS.534..655B, 2025arXiv251202954S} measurements is evident. Comparison with hydrodynamical simulations at $k \sim 1~h\,\mathrm{Mpc}^{-1}$ reveals agreement at $<1\sigma$-level with BAHAMAS, SIMBA and FLAMINGO (including the $f_\mathrm{gas}-8\sigma$ variant), while indicating feedback stronger than IllustrisTNG~\citep{2019ComAC...6....2N} ($1.4\sigma$) and weaker than Illustris ($2.6\sigma$)~\citep{2015A&C....13...12N} and OWLS-AGN ($1.8\sigma$)~\citep{2010MNRAS.406..822M}.}
\end{figure}

\newpage
\clearpage
\newpage
\newpage

\vspace{0.5cm}  
\noindent{\bfseries \LARGE References}\setlength{\parskip}{12pt}
\setlength{\parskip}{17pt}%
\bibliographystyle{naturemag}
\bibliography{manuscript}

\begin{thebibliography}{10}
\urlstyle{rm}
\expandafter\ifx\csname url\endcsname\relax
  \def\url#1{\texttt{#1}}\fi
\expandafter\ifx\csname urlprefix\endcsname\relax\def\urlprefix{URL }\fi
\expandafter\ifx\csname doiprefix\endcsname\relax\def\doiprefix{DOI: }\fi
\providecommand{\bibinfo}[2]{#2}
\providecommand{\eprint}[2][]{\url{#2}}

\bibitem{2019OJAp....2E...4C}
\bibinfo{author}{{Chisari}, N.~E.} \emph{et~al.}
\newblock \bibinfo{journal}{\bibinfo{title}{{Modelling baryonic feedback for survey cosmology}}}.
\newblock {\emph{\JournalTitle{The Open Journal of Astrophysics}}} \textbf{\bibinfo{volume}{2}}, \bibinfo{pages}{4}, \doiprefix\url{https://dx.doi.org/10.21105/astro.1905.06082} (\bibinfo{year}{2019}).
\newblock \eprint{1905.06082}.

\bibitem{2025A&A...703A.158W}
\bibinfo{author}{{Wright}, A.~H.} \emph{et~al.}
\newblock \bibinfo{journal}{\bibinfo{title}{{KiDS-Legacy: Cosmological constraints from cosmic shear with the complete Kilo-Degree Survey}}}.
\newblock {\emph{\JournalTitle{\aap}}} \textbf{\bibinfo{volume}{703}}, \bibinfo{pages}{A158}, \doiprefix\url{https://dx.doi.org/10.1051/0004-6361/202554908} (\bibinfo{year}{2025}).
\newblock \eprint{2503.19441}.

\bibitem{2026arXiv260210065D}
\bibinfo{author}{{DES Collaboration}} \emph{et~al.}
\newblock \bibinfo{journal}{\bibinfo{title}{{Dark Energy Survey Year 6 Results: Cosmological Constraints from Cosmic Shear}}}.
\newblock {\emph{\JournalTitle{arXiv e-prints}}} \bibinfo{pages}{arXiv:2602.10065}, \doiprefix\url{https://dx.doi.org/10.48550/arXiv.2602.10065} (\bibinfo{year}{2026}).
\newblock \eprint{2602.10065}.

\bibitem{2022PASJ...74..175A}
\bibinfo{author}{{Akino}, D.} \emph{et~al.}
\newblock \bibinfo{journal}{\bibinfo{title}{{HSC-XXL: Baryon budget of the 136 XXL groups and clusters}}}.
\newblock {\emph{\JournalTitle{\pasj}}} \textbf{\bibinfo{volume}{74}}, \bibinfo{pages}{175--208}, \doiprefix\url{https://dx.doi.org/10.1093/pasj/psab115} (\bibinfo{year}{2022}).
\newblock \eprint{2111.10080}.

\bibitem{2023MNRAS.526.6103K}
\bibinfo{author}{{Kugel}, R.} \emph{et~al.}
\newblock \bibinfo{journal}{\bibinfo{title}{{FLAMINGO: calibrating large cosmological hydrodynamical simulations with machine learning}}}.
\newblock {\emph{\JournalTitle{\mnras}}} \textbf{\bibinfo{volume}{526}}, \bibinfo{pages}{6103--6127}, \doiprefix\url{https://dx.doi.org/10.1093/mnras/stad2540} (\bibinfo{year}{2023}).
\newblock \eprint{2306.05492}.

\bibitem{2024arXiv241116555P}
\bibinfo{author}{{Popesso}, P.} \emph{et~al.}
\newblock \bibinfo{journal}{\bibinfo{title}{{The hot gas mass fraction in halos. From Milky Way-like groups to massive clusters}}}.
\newblock {\emph{\JournalTitle{arXiv e-prints}}} \bibinfo{pages}{arXiv:2411.16555}, \doiprefix\url{https://dx.doi.org/10.48550/arXiv.2411.16555} (\bibinfo{year}{2024}).
\newblock \eprint{2411.16555}.

\bibitem{2025arXiv250910455S}
\bibinfo{author}{{Siegel}, J.} \emph{et~al.}
\newblock \bibinfo{journal}{\bibinfo{title}{{Joint X-ray, kinetic Sunyaev-Zeldovich, and weak lensing measurements: toward a consensus picture of efficient gas expulsion from groups and clusters}}}.
\newblock {\emph{\JournalTitle{arXiv e-prints}}} \bibinfo{pages}{arXiv:2509.10455}, \doiprefix\url{https://dx.doi.org/10.48550/arXiv.2509.10455} (\bibinfo{year}{2025}).
\newblock \eprint{2509.10455}.

\bibitem{2025arXiv251202954S}
\bibinfo{author}{{Siegel}, J.} \emph{et~al.}
\newblock \bibinfo{journal}{\bibinfo{title}{{The suppression of the matter power spectrum: strong feedback from X-ray gas mass fractions, kSZ effect profiles, and galaxy-galaxy lensing}}}.
\newblock {\emph{\JournalTitle{arXiv e-prints}}} \bibinfo{pages}{arXiv:2512.02954}, \doiprefix\url{https://dx.doi.org/10.48550/arXiv.2512.02954} (\bibinfo{year}{2025}).
\newblock \eprint{2512.02954}.

\bibitem{2024MNRAS.534..655B}
\bibinfo{author}{{Bigwood}, L.} \emph{et~al.}
\newblock \bibinfo{journal}{\bibinfo{title}{{Weak lensing combined with the kinetic Sunyaev-Zel'dovich effect: a study of baryonic feedback}}}.
\newblock {\emph{\JournalTitle{\mnras}}} \textbf{\bibinfo{volume}{534}}, \bibinfo{pages}{655--682}, \doiprefix\url{https://dx.doi.org/10.1093/mnras/stae2100} (\bibinfo{year}{2024}).
\newblock \eprint{2404.06098}.

\bibitem{2025arXiv250704476D}
\bibinfo{author}{{Dalal}, N.} \emph{et~al.}
\newblock \bibinfo{journal}{\bibinfo{title}{{Deciphering Baryonic Feedback from ACT tSZ Galaxy Clusters}}}.
\newblock {\emph{\JournalTitle{arXiv e-prints}}} \bibinfo{pages}{arXiv:2507.04476}, \doiprefix\url{https://dx.doi.org/10.48550/arXiv.2507.04476} (\bibinfo{year}{2025}).
\newblock \eprint{2507.04476}.

\bibitem{2025arXiv250607432P}
\bibinfo{author}{{Pandey}, S.} \emph{et~al.}
\newblock \bibinfo{journal}{\bibinfo{title}{{Constraints on cosmology and baryonic feedback with joint analysis of Dark Energy Survey Year 3 lensing data and ACT DR6 thermal Sunyaev-Zel'dovich effect observations}}}.
\newblock {\emph{\JournalTitle{arXiv e-prints}}} \bibinfo{pages}{arXiv:2506.07432}, \doiprefix\url{https://dx.doi.org/10.48550/arXiv.2506.07432} (\bibinfo{year}{2025}).
\newblock \eprint{2506.07432}.

\bibitem{2022A&ARv..30....2P}
\bibinfo{author}{{Petroff}, E.}, \bibinfo{author}{{Hessels}, J.~W.~T.} \& \bibinfo{author}{{Lorimer}, D.~R.}
\newblock \bibinfo{journal}{\bibinfo{title}{{Fast radio bursts at the dawn of the 2020s}}}.
\newblock {\emph{\JournalTitle{\aapr}}} \textbf{\bibinfo{volume}{30}}, \bibinfo{pages}{2}, \doiprefix\url{https://dx.doi.org/10.1007/s00159-022-00139-w} (\bibinfo{year}{2022}).
\newblock \eprint{2107.10113}.

\bibitem{2014ApJ...780L..33M}
\bibinfo{author}{{McQuinn}, M.}
\newblock \bibinfo{journal}{\bibinfo{title}{{Locating the ``Missing'' Baryons with Extragalactic Dispersion Measure Estimates}}}.
\newblock {\emph{\JournalTitle{\apjl}}} \textbf{\bibinfo{volume}{780}}, \bibinfo{pages}{L33}, \doiprefix\url{https://dx.doi.org/10.1088/2041-8205/780/2/L33} (\bibinfo{year}{2014}).
\newblock \eprint{1309.4451}.

\bibitem{2020Natur.581..391M}
\bibinfo{author}{{Macquart}, J.~P.} \emph{et~al.}
\newblock \bibinfo{journal}{\bibinfo{title}{{A census of baryons in the Universe from localized fast radio bursts}}}.
\newblock {\emph{\JournalTitle{\nat}}} \textbf{\bibinfo{volume}{581}}, \bibinfo{pages}{391--395}, \doiprefix\url{https://dx.doi.org/10.1038/s41586-020-2300-2} (\bibinfo{year}{2020}).
\newblock \eprint{2005.13161}.

\bibitem{2024ApJ...973..151K}
\bibinfo{author}{{Khrykin}, I.~S.} \emph{et~al.}
\newblock \bibinfo{journal}{\bibinfo{title}{{FLIMFLAM DR1: The First Constraints on the Cosmic Baryon Distribution from Eight Fast Radio Burst Sight Lines}}}.
\newblock {\emph{\JournalTitle{\apj}}} \textbf{\bibinfo{volume}{973}}, \bibinfo{pages}{151}, \doiprefix\url{https://dx.doi.org/10.3847/1538-4357/ad6567} (\bibinfo{year}{2024}).
\newblock \eprint{2402.00505}.

\bibitem{2025NatAs...9.1226C}
\bibinfo{author}{{Connor}, L.} \emph{et~al.}
\newblock \bibinfo{journal}{\bibinfo{title}{{A gas-rich cosmic web revealed by the partitioning of the missing baryons}}}.
\newblock {\emph{\JournalTitle{Nature Astronomy}}} \textbf{\bibinfo{volume}{9}}, \bibinfo{pages}{1226--1239}, \doiprefix\url{https://dx.doi.org/10.1038/s41550-025-02566-y} (\bibinfo{year}{2025}).
\newblock \eprint{2409.16952}.

\bibitem{2015A&C....13...12N}
\bibinfo{author}{{Nelson}, D.} \emph{et~al.}
\newblock \bibinfo{journal}{\bibinfo{title}{{The illustris simulation: Public data release}}}.
\newblock {\emph{\JournalTitle{Astronomy and Computing}}} \textbf{\bibinfo{volume}{13}}, \bibinfo{pages}{12--37}, \doiprefix\url{https://dx.doi.org/10.1016/j.ascom.2015.09.003} (\bibinfo{year}{2015}).
\newblock \eprint{1504.00362}.

\bibitem{2010MNRAS.406..822M}
\bibinfo{author}{{McCarthy}, I.~G.} \emph{et~al.}
\newblock \bibinfo{journal}{\bibinfo{title}{{The case for AGN feedback in galaxy groups}}}.
\newblock {\emph{\JournalTitle{\mnras}}} \textbf{\bibinfo{volume}{406}}, \bibinfo{pages}{822--839}, \doiprefix\url{https://dx.doi.org/10.1111/j.1365-2966.2010.16750.x} (\bibinfo{year}{2010}).
\newblock \eprint{0911.2641}.

\bibitem{2019ApJ...873..111I}
\bibinfo{author}{{Ivezi{\'c}}, {\v{Z}}.} \emph{et~al.}
\newblock \bibinfo{journal}{\bibinfo{title}{{LSST: From Science Drivers to Reference Design and Anticipated Data Products}}}.
\newblock {\emph{\JournalTitle{\apj}}} \textbf{\bibinfo{volume}{873}}, \bibinfo{pages}{111}, \doiprefix\url{https://dx.doi.org/10.3847/1538-4357/ab042c} (\bibinfo{year}{2019}).
\newblock \eprint{0805.2366}.

\bibitem{2011arXiv1110.3193L}
\bibinfo{author}{{Laureijs}, R.} \emph{et~al.}
\newblock \bibinfo{journal}{\bibinfo{title}{{Euclid Definition Study Report}}}.
\newblock {\emph{\JournalTitle{arXiv e-prints}}} \bibinfo{pages}{arXiv:1110.3193}, \doiprefix\url{https://dx.doi.org/10.48550/arXiv.1110.3193} (\bibinfo{year}{2011}).
\newblock \eprint{1110.3193}.

\bibitem{2019JCAP...03..020S}
\bibinfo{author}{{Schneider}, A.} \emph{et~al.}
\newblock \bibinfo{journal}{\bibinfo{title}{{Quantifying baryon effects on the matter power spectrum and the weak lensing shear correlation}}}.
\newblock {\emph{\JournalTitle{\jcap}}} \textbf{\bibinfo{volume}{2019}}, \bibinfo{pages}{020}, \doiprefix\url{https://dx.doi.org/10.1088/1475-7516/2019/03/020} (\bibinfo{year}{2019}).
\newblock \eprint{1810.08629}.

\bibitem{2024PhRvL.133e1001F}
\bibinfo{author}{{Ferreira}, T.}, \bibinfo{author}{{Alonso}, D.}, \bibinfo{author}{{Garcia-Garcia}, C.} \& \bibinfo{author}{{Chisari}, N.~E.}
\newblock \bibinfo{journal}{\bibinfo{title}{{X-Ray-Cosmic-Shear Cross-Correlations: First Detection and Constraints on Baryonic Effects}}}.
\newblock {\emph{\JournalTitle{\prl}}} \textbf{\bibinfo{volume}{133}}, \bibinfo{pages}{051001}, \doiprefix\url{https://dx.doi.org/10.1103/PhysRevLett.133.051001} (\bibinfo{year}{2024}).
\newblock \eprint{2309.11129}.

\bibitem{2024A&A...682A..34M}
\bibinfo{author}{{Merloni}, A.} \emph{et~al.}
\newblock \bibinfo{journal}{\bibinfo{title}{{The SRG/eROSITA all-sky survey. First X-ray catalogues and data release of the western Galactic hemisphere}}}.
\newblock {\emph{\JournalTitle{\aap}}} \textbf{\bibinfo{volume}{682}}, \bibinfo{pages}{A34}, \doiprefix\url{https://dx.doi.org/10.1051/0004-6361/202347165} (\bibinfo{year}{2024}).
\newblock \eprint{2401.17274}.

\bibitem{2025JCAP...08..034A}
\bibinfo{author}{{Abitbol}, M.} \emph{et~al.}
\newblock \bibinfo{journal}{\bibinfo{title}{{The Simons Observatory: science goals and forecasts for the enhanced Large Aperture Telescope}}}.
\newblock {\emph{\JournalTitle{\jcap}}} \textbf{\bibinfo{volume}{2025}}, \bibinfo{pages}{034}, \doiprefix\url{https://dx.doi.org/10.1088/1475-7516/2025/08/034} (\bibinfo{year}{2025}).
\newblock \eprint{2503.00636}.

\bibitem{2025arXiv251204203E}
\bibinfo{author}{{Eckert}, D.} \emph{et~al.}
\newblock \bibinfo{journal}{\bibinfo{title}{{The impact of strong feedback on galaxy group scaling relations}}}.
\newblock {\emph{\JournalTitle{arXiv e-prints}}} \bibinfo{pages}{arXiv:2512.04203}, \doiprefix\url{https://dx.doi.org/10.48550/arXiv.2512.04203} (\bibinfo{year}{2025}).
\newblock \eprint{2512.04203}.

\bibitem{2026arXiv260202484M}
\bibinfo{author}{{McDonald}, W.} \emph{et~al.}
\newblock \bibinfo{journal}{\bibinfo{title}{{The FLAMINGO Project: Exploring the X-ray--cosmic-shear cross-correlation as a probe of large-scale structure}}}.
\newblock {\emph{\JournalTitle{arXiv e-prints}}} \bibinfo{pages}{arXiv:2602.02484}, \doiprefix\url{https://dx.doi.org/10.48550/arXiv.2602.02484} (\bibinfo{year}{2026}).
\newblock \eprint{2602.02484}.

\bibitem{2019PhRvD.100j3532M}
\bibinfo{author}{{Madhavacheril}, M.~S.}, \bibinfo{author}{{Battaglia}, N.}, \bibinfo{author}{{Smith}, K.~M.} \& \bibinfo{author}{{Sievers}, J.~L.}
\newblock \bibinfo{journal}{\bibinfo{title}{{Cosmology with the kinematic Sunyaev-Zeldovich effect: Breaking the optical depth degeneracy with fast radio bursts}}}.
\newblock {\emph{\JournalTitle{\prd}}} \textbf{\bibinfo{volume}{100}}, \bibinfo{pages}{103532}, \doiprefix\url{https://dx.doi.org/10.1103/PhysRevD.100.103532} (\bibinfo{year}{2019}).
\newblock \eprint{1901.02418}.

\bibitem{2025MNRAS.536..572D}
\bibinfo{author}{{Ding}, J.} \emph{et~al.}
\newblock \bibinfo{journal}{\bibinfo{title}{{Miscentring of optical galaxy clusters based on Sunyaev-Zeldovich counterparts}}}.
\newblock {\emph{\JournalTitle{\mnras}}} \textbf{\bibinfo{volume}{536}}, \bibinfo{pages}{572--591}, \doiprefix\url{https://dx.doi.org/10.1093/mnras/stae2601} (\bibinfo{year}{2025}).
\newblock \eprint{2411.12120}.

\bibitem{2025MNRAS.540.1055E}
\bibinfo{author}{{Efstathiou}, G.} \& \bibinfo{author}{{McCarthy}, F.}
\newblock \bibinfo{journal}{\bibinfo{title}{{The power spectrum of the thermal Sunyaev{\textendash}Zeldovich effect}}}.
\newblock {\emph{\JournalTitle{\mnras}}} \textbf{\bibinfo{volume}{540}}, \bibinfo{pages}{1055--1068}, \doiprefix\url{https://dx.doi.org/10.1093/mnras/staf709} (\bibinfo{year}{2025}).
\newblock \eprint{2502.10232}.

\bibitem{2020ApJ...900..170Z}
\bibinfo{author}{{Zhang}, G.~Q.}, \bibinfo{author}{{Yu}, H.}, \bibinfo{author}{{He}, J.~H.} \& \bibinfo{author}{{Wang}, F.~Y.}
\newblock \bibinfo{journal}{\bibinfo{title}{{Dispersion Measures of Fast Radio Burst Host Galaxies Derived from IllustrisTNG Simulation}}}.
\newblock {\emph{\JournalTitle{\apj}}} \textbf{\bibinfo{volume}{900}}, \bibinfo{pages}{170}, \doiprefix\url{https://dx.doi.org/10.3847/1538-4357/abaa4a} (\bibinfo{year}{2020}).
\newblock \eprint{2007.13935}.

\bibitem{2024Natur.635...61S}
\bibinfo{author}{{Sharma}, K.} \emph{et~al.}
\newblock \bibinfo{journal}{\bibinfo{title}{{Preferential occurrence of fast radio bursts in massive star-forming galaxies}}}.
\newblock {\emph{\JournalTitle{\nat}}} \textbf{\bibinfo{volume}{635}}, \bibinfo{pages}{61--66}, \doiprefix\url{https://dx.doi.org/10.1038/s41586-024-08074-9} (\bibinfo{year}{2024}).
\newblock \eprint{2409.16964}.

\bibitem{2026ApJ...999..202S}
\bibinfo{author}{{Sharma}, K.} \emph{et~al.}
\newblock \bibinfo{journal}{\bibinfo{title}{{Quantifying the Impact of Selection Effects on FRB DM-z Relation Cosmological Inference}}}.
\newblock {\emph{\JournalTitle{\apj}}} \textbf{\bibinfo{volume}{999}}, \bibinfo{pages}{202}, \doiprefix\url{https://dx.doi.org/10.3847/1538-4357/ae4696} (\bibinfo{year}{2026}).

\bibitem{2025ApJ...989...81S}
\bibinfo{author}{{Sharma}, K.} \emph{et~al.}
\newblock \bibinfo{journal}{\bibinfo{title}{{A Hydrodynamical Simulations-based Model that Connects the FRB DM{\textendash}Redshift Relation to Suppression of the Matter Power Spectrum via Feedback}}}.
\newblock {\emph{\JournalTitle{\apj}}} \textbf{\bibinfo{volume}{989}}, \bibinfo{pages}{81}, \doiprefix\url{https://dx.doi.org/10.3847/1538-4357/adeca4} (\bibinfo{year}{2025}).
\newblock \eprint{2504.18745}.

\bibitem{2022PhRvD.105b3514A}
\bibinfo{author}{{Amon}, A.} \emph{et~al.}
\newblock \bibinfo{journal}{\bibinfo{title}{{Dark Energy Survey Year 3 results: Cosmology from cosmic shear and robustness to data calibration}}}.
\newblock {\emph{\JournalTitle{\prd}}} \textbf{\bibinfo{volume}{105}}, \bibinfo{pages}{023514}, \doiprefix\url{https://dx.doi.org/10.1103/PhysRevD.105.023514} (\bibinfo{year}{2022}).
\newblock \eprint{2105.13543}.

\bibitem{2019BAAS...51g.255H}
\bibinfo{author}{{Hallinan}, G.} \emph{et~al.}
\newblock \bibinfo{title}{{The DSA-2000 {\textemdash} A Radio Survey Camera}}.
\newblock In \emph{\bibinfo{booktitle}{Bulletin of the American Astronomical Society}}, vol.~\bibinfo{volume}{51}, \bibinfo{pages}{255}, \doiprefix\url{https://dx.doi.org/10.48550/arXiv.1907.07648} (\bibinfo{year}{2019}).
\newblock \eprint{1907.07648}.

\bibitem{2004NewAR..48..979C}
\bibinfo{author}{{Carilli}, C.~L.} \& \bibinfo{author}{{Rawlings}, S.}
\newblock \bibinfo{journal}{\bibinfo{title}{{Motivation, key science projects, standards and assumptions}}}.
\newblock {\emph{\JournalTitle{\nar}}} \textbf{\bibinfo{volume}{48}}, \bibinfo{pages}{979--984}, \doiprefix\url{https://dx.doi.org/10.1016/j.newar.2004.09.001} (\bibinfo{year}{2004}).
\newblock \eprint{astro-ph/0409274}.

\bibitem{2019clrp.2020...28V}
\bibinfo{author}{{Vanderlinde}, K.} \emph{et~al.}
\newblock \bibinfo{title}{{The Canadian Hydrogen Observatory and Radio-transient Detector (CHORD)}}.
\newblock In \emph{\bibinfo{booktitle}{Canadian Long Range Plan for Astronomy and Astrophysics White Papers}}, vol. \bibinfo{volume}{2020}, \bibinfo{pages}{28}, \doiprefix\url{https://dx.doi.org/10.5281/zenodo.3765414} (\bibinfo{year}{2019}).
\newblock \eprint{1911.01777}.

\bibitem{2021JCAP...12..046G}
\bibinfo{author}{{Giri}, S.~K.} \& \bibinfo{author}{{Schneider}, A.}
\newblock \bibinfo{journal}{\bibinfo{title}{{Emulation of baryonic effects on the matter power spectrum and constraints from galaxy cluster data}}}.
\newblock {\emph{\JournalTitle{\jcap}}} \textbf{\bibinfo{volume}{2021}}, \bibinfo{pages}{046}, \doiprefix\url{https://dx.doi.org/10.1088/1475-7516/2021/12/046} (\bibinfo{year}{2021}).
\newblock \eprint{2108.08863}.

\bibitem{2018AstL...44....8K}
\bibinfo{author}{{Kravtsov}, A.~V.}, \bibinfo{author}{{Vikhlinin}, A.~A.} \& \bibinfo{author}{{Meshcheryakov}, A.~V.}
\newblock \bibinfo{journal}{\bibinfo{title}{{Stellar Mass{\textemdash}Halo Mass Relation and Star Formation Efficiency in High-Mass Halos}}}.
\newblock {\emph{\JournalTitle{Astronomy Letters}}} \textbf{\bibinfo{volume}{44}}, \bibinfo{pages}{8--34}, \doiprefix\url{https://dx.doi.org/10.1134/S1063773717120015} (\bibinfo{year}{2018}).
\newblock \eprint{1401.7329}.

\bibitem{2025ApJ...991..205D}
\bibinfo{author}{{Das}, S.}, \bibinfo{author}{{Truong}, N.}, \bibinfo{author}{{Chiang}, Y.-K.} \& \bibinfo{author}{{Mathur}, S.}
\newblock \bibinfo{journal}{\bibinfo{title}{{Thermal Sunyaev{\textendash}Zel'dovich Effect in the Circumgalactic Medium. II. Dependence on Star Formation}}}.
\newblock {\emph{\JournalTitle{\apj}}} \textbf{\bibinfo{volume}{991}}, \bibinfo{pages}{205}, \doiprefix\url{https://dx.doi.org/10.3847/1538-4357/adfdd6} (\bibinfo{year}{2025}).
\newblock \eprint{2508.09514}.

\bibitem{2025ApJ...991L..25L}
\bibinfo{author}{{Leung}, C.} \emph{et~al.}
\newblock \bibinfo{journal}{\bibinfo{title}{{Stellar Mass{\textendash}Dispersion Measure Correlations Constrain Baryonic Feedback in Fast Radio Burst Host Galaxies}}}.
\newblock {\emph{\JournalTitle{\apjl}}} \textbf{\bibinfo{volume}{991}}, \bibinfo{pages}{L25}, \doiprefix\url{https://dx.doi.org/10.3847/2041-8213/ae044d} (\bibinfo{year}{2025}).
\newblock \eprint{2507.16816}.

\bibitem{2025arXiv250717742R}
\bibinfo{author}{{Reischke}, R.} \& \bibinfo{author}{{Hagstotz}, S.}
\newblock \bibinfo{journal}{\bibinfo{title}{{A first measurement of baryonic feedback with Fast Radio Bursts}}}.
\newblock {\emph{\JournalTitle{arXiv e-prints}}} \bibinfo{pages}{arXiv:2507.17742}, \doiprefix\url{https://dx.doi.org/10.48550/arXiv.2507.17742} (\bibinfo{year}{2025}).
\newblock \eprint{2507.17742}.

\bibitem{2025arXiv250903582A}
\bibinfo{author}{{Anbajagane}, D.} \emph{et~al.}
\newblock \bibinfo{journal}{\bibinfo{title}{{The Dark Energy Camera All Data Everywhere cosmic shear project V: Constraints on cosmology and astrophysics from 270 million galaxies across 13,000 deg$^2$ of the sky}}}.
\newblock {\emph{\JournalTitle{arXiv e-prints}}} \bibinfo{pages}{arXiv:2509.03582}, \doiprefix\url{https://dx.doi.org/10.48550/arXiv.2509.03582} (\bibinfo{year}{2025}).
\newblock \eprint{2509.03582}.

\bibitem{2025arXiv251025596X}
\bibinfo{author}{{Xu}, J.} \emph{et~al.}
\newblock \bibinfo{journal}{\bibinfo{title}{{Constraining baryonic feedback and cosmology from DES Y3 and Planck PR4 6$\times$2pt data. I. $\Lambda$CDM models}}}.
\newblock {\emph{\JournalTitle{arXiv e-prints}}} \bibinfo{pages}{arXiv:2510.25596}, \doiprefix\url{https://dx.doi.org/10.48550/arXiv.2510.25596} (\bibinfo{year}{2025}).
\newblock \eprint{2510.25596}.

\bibitem{2019ComAC...6....2N}
\bibinfo{author}{{Nelson}, D.} \emph{et~al.}
\newblock \bibinfo{journal}{\bibinfo{title}{{The IllustrisTNG simulations: public data release}}}.
\newblock {\emph{\JournalTitle{Computational Astrophysics and Cosmology}}} \textbf{\bibinfo{volume}{6}}, \bibinfo{pages}{2}, \doiprefix\url{https://dx.doi.org/10.1186/s40668-019-0028-x} (\bibinfo{year}{2019}).
\newblock \eprint{1812.05609}.

\bibitem{2025OJAp....8E..66Q}
\bibinfo{author}{{Quataert}, E.} \& \bibinfo{author}{{Hopkins}, P.~F.}
\newblock \bibinfo{journal}{\bibinfo{title}{{Cosmic Ray Feedback in Massive Halos: Implications for the Distribution of Baryons}}}.
\newblock {\emph{\JournalTitle{The Open Journal of Astrophysics}}} \textbf{\bibinfo{volume}{8}}, \bibinfo{pages}{66}, \doiprefix\url{https://dx.doi.org/10.33232/001c.138772} (\bibinfo{year}{2025}).
\newblock \eprint{2502.01753}.

\bibitem{2025MNRAS.542.2698P}
\bibinfo{author}{{Preston}, C.}, \bibinfo{author}{{Rogers}, K.~K.}, \bibinfo{author}{{Amon}, A.} \& \bibinfo{author}{{Efstathiou}, G.}
\newblock \bibinfo{journal}{\bibinfo{title}{{Prospects for disentangling dark matter with weak lensing}}}.
\newblock {\emph{\JournalTitle{\mnras}}} \textbf{\bibinfo{volume}{542}}, \bibinfo{pages}{2698--2713}, \doiprefix\url{https://dx.doi.org/10.1093/mnras/staf1321} (\bibinfo{year}{2025}).
\newblock \eprint{2505.02233}.

\bibitem{2025A&A...696A..81B}
\bibinfo{author}{{Bernales-Cortes}, L.} \emph{et~al.}
\newblock \bibinfo{journal}{\bibinfo{title}{{Empirical estimation of host galaxy dispersion measure toward well-localized fast radio bursts}}}.
\newblock {\emph{\JournalTitle{\aap}}} \textbf{\bibinfo{volume}{696}}, \bibinfo{pages}{A81}, \doiprefix\url{https://dx.doi.org/10.1051/0004-6361/202452026} (\bibinfo{year}{2025}).
\newblock \eprint{2501.14063}.

\bibitem{2026ApJ...998....1M}
\bibinfo{author}{{Mas-Ribas}, L.} \& \bibinfo{author}{{James}, C.~W.}
\newblock \bibinfo{journal}{\bibinfo{title}{{A {\ensuremath{\tau}}-DM Relation for Fast Radio Burst Hosts?}}}
\newblock {\emph{\JournalTitle{\apj}}} \textbf{\bibinfo{volume}{998}}, \bibinfo{pages}{1}, \doiprefix\url{https://dx.doi.org/10.3847/1538-4357/ae36a9} (\bibinfo{year}{2026}).
\newblock \eprint{2508.13317}.

\bibitem{2026ApJ...998..252C}
\bibinfo{author}{{Cheng}, A.~Q.}, \bibinfo{author}{{Andrew}, S.~E.}, \bibinfo{author}{{Wang}, H.} \& \bibinfo{author}{{Masui}, K.~W.}
\newblock \bibinfo{journal}{\bibinfo{title}{{Exploring Selection Biases in Fast Radio Burst Dispersion-Galaxy Cross-correlations with Magnetohydrodynamical Simulations}}}.
\newblock {\emph{\JournalTitle{\apj}}} \textbf{\bibinfo{volume}{998}}, \bibinfo{pages}{252}, \doiprefix\url{https://dx.doi.org/10.3847/1538-4357/ae369f} (\bibinfo{year}{2026}).
\newblock \eprint{2506.03258}.

\bibitem{2025arXiv250919514L}
\bibinfo{author}{{Leung}, C.} \emph{et~al.}
\newblock \bibinfo{journal}{\bibinfo{title}{{Nulling baryonic feedback in weak lensing surveys using cross-correlations with fast radio bursts}}}.
\newblock {\emph{\JournalTitle{arXiv e-prints}}} \bibinfo{pages}{arXiv:2509.19514}, \doiprefix\url{https://dx.doi.org/10.48550/arXiv.2509.19514} (\bibinfo{year}{2025}).
\newblock \eprint{2509.19514}.

\bibitem{2026ApJ...998..109S}
\bibinfo{author}{{Sharma}, K.} \emph{et~al.}
\newblock \bibinfo{journal}{\bibinfo{title}{{Probing Baryonic Feedback and Cosmology with the 3 {\texttimes} 2-point Statistic of Fast Radio Bursts and Galaxies}}}.
\newblock {\emph{\JournalTitle{\apj}}} \textbf{\bibinfo{volume}{998}}, \bibinfo{pages}{109}, \doiprefix\url{https://dx.doi.org/10.3847/1538-4357/ae2ff9} (\bibinfo{year}{2026}).
\newblock \eprint{2509.05866}.

\bibitem{2025PDU....4901965D}
\bibinfo{author}{{Di Valentino}, E.} \emph{et~al.}
\newblock \bibinfo{journal}{\bibinfo{title}{{The CosmoVerse White Paper: Addressing observational tensions in cosmology with systematics and fundamental physics}}}.
\newblock {\emph{\JournalTitle{Physics of the Dark Universe}}} \textbf{\bibinfo{volume}{49}}, \bibinfo{pages}{101965}, \doiprefix\url{https://dx.doi.org/10.1016/j.dark.2025.101965} (\bibinfo{year}{2025}).
\newblock \eprint{2504.01669}.

\bibitem{2025arXiv250621657H}
\bibinfo{author}{{Hotinli}, S.~C.}, \bibinfo{author}{{Smith}, K.~M.} \& \bibinfo{author}{{Ferraro}, S.}
\newblock \bibinfo{journal}{\bibinfo{title}{{Velocity Reconstruction from KSZ: Measuring $f_{NL}$ with ACT and DESILS}}}.
\newblock {\emph{\JournalTitle{arXiv e-prints}}} \bibinfo{pages}{arXiv:2506.21657}, \doiprefix\url{https://dx.doi.org/10.48550/arXiv.2506.21657} (\bibinfo{year}{2025}).
\newblock \eprint{2506.21657}.

\end{thebibliography}


\begin{thebibliography}{100}
\urlstyle{rm}
\expandafter\ifx\csname url\endcsname\relax
  \def\url#1{\texttt{#1}}\fi
\expandafter\ifx\csname urlprefix\endcsname\relax\def\urlprefix{URL }\fi
\expandafter\ifx\csname doiprefix\endcsname\relax\def\doiprefix{DOI: }\fi
\providecommand{\bibinfo}[2]{#2}
\providecommand{\eprint}[2][]{\url{#2}}

\bibitem{2025PhRvD.112f3541L}
\bibinfo{author}{{Lucie-Smith}, L.} \emph{et~al.}
\newblock \bibinfo{journal}{\bibinfo{title}{{Cosmological feedback from a halo assembly perspective}}}.
\newblock {\emph{\JournalTitle{\prd}}} \textbf{\bibinfo{volume}{112}}, \bibinfo{pages}{063541}, \doiprefix\url{https://dx.doi.org/10.1103/vh8n-9cr2} (\bibinfo{year}{2025}).
\newblock \eprint{2505.18258}.

\bibitem{2022PhRvD.105b3515S}
\bibinfo{author}{{Secco}, L.~F.} \emph{et~al.}
\newblock \bibinfo{journal}{\bibinfo{title}{{Dark Energy Survey Year 3 results: Cosmology from cosmic shear and robustness to modeling uncertainty}}}.
\newblock {\emph{\JournalTitle{\prd}}} \textbf{\bibinfo{volume}{105}}, \bibinfo{pages}{023515}, \doiprefix\url{https://dx.doi.org/10.1103/PhysRevD.105.023515} (\bibinfo{year}{2022}).
\newblock \eprint{2105.13544}.

\bibitem{2008ApJ...688..709T}
\bibinfo{author}{{Tinker}, J.} \emph{et~al.}
\newblock \bibinfo{journal}{\bibinfo{title}{{Toward a Halo Mass Function for Precision Cosmology: The Limits of Universality}}}.
\newblock {\emph{\JournalTitle{\apj}}} \textbf{\bibinfo{volume}{688}}, \bibinfo{pages}{709--728}, \doiprefix\url{https://dx.doi.org/10.1086/591439} (\bibinfo{year}{2008}).
\newblock \eprint{0803.2706}.

\bibitem{2010ApJ...724..878T}
\bibinfo{author}{{Tinker}, J.~L.} \emph{et~al.}
\newblock \bibinfo{journal}{\bibinfo{title}{{The Large-scale Bias of Dark Matter Halos: Numerical Calibration and Model Tests}}}.
\newblock {\emph{\JournalTitle{\apj}}} \textbf{\bibinfo{volume}{724}}, \bibinfo{pages}{878--886}, \doiprefix\url{https://dx.doi.org/10.1088/0004-637X/724/2/878} (\bibinfo{year}{2010}).
\newblock \eprint{1001.3162}.

\bibitem{2015ApJ...799..108D}
\bibinfo{author}{{Diemer}, B.} \& \bibinfo{author}{{Kravtsov}, A.~V.}
\newblock \bibinfo{journal}{\bibinfo{title}{{A Universal Model for Halo Concentrations}}}.
\newblock {\emph{\JournalTitle{\apj}}} \textbf{\bibinfo{volume}{799}}, \bibinfo{pages}{108}, \doiprefix\url{https://dx.doi.org/10.1088/0004-637X/799/1/108} (\bibinfo{year}{2015}).
\newblock \eprint{1407.4730}.

\bibitem{2021MNRAS.504.4312G}
\bibinfo{author}{{Gatti}, M.} \emph{et~al.}
\newblock \bibinfo{journal}{\bibinfo{title}{{Dark energy survey year 3 results: weak lensing shape catalogue}}}.
\newblock {\emph{\JournalTitle{\mnras}}} \textbf{\bibinfo{volume}{504}}, \bibinfo{pages}{4312--4336}, \doiprefix\url{https://dx.doi.org/10.1093/mnras/stab918} (\bibinfo{year}{2021}).
\newblock \eprint{2011.03408}.

\bibitem{2021ApJS..253....3H}
\bibinfo{author}{{Hilton}, M.} \emph{et~al.}
\newblock \bibinfo{journal}{\bibinfo{title}{{The Atacama Cosmology Telescope: A Catalog of >4000 Sunyaev-Zel{\textquoteright}dovich Galaxy Clusters}}}.
\newblock {\emph{\JournalTitle{\apjs}}} \textbf{\bibinfo{volume}{253}}, \bibinfo{pages}{3}, \doiprefix\url{https://dx.doi.org/10.3847/1538-4365/abd023} (\bibinfo{year}{2021}).
\newblock \eprint{2009.11043}.

\bibitem{2020A&A...643A..42P}
\bibinfo{author}{{Planck Collaboration}} \emph{et~al.}
\newblock \bibinfo{journal}{\bibinfo{title}{{Planck intermediate results. LVII. Joint Planck LFI and HFI data processing}}}.
\newblock {\emph{\JournalTitle{\aap}}} \textbf{\bibinfo{volume}{643}}, \bibinfo{pages}{A42}, \doiprefix\url{https://dx.doi.org/10.1051/0004-6361/202038073} (\bibinfo{year}{2020}).
\newblock \eprint{2007.04997}.

\bibitem{2020JCAP...12..047A}
\bibinfo{author}{{Aiola}, S.} \emph{et~al.}
\newblock \bibinfo{journal}{\bibinfo{title}{{The Atacama Cosmology Telescope: DR4 maps and cosmological parameters}}}.
\newblock {\emph{\JournalTitle{\jcap}}} \textbf{\bibinfo{volume}{2020}}, \bibinfo{pages}{047}, \doiprefix\url{https://dx.doi.org/10.1088/1475-7516/2020/12/047} (\bibinfo{year}{2020}).
\newblock \eprint{2007.07288}.

\bibitem{2024PhRvD.109f3530C}
\bibinfo{author}{{Coulton}, W.} \emph{et~al.}
\newblock \bibinfo{journal}{\bibinfo{title}{{Atacama Cosmology Telescope: High-resolution component-separated maps across one third of the sky}}}.
\newblock {\emph{\JournalTitle{\prd}}} \textbf{\bibinfo{volume}{109}}, \bibinfo{pages}{063530}, \doiprefix\url{https://dx.doi.org/10.1103/PhysRevD.109.063530} (\bibinfo{year}{2024}).
\newblock \eprint{2307.01258}.

\bibitem{2025JCAP...11..061N}
\bibinfo{author}{{Naess}, S.} \emph{et~al.}
\newblock \bibinfo{journal}{\bibinfo{title}{{The Atacama Cosmology Telescope: DR6 maps}}}.
\newblock {\emph{\JournalTitle{\jcap}}} \textbf{\bibinfo{volume}{2025}}, \bibinfo{pages}{061}, \doiprefix\url{https://dx.doi.org/10.1088/1475-7516/2025/11/061} (\bibinfo{year}{2025}).
\newblock \eprint{2503.14451}.

\bibitem{2021PhRvD.103f3513S}
\bibinfo{author}{{Schaan}, E.} \emph{et~al.}
\newblock \bibinfo{journal}{\bibinfo{title}{{Atacama Cosmology Telescope: Combined kinematic and thermal Sunyaev-Zel'dovich measurements from BOSS CMASS and LOWZ halos}}}.
\newblock {\emph{\JournalTitle{\prd}}} \textbf{\bibinfo{volume}{103}}, \bibinfo{pages}{063513}, \doiprefix\url{https://dx.doi.org/10.1103/PhysRevD.103.063513} (\bibinfo{year}{2021}).
\newblock \eprint{2009.05557}.

\bibitem{2025PhRvD.112j3512R}
\bibinfo{author}{{Ried Guachalla}, B.} \emph{et~al.}
\newblock \bibinfo{journal}{\bibinfo{title}{{Backlighting extended gas halos around luminous red galaxies: Kinematic Sunyaev-Zel'dovich effect from DESI Y1 and ACT data}}}.
\newblock {\emph{\JournalTitle{\prd}}} \textbf{\bibinfo{volume}{112}}, \bibinfo{pages}{103512}, \doiprefix\url{https://dx.doi.org/10.1103/lqbj-wcqj} (\bibinfo{year}{2025}).
\newblock \eprint{2503.19870}.

\bibitem{2023AJ....165..253H}
\bibinfo{author}{{Hahn}, C.} \emph{et~al.}
\newblock \bibinfo{journal}{\bibinfo{title}{{The DESI Bright Galaxy Survey: Final Target Selection, Design, and Validation}}}.
\newblock {\emph{\JournalTitle{\aj}}} \textbf{\bibinfo{volume}{165}}, \bibinfo{pages}{253}, \doiprefix\url{https://dx.doi.org/10.3847/1538-3881/accff8} (\bibinfo{year}{2023}).
\newblock \eprint{2208.08512}.

\bibitem{2023AJ....165...58Z}
\bibinfo{author}{{Zhou}, R.} \emph{et~al.}
\newblock \bibinfo{journal}{\bibinfo{title}{{Target Selection and Validation of DESI Luminous Red Galaxies}}}.
\newblock {\emph{\JournalTitle{\aj}}} \textbf{\bibinfo{volume}{165}}, \bibinfo{pages}{58}, \doiprefix\url{https://dx.doi.org/10.3847/1538-3881/aca5fb} (\bibinfo{year}{2023}).
\newblock \eprint{2208.08515}.

\bibitem{2024A&A...685A.106B}
\bibinfo{author}{{Bulbul}, E.} \emph{et~al.}
\newblock \bibinfo{journal}{\bibinfo{title}{{The SRG/eROSITA All-Sky Survey. The first catalog of galaxy clusters and groups in the Western Galactic Hemisphere}}}.
\newblock {\emph{\JournalTitle{\aap}}} \textbf{\bibinfo{volume}{685}}, \bibinfo{pages}{A106}, \doiprefix\url{https://dx.doi.org/10.1051/0004-6361/202348264} (\bibinfo{year}{2024}).
\newblock \eprint{2402.08452}.

\bibitem{2024A&A...688A.210K}
\bibinfo{author}{{Kluge}, M.} \emph{et~al.}
\newblock \bibinfo{journal}{\bibinfo{title}{{The SRG/eROSITA All-Sky Survey. Optical identification and properties of galaxy clusters and groups in the western galactic hemisphere}}}.
\newblock {\emph{\JournalTitle{\aap}}} \textbf{\bibinfo{volume}{688}}, \bibinfo{pages}{A210}, \doiprefix\url{https://dx.doi.org/10.1051/0004-6361/202349031} (\bibinfo{year}{2024}).
\newblock \eprint{2402.08453}.

\bibitem{2022ApJ...928....9L}
\bibinfo{author}{{Lee}, K.-G.} \emph{et~al.}
\newblock \bibinfo{journal}{\bibinfo{title}{{Constraining the Cosmic Baryon Distribution with Fast Radio Burst Foreground Mapping}}}.
\newblock {\emph{\JournalTitle{\apj}}} \textbf{\bibinfo{volume}{928}}, \bibinfo{pages}{9}, \doiprefix\url{https://dx.doi.org/10.3847/1538-4357/ac4f62} (\bibinfo{year}{2022}).
\newblock \eprint{2109.00386}.

\bibitem{2023arXiv230909766R}
\bibinfo{author}{{Reischke}, R.}, \bibinfo{author}{{Neumann}, D.}, \bibinfo{author}{{Bertmann}, K.~A.}, \bibinfo{author}{{Hagstotz}, S.} \& \bibinfo{author}{{Hildebrandt}, H.}
\newblock \bibinfo{journal}{\bibinfo{title}{{Calibrating baryonic feedback with weak lensing and fast radio bursts}}}.
\newblock {\emph{\JournalTitle{arXiv e-prints}}} \bibinfo{pages}{arXiv:2309.09766}, \doiprefix\url{https://dx.doi.org/10.48550/arXiv.2309.09766} (\bibinfo{year}{2023}).
\newblock \eprint{2309.09766}.

\bibitem{2024ApJ...965...57B}
\bibinfo{author}{{Baptista}, J.} \emph{et~al.}
\newblock \bibinfo{journal}{\bibinfo{title}{{Measuring the Variance of the Macquart Relation in Redshift{\textendash}Extragalactic Dispersion Measure Modeling}}}.
\newblock {\emph{\JournalTitle{\apj}}} \textbf{\bibinfo{volume}{965}}, \bibinfo{pages}{57}, \doiprefix\url{https://dx.doi.org/10.3847/1538-4357/ad2705} (\bibinfo{year}{2024}).
\newblock \eprint{2305.07022}.

\bibitem{2025ApJ...983...46M}
\bibinfo{author}{{Medlock}, I.}, \bibinfo{author}{{Nagai}, D.}, \bibinfo{author}{{Angl{\'e}s-Alc{\'a}zar}, D.} \& \bibinfo{author}{{Gebhardt}, M.}
\newblock \bibinfo{journal}{\bibinfo{title}{{Constraining Baryonic Feedback Effects on the Matter Power Spectrum with Fast Radio Bursts}}}.
\newblock {\emph{\JournalTitle{\apj}}} \textbf{\bibinfo{volume}{983}}, \bibinfo{pages}{46}, \doiprefix\url{https://dx.doi.org/10.3847/1538-4357/adbc9c} (\bibinfo{year}{2025}).
\newblock \eprint{2501.17922}.

\bibitem{2002astro.ph..7156C}
\bibinfo{author}{{Cordes}, J.~M.} \& \bibinfo{author}{{Lazio}, T.~J.~W.}
\newblock \bibinfo{journal}{\bibinfo{title}{{NE2001.I. A New Model for the Galactic Distribution of Free Electrons and its Fluctuations}}}.
\newblock {\emph{\JournalTitle{arXiv e-prints}}} \bibinfo{pages}{astro--ph/0207156}, \doiprefix\url{https://dx.doi.org/10.48550/arXiv.astro-ph/0207156} (\bibinfo{year}{2002}).
\newblock \eprint{astro-ph/0207156}.

\bibitem{2003astro.ph..1598C}
\bibinfo{author}{{Cordes}, J.~M.} \& \bibinfo{author}{{Lazio}, T.~J.~W.}
\newblock \bibinfo{journal}{\bibinfo{title}{{NE2001. II. Using Radio Propagation Data to Construct a Model for the Galactic Distribution of Free Electrons}}}.
\newblock {\emph{\JournalTitle{arXiv e-prints}}} \bibinfo{pages}{astro--ph/0301598}, \doiprefix\url{https://dx.doi.org/10.48550/arXiv.astro-ph/0301598} (\bibinfo{year}{2003}).
\newblock \eprint{astro-ph/0301598}.

\bibitem{2019ascl.soft08022Y}
\bibinfo{author}{{Yao}, J.}, \bibinfo{author}{{Manchester}, R.~N.} \& \bibinfo{author}{{Wang}, N.}
\newblock \bibinfo{title}{{YMW16: Electron-density model}}.
\newblock \bibinfo{howpublished}{Astrophysics Source Code Library, record ascl:1908.022} (\bibinfo{year}{2019}).

\bibitem{2019MNRAS.485..648P}
\bibinfo{author}{{Prochaska}, J.~X.} \& \bibinfo{author}{{Zheng}, Y.}
\newblock \bibinfo{journal}{\bibinfo{title}{{Probing Galactic haloes with fast radio bursts}}}.
\newblock {\emph{\JournalTitle{\mnras}}} \textbf{\bibinfo{volume}{485}}, \bibinfo{pages}{648--665}, \doiprefix\url{https://dx.doi.org/10.1093/mnras/stz261} (\bibinfo{year}{2019}).
\newblock \eprint{1901.11051}.

\bibitem{2020ApJ...888..105Y}
\bibinfo{author}{{Yamasaki}, S.} \& \bibinfo{author}{{Totani}, T.}
\newblock \bibinfo{journal}{\bibinfo{title}{{The Galactic Halo Contribution to the Dispersion Measure of Extragalactic Fast Radio Bursts}}}.
\newblock {\emph{\JournalTitle{\apj}}} \textbf{\bibinfo{volume}{888}}, \bibinfo{pages}{105}, \doiprefix\url{https://dx.doi.org/10.3847/1538-4357/ab58c4} (\bibinfo{year}{2020}).
\newblock \eprint{1909.00849}.

\bibitem{2025AJ....169..330R}
\bibinfo{author}{{Ravi}, V.} \emph{et~al.}
\newblock \bibinfo{journal}{\bibinfo{title}{{Deep Synoptic Array Science: A 50 Mpc Fast Radio Burst Constrains the Mass of the Milky Way Circumgalactic Medium}}}.
\newblock {\emph{\JournalTitle{\aj}}} \textbf{\bibinfo{volume}{169}}, \bibinfo{pages}{330}, \doiprefix\url{https://dx.doi.org/10.3847/1538-3881/adc725} (\bibinfo{year}{2025}).
\newblock \eprint{2301.01000}.

\bibitem{2023ApJ...946...58C}
\bibinfo{author}{{Cook}, A.~M.} \emph{et~al.}
\newblock \bibinfo{journal}{\bibinfo{title}{{An FRB Sent Me a DM: Constraining the Electron Column of the Milky Way Halo with Fast Radio Burst Dispersion Measures from CHIME/FRB}}}.
\newblock {\emph{\JournalTitle{\apj}}} \textbf{\bibinfo{volume}{946}}, \bibinfo{pages}{58}, \doiprefix\url{https://dx.doi.org/10.3847/1538-4357/acbbd0} (\bibinfo{year}{2023}).
\newblock \eprint{2301.03502}.

\bibitem{2025A&A...695A.163B}
\bibinfo{author}{{Bollo}, V.} \emph{et~al.}
\newblock \bibinfo{journal}{\bibinfo{title}{{ALMACAL: XIII. Evolution of the CO luminosity function and the molecular gas mass density out to z {\ensuremath{\sim}} 6}}}.
\newblock {\emph{\JournalTitle{\aap}}} \textbf{\bibinfo{volume}{695}}, \bibinfo{pages}{A163}, \doiprefix\url{https://dx.doi.org/10.1051/0004-6361/202453223} (\bibinfo{year}{2025}).
\newblock \eprint{2502.06778}.

\bibitem{2020ARA&A..58..363P}
\bibinfo{author}{{P{\'e}roux}, C.} \& \bibinfo{author}{{Howk}, J.~C.}
\newblock \bibinfo{journal}{\bibinfo{title}{{The Cosmic Baryon and Metal Cycles}}}.
\newblock {\emph{\JournalTitle{\araa}}} \textbf{\bibinfo{volume}{58}}, \bibinfo{pages}{363--406}, \doiprefix\url{https://dx.doi.org/10.1146/annurev-astro-021820-120014} (\bibinfo{year}{2020}).
\newblock \eprint{2011.01935}.

\bibitem{2020A&A...641A...6P}
\bibinfo{author}{{Planck Collaboration}} \emph{et~al.}
\newblock \bibinfo{journal}{\bibinfo{title}{{Planck 2018 results. VI. Cosmological parameters}}}.
\newblock {\emph{\JournalTitle{\aap}}} \textbf{\bibinfo{volume}{641}}, \bibinfo{pages}{A6}, \doiprefix\url{https://dx.doi.org/10.1051/0004-6361/201833910} (\bibinfo{year}{2020}).
\newblock \eprint{1807.06209}.

\bibitem{2023MNRAS.524.2237R}
\bibinfo{author}{{Reischke}, R.} \& \bibinfo{author}{{Hagstotz}, S.}
\newblock \bibinfo{journal}{\bibinfo{title}{{Cosmological covariance of fast radio burst dispersions}}}.
\newblock {\emph{\JournalTitle{\mnras}}} \textbf{\bibinfo{volume}{524}}, \bibinfo{pages}{2237--2243}, \doiprefix\url{https://dx.doi.org/10.1093/mnras/stad1645} (\bibinfo{year}{2023}).
\newblock \eprint{2301.03527}.

\bibitem{2025arXiv250707090K}
\bibinfo{author}{{Konietzka}, R.~M.} \emph{et~al.}
\newblock \bibinfo{journal}{\bibinfo{title}{{Ray-tracing Fast Radio Bursts Through IllustrisTNG: Cosmological Dispersion Measures from Redshift 0 to 5.5}}}.
\newblock {\emph{\JournalTitle{arXiv e-prints}}} \bibinfo{pages}{arXiv:2507.07090}, \doiprefix\url{https://dx.doi.org/10.48550/arXiv.2507.07090} (\bibinfo{year}{2025}).
\newblock \eprint{2507.07090}.

\bibitem{2013PASP..125..306F}
\bibinfo{author}{{Foreman-Mackey}, D.}, \bibinfo{author}{{Hogg}, D.~W.}, \bibinfo{author}{{Lang}, D.} \& \bibinfo{author}{{Goodman}, J.}
\newblock \bibinfo{journal}{\bibinfo{title}{{emcee: The MCMC Hammer}}}.
\newblock {\emph{\JournalTitle{\pasp}}} \textbf{\bibinfo{volume}{125}}, \bibinfo{pages}{306}, \doiprefix\url{https://dx.doi.org/10.1086/670067} (\bibinfo{year}{2013}).
\newblock \eprint{1202.3665}.

\bibitem{2020A&A...641A.130M}
\bibinfo{author}{{Mead}, A.~J.}, \bibinfo{author}{{Tr{\"o}ster}, T.}, \bibinfo{author}{{Heymans}, C.}, \bibinfo{author}{{Van Waerbeke}, L.} \& \bibinfo{author}{{McCarthy}, I.~G.}
\newblock \bibinfo{journal}{\bibinfo{title}{{A hydrodynamical halo model for weak-lensing cross correlations}}}.
\newblock {\emph{\JournalTitle{\aap}}} \textbf{\bibinfo{volume}{641}}, \bibinfo{pages}{A130}, \doiprefix\url{https://dx.doi.org/10.1051/0004-6361/202038308} (\bibinfo{year}{2020}).
\newblock \eprint{2005.00009}.

\bibitem{2021MNRAS.502.1401M}
\bibinfo{author}{{Mead}, A.~J.}, \bibinfo{author}{{Brieden}, S.}, \bibinfo{author}{{Tr{\"o}ster}, T.} \& \bibinfo{author}{{Heymans}, C.}
\newblock \bibinfo{journal}{\bibinfo{title}{{HMCODE-2020: improved modelling of non-linear cosmological power spectra with baryonic feedback}}}.
\newblock {\emph{\JournalTitle{\mnras}}} \textbf{\bibinfo{volume}{502}}, \bibinfo{pages}{1401--1422}, \doiprefix\url{https://dx.doi.org/10.1093/mnras/stab082} (\bibinfo{year}{2021}).
\newblock \eprint{2009.01858}.

\bibitem{2015JCAP...12..049S}
\bibinfo{author}{{Schneider}, A.} \& \bibinfo{author}{{Teyssier}, R.}
\newblock \bibinfo{journal}{\bibinfo{title}{{A new method to quantify the effects of baryons on the matter power spectrum}}}.
\newblock {\emph{\JournalTitle{\jcap}}} \textbf{\bibinfo{volume}{2015}}, \bibinfo{pages}{049--049}, \doiprefix\url{https://dx.doi.org/10.1088/1475-7516/2015/12/049} (\bibinfo{year}{2015}).
\newblock \eprint{1510.06034}.

\bibitem{2024OJAp....7E.108A}
\bibinfo{author}{{Anbajagane}, D.}, \bibinfo{author}{{Pandey}, S.} \& \bibinfo{author}{{Chang}, C.}
\newblock \bibinfo{journal}{\bibinfo{title}{{Map-level baryonification: Efficient modelling of higher-order correlations in the weak lensing and thermal Sunyaev-Zeldovich fields}}}.
\newblock {\emph{\JournalTitle{The Open Journal of Astrophysics}}} \textbf{\bibinfo{volume}{7}}, \bibinfo{pages}{108}, \doiprefix\url{https://dx.doi.org/10.33232/001c.126788} (\bibinfo{year}{2024}).
\newblock \eprint{2409.03822}.

\bibitem{2019MNRAS.488.3143B}
\bibinfo{author}{{Behroozi}, P.}, \bibinfo{author}{{Wechsler}, R.~H.}, \bibinfo{author}{{Hearin}, A.~P.} \& \bibinfo{author}{{Conroy}, C.}
\newblock \bibinfo{journal}{\bibinfo{title}{{UNIVERSEMACHINE: The correlation between galaxy growth and dark matter halo assembly from z = 0-10}}}.
\newblock {\emph{\JournalTitle{\mnras}}} \textbf{\bibinfo{volume}{488}}, \bibinfo{pages}{3143--3194}, \doiprefix\url{https://dx.doi.org/10.1093/mnras/stz1182} (\bibinfo{year}{2019}).
\newblock \eprint{1806.07893}.

\bibitem{2013MNRAS.428.3121M}
\bibinfo{author}{{Moster}, B.~P.}, \bibinfo{author}{{Naab}, T.} \& \bibinfo{author}{{White}, S. D.~M.}
\newblock \bibinfo{journal}{\bibinfo{title}{{Galactic star formation and accretion histories from matching galaxies to dark matter haloes}}}.
\newblock {\emph{\JournalTitle{\mnras}}} \textbf{\bibinfo{volume}{428}}, \bibinfo{pages}{3121--3138}, \doiprefix\url{https://dx.doi.org/10.1093/mnras/sts261} (\bibinfo{year}{2013}).
\newblock \eprint{1205.5807}.

\bibitem{2025arXiv250713317P}
\bibinfo{author}{{Pranjal R.}, S.}, \bibinfo{author}{{Pandey}, S.}, \bibinfo{author}{{Anbajagane}, D.}, \bibinfo{author}{{Krause}, E.} \& \bibinfo{author}{{Dolag}, K.}
\newblock \bibinfo{journal}{\bibinfo{title}{{Testing halo models for constraining astrophysical feedback with multi-probe modeling: I. 3D Power spectra and mass fractions}}}.
\newblock {\emph{\JournalTitle{arXiv e-prints}}} \bibinfo{pages}{arXiv:2507.13317}, \doiprefix\url{https://dx.doi.org/10.48550/arXiv.2507.13317} (\bibinfo{year}{2025}).
\newblock \eprint{2507.13317}.

\bibitem{2021MNRAS.506.4070A}
\bibinfo{author}{{Aric{\`o}}, G.} \emph{et~al.}
\newblock \bibinfo{journal}{\bibinfo{title}{{The BACCO simulation project: a baryonification emulator with neural networks}}}.
\newblock {\emph{\JournalTitle{\mnras}}} \textbf{\bibinfo{volume}{506}}, \bibinfo{pages}{4070--4082}, \doiprefix\url{https://dx.doi.org/10.1093/mnras/stab1911} (\bibinfo{year}{2021}).
\newblock \eprint{2011.15018}.

\bibitem{2025ApJ...993L..27H}
\bibinfo{author}{{Hussaini}, M.} \emph{et~al.}
\newblock \bibinfo{journal}{\bibinfo{title}{{A Correlation between Fast Radio Burst Dispersion Measure and Foreground Large-scale Structure}}}.
\newblock {\emph{\JournalTitle{\apjl}}} \textbf{\bibinfo{volume}{993}}, \bibinfo{pages}{L27}, \doiprefix\url{https://dx.doi.org/10.3847/2041-8213/ae0a49} (\bibinfo{year}{2025}).
\newblock \eprint{2506.04186}.

\bibitem{2025arXiv250608932W}
\bibinfo{author}{{Wang}, H.} \emph{et~al.}
\newblock \bibinfo{journal}{\bibinfo{title}{{Measurement of the Dispersion-Galaxy Cross-Power Spectrum with the Second CHIME/FRB Catalog}}}.
\newblock {\emph{\JournalTitle{arXiv e-prints}}} \bibinfo{pages}{arXiv:2506.08932}, \doiprefix\url{https://dx.doi.org/10.48550/arXiv.2506.08932} (\bibinfo{year}{2025}).
\newblock \eprint{2506.08932}.

\bibitem{2023ApJ...957L...8S}
\bibinfo{author}{{Sherman}, M.~B.} \emph{et~al.}
\newblock \bibinfo{journal}{\bibinfo{title}{{Deep Synoptic Array Science: Implications of Faraday Rotation Measures of Fast Radio Bursts Localized to Host Galaxies}}}.
\newblock {\emph{\JournalTitle{\apjl}}} \textbf{\bibinfo{volume}{957}}, \bibinfo{pages}{L8}, \doiprefix\url{https://dx.doi.org/10.3847/2041-8213/ad0380} (\bibinfo{year}{2023}).
\newblock \eprint{2308.06816}.

\bibitem{2022ApJ...934...71O}
\bibinfo{author}{{Ocker}, S.~K.}, \bibinfo{author}{{Cordes}, J.~M.}, \bibinfo{author}{{Chatterjee}, S.} \& \bibinfo{author}{{Gorsuch}, M.~R.}
\newblock \bibinfo{journal}{\bibinfo{title}{{Radio Scattering Horizons for Galactic and Extragalactic Transients}}}.
\newblock {\emph{\JournalTitle{\apj}}} \textbf{\bibinfo{volume}{934}}, \bibinfo{pages}{71}, \doiprefix\url{https://dx.doi.org/10.3847/1538-4357/ac75ba} (\bibinfo{year}{2022}).
\newblock \eprint{2203.16716}.

\bibitem{2023ApJ...954...80G}
\bibinfo{author}{{Gordon}, A.~C.} \emph{et~al.}
\newblock \bibinfo{journal}{\bibinfo{title}{{The Demographics, Stellar Populations, and Star Formation Histories of Fast Radio Burst Host Galaxies: Implications for the Progenitors}}}.
\newblock {\emph{\JournalTitle{\apj}}} \textbf{\bibinfo{volume}{954}}, \bibinfo{pages}{80}, \doiprefix\url{https://dx.doi.org/10.3847/1538-4357/ace5aa} (\bibinfo{year}{2023}).
\newblock \eprint{2302.05465}.

\bibitem{2014ARA&A..52..415M}
\bibinfo{author}{{Madau}, P.} \& \bibinfo{author}{{Dickinson}, M.}
\newblock \bibinfo{journal}{\bibinfo{title}{{Cosmic Star-Formation History}}}.
\newblock {\emph{\JournalTitle{\araa}}} \textbf{\bibinfo{volume}{52}}, \bibinfo{pages}{415--486}, \doiprefix\url{https://dx.doi.org/10.1146/annurev-astro-081811-125615} (\bibinfo{year}{2014}).
\newblock \eprint{1403.0007}.

\bibitem{2019PhRvD..99d3506R}
\bibinfo{author}{{Raveri}, M.} \& \bibinfo{author}{{Hu}, W.}
\newblock \bibinfo{journal}{\bibinfo{title}{{Concordance and discordance in cosmology}}}.
\newblock {\emph{\JournalTitle{\prd}}} \textbf{\bibinfo{volume}{99}}, \bibinfo{pages}{043506}, \doiprefix\url{https://dx.doi.org/10.1103/PhysRevD.99.043506} (\bibinfo{year}{2019}).
\newblock \eprint{1806.04649}.

\bibitem{2006ApJ...640..691V}
\bibinfo{author}{{Vikhlinin}, A.} \emph{et~al.}
\newblock \bibinfo{journal}{\bibinfo{title}{{Chandra Sample of Nearby Relaxed Galaxy Clusters: Mass, Gas Fraction, and Mass-Temperature Relation}}}.
\newblock {\emph{\JournalTitle{\apj}}} \textbf{\bibinfo{volume}{640}}, \bibinfo{pages}{691--709}, \doiprefix\url{https://dx.doi.org/10.1086/500288} (\bibinfo{year}{2006}).
\newblock \eprint{astro-ph/0507092}.

\bibitem{2013ApJ...778...14G}
\bibinfo{author}{{Gonzalez}, A.~H.}, \bibinfo{author}{{Sivanandam}, S.}, \bibinfo{author}{{Zabludoff}, A.~I.} \& \bibinfo{author}{{Zaritsky}, D.}
\newblock \bibinfo{journal}{\bibinfo{title}{{Galaxy Cluster Baryon Fractions Revisited}}}.
\newblock {\emph{\JournalTitle{\apj}}} \textbf{\bibinfo{volume}{778}}, \bibinfo{pages}{14}, \doiprefix\url{https://dx.doi.org/10.1088/0004-637X/778/1/14} (\bibinfo{year}{2013}).
\newblock \eprint{1309.3565}.

\bibitem{2013MNRAS.429.3288S}
\bibinfo{author}{{Sanderson}, A. J.~R.} \emph{et~al.}
\newblock \bibinfo{journal}{\bibinfo{title}{{The baryon budget on the galaxy group/cluster boundary}}}.
\newblock {\emph{\JournalTitle{\mnras}}} \textbf{\bibinfo{volume}{429}}, \bibinfo{pages}{3288--3304}, \doiprefix\url{https://dx.doi.org/10.1093/mnras/sts586} (\bibinfo{year}{2013}).
\newblock \eprint{1212.1613}.

\bibitem{2018MNRAS.478.3072C}
\bibinfo{author}{{Chiu}, I.} \emph{et~al.}
\newblock \bibinfo{journal}{\bibinfo{title}{{Baryon content in a sample of 91 galaxy clusters selected by the South Pole Telescope at 0.2 <z < 1.25}}}.
\newblock {\emph{\JournalTitle{\mnras}}} \textbf{\bibinfo{volume}{478}}, \bibinfo{pages}{3072--3099}, \doiprefix\url{https://dx.doi.org/10.1093/mnras/sty1284} (\bibinfo{year}{2018}).
\newblock \eprint{1711.00917}.

\bibitem{2003PASP..115..763C}
\bibinfo{author}{{Chabrier}, G.}
\newblock \bibinfo{journal}{\bibinfo{title}{{Galactic Stellar and Substellar Initial Mass Function}}}.
\newblock {\emph{\JournalTitle{\pasp}}} \textbf{\bibinfo{volume}{115}}, \bibinfo{pages}{763--795}, \doiprefix\url{https://dx.doi.org/10.1086/376392} (\bibinfo{year}{2003}).
\newblock \eprint{astro-ph/0304382}.

\bibitem{2024ApJ...973L..32V}
\bibinfo{author}{{van Dokkum}, P.} \& \bibinfo{author}{{Conroy}, C.}
\newblock \bibinfo{journal}{\bibinfo{title}{{Reconciling M/L Ratios Across Cosmic Time: a Concordance IMF for Massive Galaxies}}}.
\newblock {\emph{\JournalTitle{\apjl}}} \textbf{\bibinfo{volume}{973}}, \bibinfo{pages}{L32}, \doiprefix\url{https://dx.doi.org/10.3847/2041-8213/ad77b8} (\bibinfo{year}{2024}).
\newblock \eprint{2407.06281}.

\bibitem{1955ApJ...121..161S}
\bibinfo{author}{{Salpeter}, E.~E.}
\newblock \bibinfo{journal}{\bibinfo{title}{{The Luminosity Function and Stellar Evolution.}}}
\newblock {\emph{\JournalTitle{\apj}}} \textbf{\bibinfo{volume}{121}}, \bibinfo{pages}{161}, \doiprefix\url{https://dx.doi.org/10.1086/145971} (\bibinfo{year}{1955}).

\bibitem{2001MNRAS.322..231K}
\bibinfo{author}{{Kroupa}, P.}
\newblock \bibinfo{journal}{\bibinfo{title}{{On the variation of the initial mass function}}}.
\newblock {\emph{\JournalTitle{\mnras}}} \textbf{\bibinfo{volume}{322}}, \bibinfo{pages}{231--246}, \doiprefix\url{https://dx.doi.org/10.1046/j.1365-8711.2001.04022.x} (\bibinfo{year}{2001}).
\newblock \eprint{astro-ph/0009005}.

\bibitem{2009ApJ...693.1142S}
\bibinfo{author}{{Sun}, M.} \emph{et~al.}
\newblock \bibinfo{journal}{\bibinfo{title}{{Chandra Studies of the X-Ray Gas Properties of Galaxy Groups}}}.
\newblock {\emph{\JournalTitle{\apj}}} \textbf{\bibinfo{volume}{693}}, \bibinfo{pages}{1142--1172}, \doiprefix\url{https://dx.doi.org/10.1088/0004-637X/693/2/1142} (\bibinfo{year}{2009}).
\newblock \eprint{0805.2320}.

\bibitem{2010ApJ...719..119D}
\bibinfo{author}{{Dai}, X.}, \bibinfo{author}{{Bregman}, J.~N.}, \bibinfo{author}{{Kochanek}, C.~S.} \& \bibinfo{author}{{Rasia}, E.}
\newblock \bibinfo{journal}{\bibinfo{title}{{On the Baryon Fractions in Clusters and Groups of Galaxies}}}.
\newblock {\emph{\JournalTitle{\apj}}} \textbf{\bibinfo{volume}{719}}, \bibinfo{pages}{119--125}, \doiprefix\url{https://dx.doi.org/10.1088/0004-637X/719/1/119} (\bibinfo{year}{2010}).
\newblock \eprint{0911.2230}.

\bibitem{2014PhRvD..89h3501A}
\bibinfo{author}{{Amara}, A.} \& \bibinfo{author}{{Refregier}, A.}
\newblock \bibinfo{journal}{\bibinfo{title}{{Model breaking measure for cosmological surveys}}}.
\newblock {\emph{\JournalTitle{\prd}}} \textbf{\bibinfo{volume}{89}}, \bibinfo{pages}{083501}, \doiprefix\url{https://dx.doi.org/10.1103/PhysRevD.89.083501} (\bibinfo{year}{2014}).
\newblock \eprint{1309.5955}.

\bibitem{2017JCAP...10..045N}
\bibinfo{author}{{Nicola}, A.}, \bibinfo{author}{{Amara}, A.} \& \bibinfo{author}{{Refregier}, A.}
\newblock \bibinfo{journal}{\bibinfo{title}{{Integrated cosmological probes: concordance quantified}}}.
\newblock {\emph{\JournalTitle{\jcap}}} \textbf{\bibinfo{volume}{2017}}, \bibinfo{pages}{045}, \doiprefix\url{https://dx.doi.org/10.1088/1475-7516/2017/10/045} (\bibinfo{year}{2017}).
\newblock \eprint{1706.06593}.

\bibitem{2017MNRAS.470.2100K}
\bibinfo{author}{{Krause}, E.} \& \bibinfo{author}{{Eifler}, T.}
\newblock \bibinfo{journal}{\bibinfo{title}{{cosmolike - cosmological likelihood analyses for photometric galaxy surveys}}}.
\newblock {\emph{\JournalTitle{\mnras}}} \textbf{\bibinfo{volume}{470}}, \bibinfo{pages}{2100--2112}, \doiprefix\url{https://dx.doi.org/10.1093/mnras/stx1261} (\bibinfo{year}{2017}).
\newblock \eprint{1601.05779}.

\bibitem{2018arXiv180901669T}
\bibinfo{author}{{The LSST Dark Energy Science Collaboration}} \emph{et~al.}
\newblock \bibinfo{journal}{\bibinfo{title}{{The LSST Dark Energy Science Collaboration (DESC) Science Requirements Document}}}.
\newblock {\emph{\JournalTitle{arXiv e-prints}}} \bibinfo{pages}{arXiv:1809.01669}, \doiprefix\url{https://dx.doi.org/10.48550/arXiv.1809.01669} (\bibinfo{year}{2018}).
\newblock \eprint{1809.01669}.

\bibitem{2024JCAP...11..018Z}
\bibinfo{author}{{Zubeldia}, {\'I}.} \& \bibinfo{author}{{Bolliet}, B.}
\newblock \bibinfo{journal}{\bibinfo{title}{{cosmocnc: A fast, flexible, and accurate framework for galaxy cluster number count likelihood computation}}}.
\newblock {\emph{\JournalTitle{\jcap}}} \textbf{\bibinfo{volume}{2024}}, \bibinfo{pages}{018}, \doiprefix\url{https://dx.doi.org/10.1088/1475-7516/2024/11/018} (\bibinfo{year}{2024}).
\newblock \eprint{2403.09589}.

\bibitem{2026arXiv260212174W}
\bibinfo{author}{{Wayland}, A.}, \bibinfo{author}{{Alonso}, D.} \& \bibinfo{author}{{Reischke}, R.}
\newblock \bibinfo{journal}{\bibinfo{title}{{Probing baryonic feedback with fast radio bursts: joint analyses with cosmic shear and galaxy clustering}}}.
\newblock {\emph{\JournalTitle{arXiv e-prints}}} \bibinfo{pages}{arXiv:2602.12174}, \doiprefix\url{https://dx.doi.org/10.48550/arXiv.2602.12174} (\bibinfo{year}{2026}).
\newblock \eprint{2602.12174}.

\bibitem{2025arXiv250801648C}
\bibinfo{author}{{Caleb}, M.} \emph{et~al.}
\newblock \bibinfo{journal}{\bibinfo{title}{{A fast radio burst from the first 3 billion years of the Universe}}}.
\newblock {\emph{\JournalTitle{arXiv e-prints}}} \bibinfo{pages}{arXiv:2508.01648}, \doiprefix\url{https://dx.doi.org/10.48550/arXiv.2508.01648} (\bibinfo{year}{2025}).
\newblock \eprint{2508.01648}.

\bibitem{2023Sci...382..294R}
\bibinfo{author}{{Ryder}, S.~D.} \emph{et~al.}
\newblock \bibinfo{journal}{\bibinfo{title}{{A luminous fast radio burst that probes the Universe at redshift 1}}}.
\newblock {\emph{\JournalTitle{Science}}} \textbf{\bibinfo{volume}{382}}, \bibinfo{pages}{294--299}, \doiprefix\url{https://dx.doi.org/10.1126/science.adf2678} (\bibinfo{year}{2023}).
\newblock \eprint{2210.04680}.

\bibitem{2024ApJ...963L..34G}
\bibinfo{author}{{Gordon}, A.~C.} \emph{et~al.}
\newblock \bibinfo{journal}{\bibinfo{title}{{A Fast Radio Burst in a Compact Galaxy Group at z {\ensuremath{\sim}} 1}}}.
\newblock {\emph{\JournalTitle{\apjl}}} \textbf{\bibinfo{volume}{963}}, \bibinfo{pages}{L34}, \doiprefix\url{https://dx.doi.org/10.3847/2041-8213/ad2773} (\bibinfo{year}{2024}).
\newblock \eprint{2311.10815}.

\bibitem{2025PASA...42...36S}
\bibinfo{author}{{Shannon}, R.~M.} \emph{et~al.}
\newblock \bibinfo{journal}{\bibinfo{title}{{The commensal real-time ASKAP fast transient incoherent-sum survey}}}.
\newblock {\emph{\JournalTitle{\pasa}}} \textbf{\bibinfo{volume}{42}}, \bibinfo{pages}{e036}, \doiprefix\url{https://dx.doi.org/10.1017/pasa.2025.8} (\bibinfo{year}{2025}).
\newblock \eprint{2408.02083}.

\bibitem{2024ApJ...964..131S}
\bibinfo{author}{{Sherman}, M.~B.} \emph{et~al.}
\newblock \bibinfo{journal}{\bibinfo{title}{{Deep Synoptic Array Science: Polarimetry of 25 New Fast Radio Bursts Provides Insights into Their Origins}}}.
\newblock {\emph{\JournalTitle{\apj}}} \textbf{\bibinfo{volume}{964}}, \bibinfo{pages}{131}, \doiprefix\url{https://dx.doi.org/10.3847/1538-4357/ad275e} (\bibinfo{year}{2024}).
\newblock \eprint{2308.06813}.

\bibitem{2025MNRAS.tmp.2025P}
\bibinfo{author}{{Pastor-Marazuela}, I.} \emph{et~al.}
\newblock \bibinfo{journal}{\bibinfo{title}{{Localisation and host galaxy identification of new Fast Radio Bursts with MeerKAT}}}.
\newblock {\emph{\JournalTitle{\mnras}}} \doiprefix\url{https://dx.doi.org/10.1093/mnras/staf2144} (\bibinfo{year}{2025}).
\newblock \eprint{2507.05982}.

\bibitem{2019Natur.572..352R}
\bibinfo{author}{{Ravi}, V.} \emph{et~al.}
\newblock \bibinfo{journal}{\bibinfo{title}{{A fast radio burst localized to a massive galaxy}}}.
\newblock {\emph{\JournalTitle{\nat}}} \textbf{\bibinfo{volume}{572}}, \bibinfo{pages}{352--354}, \doiprefix\url{https://dx.doi.org/10.1038/s41586-019-1389-7} (\bibinfo{year}{2019}).
\newblock \eprint{1907.01542}.

\bibitem{2024ApJ...967...29L}
\bibinfo{author}{{Law}, C.~J.} \emph{et~al.}
\newblock \bibinfo{journal}{\bibinfo{title}{{Deep Synoptic Array Science: First FRB and Host Galaxy Catalog}}}.
\newblock {\emph{\JournalTitle{\apj}}} \textbf{\bibinfo{volume}{967}}, \bibinfo{pages}{29}, \doiprefix\url{https://dx.doi.org/10.3847/1538-4357/ad3736} (\bibinfo{year}{2024}).
\newblock \eprint{2307.03344}.

\bibitem{2020ApJ...899..161L}
\bibinfo{author}{{Law}, C.~J.} \emph{et~al.}
\newblock \bibinfo{journal}{\bibinfo{title}{{A Distant Fast Radio Burst Associated with Its Host Galaxy by the Very Large Array}}}.
\newblock {\emph{\JournalTitle{\apj}}} \textbf{\bibinfo{volume}{899}}, \bibinfo{pages}{161}, \doiprefix\url{https://dx.doi.org/10.3847/1538-4357/aba4ac} (\bibinfo{year}{2020}).
\newblock \eprint{2007.02155}.

\bibitem{2024arXiv240514182F}
\bibinfo{author}{{Faber}, J.~T.} \emph{et~al.}
\newblock \bibinfo{journal}{\bibinfo{title}{{A Heavily Scattered Fast Radio Burst Is Viewed Through Multiple Galaxy Halos}}}.
\newblock {\emph{\JournalTitle{arXiv e-prints}}} \bibinfo{pages}{arXiv:2405.14182}, \doiprefix\url{https://dx.doi.org/10.48550/arXiv.2405.14182} (\bibinfo{year}{2024}).
\newblock \eprint{2405.14182}.

\bibitem{2020ApJ...903..152H}
\bibinfo{author}{{Heintz}, K.~E.} \emph{et~al.}
\newblock \bibinfo{journal}{\bibinfo{title}{{Host Galaxy Properties and Offset Distributions of Fast Radio Bursts: Implications for Their Progenitors}}}.
\newblock {\emph{\JournalTitle{\apj}}} \textbf{\bibinfo{volume}{903}}, \bibinfo{pages}{152}, \doiprefix\url{https://dx.doi.org/10.3847/1538-4357/abb6fb} (\bibinfo{year}{2020}).
\newblock \eprint{2009.10747}.

\bibitem{2021ApJ...917...75M}
\bibinfo{author}{{Mannings}, A.~G.} \emph{et~al.}
\newblock \bibinfo{journal}{\bibinfo{title}{{A High-resolution View of Fast Radio Burst Host Environments}}}.
\newblock {\emph{\JournalTitle{\apj}}} \textbf{\bibinfo{volume}{917}}, \bibinfo{pages}{75}, \doiprefix\url{https://dx.doi.org/10.3847/1538-4357/abff56} (\bibinfo{year}{2021}).
\newblock \eprint{2012.11617}.

\bibitem{2025ApJ...993..119G}
\bibinfo{author}{{Gordon}, A.~C.} \emph{et~al.}
\newblock \bibinfo{journal}{\bibinfo{title}{{Mapping the Spatial Distribution of Fast Radio Bursts within their Host Galaxies}}}.
\newblock {\emph{\JournalTitle{\apj}}} \textbf{\bibinfo{volume}{993}}, \bibinfo{pages}{119}, \doiprefix\url{https://dx.doi.org/10.3847/1538-4357/ae0298} (\bibinfo{year}{2025}).
\newblock \eprint{2506.06453}.

\bibitem{2022AJ....163...69B}
\bibinfo{author}{{Bhandari}, S.} \emph{et~al.}
\newblock \bibinfo{journal}{\bibinfo{title}{{Characterizing the Fast Radio Burst Host Galaxy Population and its Connection to Transients in the Local and Extragalactic Universe}}}.
\newblock {\emph{\JournalTitle{\aj}}} \textbf{\bibinfo{volume}{163}}, \bibinfo{pages}{69}, \doiprefix\url{https://dx.doi.org/10.3847/1538-3881/ac3aec} (\bibinfo{year}{2022}).
\newblock \eprint{2108.01282}.

\bibitem{2024MNRAS.532.3881R}
\bibinfo{author}{{Rajwade}, K.~M.} \emph{et~al.}
\newblock \bibinfo{journal}{\bibinfo{title}{{A study of two FRBs with low polarization fractions localized with the MeerTRAP transient buffer system}}}.
\newblock {\emph{\JournalTitle{\mnras}}} \textbf{\bibinfo{volume}{532}}, \bibinfo{pages}{3881--3892}, \doiprefix\url{https://dx.doi.org/10.1093/mnras/stae1652} (\bibinfo{year}{2024}).
\newblock \eprint{2407.02173}.

\bibitem{2019Sci...365..565B}
\bibinfo{author}{{Bannister}, K.~W.} \emph{et~al.}
\newblock \bibinfo{journal}{\bibinfo{title}{{A single fast radio burst localized to a massive galaxy at cosmological distance}}}.
\newblock {\emph{\JournalTitle{Science}}} \textbf{\bibinfo{volume}{365}}, \bibinfo{pages}{565--570}, \doiprefix\url{https://dx.doi.org/10.1126/science.aaw5903} (\bibinfo{year}{2019}).
\newblock \eprint{1906.11476}.

\bibitem{2024ApJ...973...64W}
\bibinfo{author}{{Woodland}, M.~N.} \emph{et~al.}
\newblock \bibinfo{journal}{\bibinfo{title}{{The Environments of Fast Radio Bursts Viewed Using Adaptive Optics}}}.
\newblock {\emph{\JournalTitle{\apj}}} \textbf{\bibinfo{volume}{973}}, \bibinfo{pages}{64}, \doiprefix\url{https://dx.doi.org/10.3847/1538-4357/ad643c} (\bibinfo{year}{2024}).
\newblock \eprint{2312.01578}.

\bibitem{2017ApJ...843L...8B}
\bibinfo{author}{{Bassa}, C.~G.} \emph{et~al.}
\newblock \bibinfo{journal}{\bibinfo{title}{{FRB 121102 Is Coincident with a Star-forming Region in Its Host Galaxy}}}.
\newblock {\emph{\JournalTitle{\apjl}}} \textbf{\bibinfo{volume}{843}}, \bibinfo{pages}{L8}, \doiprefix\url{https://dx.doi.org/10.3847/2041-8213/aa7a0c10.48550/arXiv.1705.07698} (\bibinfo{year}{2017}).
\newblock \eprint{1705.07698}.

\bibitem{2026Sci...391..280L}
\bibinfo{author}{{Li}, Y.} \emph{et~al.}
\newblock \bibinfo{journal}{\bibinfo{title}{{A sudden change and recovery in the magnetic environment around a repeating fast radio burst}}}.
\newblock {\emph{\JournalTitle{Science}}} \textbf{\bibinfo{volume}{391}}, \bibinfo{pages}{280--284}, \doiprefix\url{https://dx.doi.org/10.1126/science.adq3225} (\bibinfo{year}{2026}).
\newblock \eprint{2503.04727}.

\bibitem{2024NatAs...8.1429C}
\bibinfo{author}{{Cassanelli}, T.} \emph{et~al.}
\newblock \bibinfo{journal}{\bibinfo{title}{{A fast radio burst localized at detection to an edge-on galaxy using very-long-baseline interferometry}}}.
\newblock {\emph{\JournalTitle{Nature Astronomy}}} \textbf{\bibinfo{volume}{8}}, \bibinfo{pages}{1429--1442}, \doiprefix\url{https://dx.doi.org/10.1038/s41550-024-02357-x} (\bibinfo{year}{2024}).
\newblock \eprint{2307.09502}.

\bibitem{2023MNRAS.524.2064C}
\bibinfo{author}{{Caleb}, M.} \emph{et~al.}
\newblock \bibinfo{journal}{\bibinfo{title}{{A subarcsec localized fast radio burst with a significant host galaxy dispersion measure contribution}}}.
\newblock {\emph{\JournalTitle{\mnras}}} \textbf{\bibinfo{volume}{524}}, \bibinfo{pages}{2064--2077}, \doiprefix\url{https://dx.doi.org/10.1093/mnras/stad1839} (\bibinfo{year}{2023}).
\newblock \eprint{2302.09754}.

\bibitem{2025ApJ...979L..22E}
\bibinfo{author}{{Eftekhari}, T.} \emph{et~al.}
\newblock \bibinfo{journal}{\bibinfo{title}{{The Massive and Quiescent Elliptical Host Galaxy of the Repeating Fast Radio Burst FRB 20240209A}}}.
\newblock {\emph{\JournalTitle{\apjl}}} \textbf{\bibinfo{volume}{979}}, \bibinfo{pages}{L22}, \doiprefix\url{https://dx.doi.org/10.3847/2041-8213/ad9de2} (\bibinfo{year}{2025}).
\newblock \eprint{2410.23336}.

\bibitem{2025ApJ...979L..21S}
\bibinfo{author}{{Shah}, V.} \emph{et~al.}
\newblock \bibinfo{journal}{\bibinfo{title}{{A Repeating Fast Radio Burst Source in the Outskirts of a Quiescent Galaxy}}}.
\newblock {\emph{\JournalTitle{\apjl}}} \textbf{\bibinfo{volume}{979}}, \bibinfo{pages}{L21}, \doiprefix\url{https://dx.doi.org/10.3847/2041-8213/ad9ddc} (\bibinfo{year}{2025}).
\newblock \eprint{2410.23374}.

\bibitem{2021ApJ...922..173C}
\bibinfo{author}{{Chittidi}, J.~S.} \emph{et~al.}
\newblock \bibinfo{journal}{\bibinfo{title}{{Dissecting the Local Environment of FRB 190608 in the Spiral Arm of its Host Galaxy}}}.
\newblock {\emph{\JournalTitle{\apj}}} \textbf{\bibinfo{volume}{922}}, \bibinfo{pages}{173}, \doiprefix\url{https://dx.doi.org/10.3847/1538-4357/ac2818} (\bibinfo{year}{2021}).
\newblock \eprint{2005.13158}.

\bibitem{2022MNRAS.513..982R}
\bibinfo{author}{{Ravi}, V.} \emph{et~al.}
\newblock \bibinfo{journal}{\bibinfo{title}{{The host galaxy and persistent radio counterpart of FRB 20201124A}}}.
\newblock {\emph{\JournalTitle{\mnras}}} \textbf{\bibinfo{volume}{513}}, \bibinfo{pages}{982--990}, \doiprefix\url{https://dx.doi.org/10.1093/mnras/stac46510.48550/arXiv.2106.09710} (\bibinfo{year}{2022}).
\newblock \eprint{2106.09710}.

\bibitem{2022ApJ...927L...3N}
\bibinfo{author}{{Nimmo}, K.} \emph{et~al.}
\newblock \bibinfo{journal}{\bibinfo{title}{{Milliarcsecond Localization of the Repeating FRB 20201124A}}}.
\newblock {\emph{\JournalTitle{\apjl}}} \textbf{\bibinfo{volume}{927}}, \bibinfo{pages}{L3}, \doiprefix\url{https://dx.doi.org/10.3847/2041-8213/ac540f} (\bibinfo{year}{2022}).
\newblock \eprint{2111.01600}.

\bibitem{2025ApJ...993..221A}
\bibinfo{author}{{Anna-Thomas}, R.} \emph{et~al.}
\newblock \bibinfo{journal}{\bibinfo{title}{{Evidence for a Hot Galactic Halo around the Andromeda Galaxy Using Fast Radio Bursts along Two Sightlines}}}.
\newblock {\emph{\JournalTitle{\apj}}} \textbf{\bibinfo{volume}{993}}, \bibinfo{pages}{221}, \doiprefix\url{https://dx.doi.org/10.3847/1538-4357/ae1014} (\bibinfo{year}{2025}).
\newblock \eprint{2503.02947}.

\bibitem{2023ApJ...950..175S}
\bibinfo{author}{{Sharma}, K.} \emph{et~al.}
\newblock \bibinfo{journal}{\bibinfo{title}{{Deep Synoptic Array Science: A Massive Elliptical Host Among Two Galaxy-cluster Fast Radio Bursts}}}.
\newblock {\emph{\JournalTitle{\apj}}} \textbf{\bibinfo{volume}{950}}, \bibinfo{pages}{175}, \doiprefix\url{https://dx.doi.org/10.3847/1538-4357/accf1d} (\bibinfo{year}{2023}).
\newblock \eprint{2302.14782}.

\bibitem{2023ApJ...949L..26C}
\bibinfo{author}{{Connor}, L.} \emph{et~al.}
\newblock \bibinfo{journal}{\bibinfo{title}{{Deep Synoptic Array Science: Two Fast Radio Burst Sources in Massive Galaxy Clusters}}}.
\newblock {\emph{\JournalTitle{\apjl}}} \textbf{\bibinfo{volume}{949}}, \bibinfo{pages}{L26}, \doiprefix\url{https://dx.doi.org/10.3847/2041-8213/acd3ea} (\bibinfo{year}{2023}).
\newblock \eprint{2302.14788}.

\bibitem{2023ApJ...949L...3R}
\bibinfo{author}{{Ravi}, V.} \emph{et~al.}
\newblock \bibinfo{journal}{\bibinfo{title}{{Deep Synoptic Array Science: Discovery of the Host Galaxy of FRB 20220912A}}}.
\newblock {\emph{\JournalTitle{\apjl}}} \textbf{\bibinfo{volume}{949}}, \bibinfo{pages}{L3}, \doiprefix\url{https://dx.doi.org/10.3847/2041-8213/acc4b6} (\bibinfo{year}{2023}).
\newblock \eprint{2211.09049}.

\bibitem{2025ApJS..280....6C}
\bibinfo{author}{{Chime/Frb Collaboration}} \emph{et~al.}
\newblock \bibinfo{journal}{\bibinfo{title}{{A Catalog of Local Universe Fast Radio Bursts from CHIME/FRB and the KKO}}}.
\newblock {\emph{\JournalTitle{\apjs}}} \textbf{\bibinfo{volume}{280}}, \bibinfo{pages}{6}, \doiprefix\url{https://dx.doi.org/10.3847/1538-4365/addbda} (\bibinfo{year}{2025}).
\newblock \eprint{2502.11217}.

\bibitem{2023ApJ...950..134M}
\bibinfo{author}{{Michilli}, D.} \emph{et~al.}
\newblock \bibinfo{journal}{\bibinfo{title}{{Subarcminute Localization of 13 Repeating Fast Radio Bursts Detected by CHIME/FRB}}}.
\newblock {\emph{\JournalTitle{\apj}}} \textbf{\bibinfo{volume}{950}}, \bibinfo{pages}{134}, \doiprefix\url{https://dx.doi.org/10.3847/1538-4357/accf89} (\bibinfo{year}{2023}).
\newblock \eprint{2212.11941}.

\bibitem{2022MNRAS.514.1961R}
\bibinfo{author}{{Rajwade}, K.~M.} \emph{et~al.}
\newblock \bibinfo{journal}{\bibinfo{title}{{First discoveries and localizations of Fast Radio Bursts with MeerTRAP: real-time, commensal MeerKAT survey}}}.
\newblock {\emph{\JournalTitle{\mnras}}} \textbf{\bibinfo{volume}{514}}, \bibinfo{pages}{1961--1974}, \doiprefix\url{https://dx.doi.org/10.1093/mnras/stac1450} (\bibinfo{year}{2022}).
\newblock \eprint{2205.14600}.

\bibitem{2021ApJ...908L..12T}
\bibinfo{author}{{Tendulkar}, S.~P.} \emph{et~al.}
\newblock \bibinfo{journal}{\bibinfo{title}{{The 60 pc Environment of FRB 20180916B}}}.
\newblock {\emph{\JournalTitle{\apjl}}} \textbf{\bibinfo{volume}{908}}, \bibinfo{pages}{L12}, \doiprefix\url{https://dx.doi.org/10.3847/2041-8213/abdb3810.48550/arXiv.2011.03257} (\bibinfo{year}{2021}).
\newblock \eprint{2011.03257}.

\bibitem{2022evlb.confE..35M}
\bibinfo{author}{{Marcote}, B.} \emph{et~al.}
\newblock \bibinfo{title}{{PRECISE localizations of repeating Fast Radio Bursts}}.
\newblock In \emph{\bibinfo{booktitle}{European VLBI Network Mini-Symposium and Users' Meeting 2021}}, vol. \bibinfo{volume}{2021}, \bibinfo{pages}{35}, \doiprefix\url{https://dx.doi.org/10.22323/1.399.0035} (\bibinfo{year}{2022}).
\newblock \eprint{2202.11644}.

\bibitem{2023PASA...40...29L}
\bibinfo{author}{{Lee-Waddell}, K.} \emph{et~al.}
\newblock \bibinfo{journal}{\bibinfo{title}{{The host galaxy of FRB 20171020A revisited}}}.
\newblock {\emph{\JournalTitle{\pasa}}} \textbf{\bibinfo{volume}{40}}, \bibinfo{pages}{e029}, \doiprefix\url{https://dx.doi.org/10.1017/pasa.2023.27} (\bibinfo{year}{2023}).
\newblock \eprint{2305.17960}.

\bibitem{2018ApJ...867L..10M}
\bibinfo{author}{{Mahony}, E.~K.} \emph{et~al.}
\newblock \bibinfo{journal}{\bibinfo{title}{{A Search for the Host Galaxy of FRB 171020}}}.
\newblock {\emph{\JournalTitle{\apjl}}} \textbf{\bibinfo{volume}{867}}, \bibinfo{pages}{L10}, \doiprefix\url{https://dx.doi.org/10.3847/2041-8213/aae7cb} (\bibinfo{year}{2018}).
\newblock \eprint{1810.04354}.

\bibitem{2024ApJ...971L..51B}
\bibinfo{author}{{Bhardwaj}, M.} \emph{et~al.}
\newblock \bibinfo{journal}{\bibinfo{title}{{Host Galaxies for Four Nearby CHIME/FRB Sources and the Local Universe FRB Host Galaxy Population}}}.
\newblock {\emph{\JournalTitle{\apjl}}} \textbf{\bibinfo{volume}{971}}, \bibinfo{pages}{L51}, \doiprefix\url{https://dx.doi.org/10.3847/2041-8213/ad64d1} (\bibinfo{year}{2024}).
\newblock \eprint{2310.10018}.

\bibitem{2024ApJ...961...99I}
\bibinfo{author}{{Ibik}, A.~L.} \emph{et~al.}
\newblock \bibinfo{journal}{\bibinfo{title}{{Proposed Host Galaxies of Repeating Fast Radio Burst Sources Detected by CHIME/FRB}}}.
\newblock {\emph{\JournalTitle{\apj}}} \textbf{\bibinfo{volume}{961}}, \bibinfo{pages}{99}, \doiprefix\url{https://dx.doi.org/10.3847/1538-4357/ad0893} (\bibinfo{year}{2024}).
\newblock \eprint{2304.02638}.

\bibitem{2021ApJ...919L..24B}
\bibinfo{author}{{Bhardwaj}, M.} \emph{et~al.}
\newblock \bibinfo{journal}{\bibinfo{title}{{A Local Universe Host for the Repeating Fast Radio Burst FRB 20181030A}}}.
\newblock {\emph{\JournalTitle{\apjl}}} \textbf{\bibinfo{volume}{919}}, \bibinfo{pages}{L24}, \doiprefix\url{https://dx.doi.org/10.3847/2041-8213/ac223b} (\bibinfo{year}{2021}).
\newblock \eprint{2108.12122}.

\bibitem{2019Sci...366..231P}
\bibinfo{author}{{Prochaska}, J.~X.} \emph{et~al.}
\newblock \bibinfo{journal}{\bibinfo{title}{{The low density and magnetization of a massive galaxy halo exposed by a fast radio burst}}}.
\newblock {\emph{\JournalTitle{Science}}} \textbf{\bibinfo{volume}{366}}, \bibinfo{pages}{231--234}, \doiprefix\url{https://dx.doi.org/10.1126/science.aay0073} (\bibinfo{year}{2019}).
\newblock \eprint{1909.11681}.

\bibitem{2025MNRAS.538.1800H}
\bibinfo{author}{{Hanmer}, K.~Y.} \emph{et~al.}
\newblock \bibinfo{journal}{\bibinfo{title}{{Contemporaneous optical-radio observations of a fast radio burst in a close galaxy pair}}}.
\newblock {\emph{\JournalTitle{\mnras}}} \textbf{\bibinfo{volume}{538}}, \bibinfo{pages}{1800--1815}, \doiprefix\url{https://dx.doi.org/10.1093/mnras/staf289} (\bibinfo{year}{2025}).
\newblock \eprint{2502.10153}.

\bibitem{2023MNRAS.519.2235P}
\bibinfo{author}{{Panther}, F.~H.} \emph{et~al.}
\newblock \bibinfo{journal}{\bibinfo{title}{{The most probable host of CHIME FRB 190425A, associated with binary neutron star merger GW190425, and a late-time transient search}}}.
\newblock {\emph{\JournalTitle{\mnras}}} \textbf{\bibinfo{volume}{519}}, \bibinfo{pages}{2235--2250}, \doiprefix\url{https://dx.doi.org/10.1093/mnras/stac3597} (\bibinfo{year}{2023}).
\newblock \eprint{2212.00954}.

\bibitem{2025ApJ...979...95Q}
\bibinfo{author}{{Qiang}, D.-C.}, \bibinfo{author}{{You}, Z.-Q.}, \bibinfo{author}{{Yang}, S.}, \bibinfo{author}{{Zhu}, Z.-H.} \& \bibinfo{author}{{Chen}, T.-W.}
\newblock \bibinfo{journal}{\bibinfo{title}{{3D Localization of FRB 20190425A for Its Potential Host Galaxy and Implications}}}.
\newblock {\emph{\JournalTitle{\apj}}} \textbf{\bibinfo{volume}{979}}, \bibinfo{pages}{95}, \doiprefix\url{https://dx.doi.org/10.3847/1538-4357/ada28c} (\bibinfo{year}{2025}).
\newblock \eprint{2411.13973}.

\bibitem{2022Natur.606..873N}
\bibinfo{author}{{Niu}, C.-H.} \emph{et~al.}
\newblock \bibinfo{journal}{\bibinfo{title}{{A repeating fast radio burst associated with a persistent radio source}}}.
\newblock {\emph{\JournalTitle{\nat}}} \textbf{\bibinfo{volume}{606}}, \bibinfo{pages}{873--877}, \doiprefix\url{https://dx.doi.org/10.1038/s41586-022-04755-5} (\bibinfo{year}{2022}).
\newblock \eprint{2110.07418}.

\bibitem{2023ApJ...948...67B}
\bibinfo{author}{{Bhandari}, S.} \emph{et~al.}
\newblock \bibinfo{journal}{\bibinfo{title}{{A Nonrepeating Fast Radio Burst in a Dwarf Host Galaxy}}}.
\newblock {\emph{\JournalTitle{\apj}}} \textbf{\bibinfo{volume}{948}}, \bibinfo{pages}{67}, \doiprefix\url{https://dx.doi.org/10.3847/1538-4357/acc178} (\bibinfo{year}{2023}).
\newblock \eprint{2211.16790}.

\bibitem{2025ApJ...980L..24C}
\bibinfo{author}{{Chen}, X.-L.} \emph{et~al.}
\newblock \bibinfo{journal}{\bibinfo{title}{{The Host Galaxy of the Hyperactive Repeating FRB 20240114A: Behind a Galaxy Cluster}}}.
\newblock {\emph{\JournalTitle{\apjl}}} \textbf{\bibinfo{volume}{980}}, \bibinfo{pages}{L24}, \doiprefix\url{https://dx.doi.org/10.3847/2041-8213/adadfd} (\bibinfo{year}{2025}).
\newblock \eprint{2502.05587}.

\bibitem{2024arXiv241201478B}
\bibinfo{author}{{Bruni}, G.} \emph{et~al.}
\newblock \bibinfo{journal}{\bibinfo{title}{{Discovery of a PRS associated with FRB 20240114A}}}.
\newblock {\emph{\JournalTitle{arXiv e-prints}}} \bibinfo{pages}{arXiv:2412.01478}, \doiprefix\url{https://dx.doi.org/10.48550/arXiv.2412.01478} (\bibinfo{year}{2024}).
\newblock \eprint{2412.01478}.

\bibitem{2025arXiv250620774M}
\bibinfo{author}{{Muller}, A.~R.} \emph{et~al.}
\newblock \bibinfo{journal}{\bibinfo{title}{{The Low Mass Dwarf Host Galaxy of Non-Repeating FRB 20230708A}}}.
\newblock {\emph{\JournalTitle{arXiv e-prints}}} \bibinfo{pages}{arXiv:2506.20774}, \doiprefix\url{https://dx.doi.org/10.48550/arXiv.2506.20774} (\bibinfo{year}{2025}).
\newblock \eprint{2506.20774}.

\bibitem{2024MNRAS.527.3659D}
\bibinfo{author}{{Driessen}, L.~N.} \emph{et~al.}
\newblock \bibinfo{journal}{\bibinfo{title}{{FRB 20210405I: a nearby Fast Radio Burst localized to sub-arcsecond precision with MeerKAT}}}.
\newblock {\emph{\JournalTitle{\mnras}}} \textbf{\bibinfo{volume}{527}}, \bibinfo{pages}{3659--3673}, \doiprefix\url{https://dx.doi.org/10.1093/mnras/stad3329} (\bibinfo{year}{2024}).
\newblock \eprint{2302.09787}.

\bibitem{2024ApJ...962L..13G}
\bibinfo{author}{{Glowacki}, M.} \emph{et~al.}
\newblock \bibinfo{journal}{\bibinfo{title}{{H I, FRB, What's Your z: The First FRB Host Galaxy Redshift from Radio Observations}}}.
\newblock {\emph{\JournalTitle{\apjl}}} \textbf{\bibinfo{volume}{962}}, \bibinfo{pages}{L13}, \doiprefix\url{https://dx.doi.org/10.3847/2041-8213/ad1f62} (\bibinfo{year}{2024}).
\newblock \eprint{2311.16808}.

\bibitem{2021ApJ...910L..18B}
\bibinfo{author}{{Bhardwaj}, M.} \emph{et~al.}
\newblock \bibinfo{journal}{\bibinfo{title}{{A Nearby Repeating Fast Radio Burst in the Direction of M81}}}.
\newblock {\emph{\JournalTitle{\apjl}}} \textbf{\bibinfo{volume}{910}}, \bibinfo{pages}{L18}, \doiprefix\url{https://dx.doi.org/10.3847/2041-8213/abeaa6} (\bibinfo{year}{2021}).
\newblock \eprint{2103.01295}.

\bibitem{2022Natur.602..585K}
\bibinfo{author}{{Kirsten}, F.} \emph{et~al.}
\newblock \bibinfo{journal}{\bibinfo{title}{{A repeating fast radio burst source in a globular cluster}}}.
\newblock {\emph{\JournalTitle{\nat}}} \textbf{\bibinfo{volume}{602}}, \bibinfo{pages}{585--589}, \doiprefix\url{https://dx.doi.org/10.1038/s41586-021-04354-w10.48550/arXiv.2105.11445} (\bibinfo{year}{2022}).
\newblock \eprint{2105.11445}.

\end{thebibliography}
\vspace{-0.5cm}  

\begin{addendum}
\item [Acknowledgments] During the preparation of this work, KS and EK were supported in part by grant NSF PHY-2309135 to the Kavli Institute for Theoretical Physics (KITP) and David \& Lucile Packard Foundation grant 2020-71384. This material is based upon work supported in part by the National Science Foundation under CAREER Grant Number 2240032. We thank Alexandra Amon, Andrew Pontzen, Daisuke Nagai, Gil Holder, Hiranya Peiris, Joel Leja and Sunil Simha for insightful discussions. We thank Chun-Hao To, Inigo Zubeldia, Nihar Dalal, Matthew McQuinn and Shivam Pandey for supporting the sensitivity analysis of Compton $y$-parameter--cosmic shear cross-correlations, tSZ Y-M relation, Simons Observatory tSZ galaxy clusters and kSZ stacks. We thank Jared Siegel for sharing $M_{200}$ and $z$ distributions for X-ray observations and kSZ effect. We thank Jiachuan Xu and Leah Bigwood for sharing matter power spectrum suppression measurements from the literature. 

\item[Author Contributions] K.S. conceived the study, developed the analysis framework, implemented the halo-model inference pipeline and performed the statistical analysis to model the FRB dispersion measure-redshift distribution and its connection to feedback-dependent matter power spectrum suppression. K.S. led the writing of the manuscript, with guidance from E.K. and V.R. L.C., D.A. and P.R.S. contributed to the interpretation of the results and supported the analysis by advising robustness tests to validate the methodology and conclusions.

\item[Competing Interests] The authors declare no competing interests.

\item[Correspondence] Correspondence and requests for materials should be addressed to K.S. and V.R.

\item[Code Availability Statement] We have created a reproduction package for our work that includes all code used for our data analysis. We have placed this code on GitHub at \url{https://github.com/krittisharma/dmz_spk_sharma2026}, which will be publicly available upon acceptance of this work.

\item[Data Availability Statement] The FRB data presented here is available in a CSV file in the same GitHub repository.

\end{addendum}

\newpage
\clearpage
\newpage
\newpage

\noindent{\bfseries \LARGE Methods}\setlength{\parskip}{12pt}

\setlength{\parskip}{12pt}

\noindent{\bfseries \large Sensitivity Analysis}
\setlength{\parskip}{3pt}

For a data vector $\hat{\mathbf{d}}$ with Gaussian likelihood $\mathcal{L}$ and theoretical model $\mathbf{m}$, we define the sensitivity with respect to halo mass and redshift as 
\begin{equation}
    S_{\hat{\mathbf{d}}}(\log M_{200}, z) = \left|\frac{\partial^2 \mathcal{L}}{\partial \log M_{200} \partial z}\right| \propto  \left|\frac{\partial \mathbf{m}}{\partial \log M_{200}} C_{\hat{\mathbf{d}}}^{-1} \frac{\partial \mathbf{m}}{\partial z}\right|\,
    \label{eqn:sensitivity_estimator_correlation_functions}
\end{equation}
where we expand around the maximum likelihood point (assuming zero model miss-specification), which causes the second derivative term to vanish. Likewise, we also compute the scale sensitivity. This definition combines the sensitivity of different angular and tomographic bins weighted by the data covariance $C_{\hat{\mathbf{d}}}$. This is in contrast to previous works~\protect\citemethods{2025PhRvD.112f3541L}, where the sensitivity is defined in terms of the second derivative of the observable in a single angular/tomographic bin. Despite the different definitions, our analysis finds qualitatively similar sensitivity of different probes. In the halo model framework, the model is written as an integral over redshift (aggregating contributions along the line of sight) and halo mass, making the partial derivatives straightforward to evaluate. To isolate the component of model sensitive to baryonic feedback, we elect to use only the one-halo term. We note that the source/cluster redshift distribution is held fixed in this analysis.
 
The cosmic shear ($\xi_\pm$) and shear-$y$ cross-correlation ($\xi^{\gamma y}$) data vectors includes correlations across 4 tomography bins with median redshifts 0.28, 0.47, 0.74 and 0.94, and 20 angular bins in DES-Y3 public data release~\protect\citemethods{2022PhRvD.105b3515S}$^,$\citep{2025arXiv250607432P, 2022PhRvD.105b3514A}. We account for the experimental noise in observational constraints by using the DES-Y3 and ACT-DR5 analyses covariance matrices $C$~\protect\citemethods{2022PhRvD.105b3515S}$^,$\citep{2025arXiv250607432P, 2022PhRvD.105b3514A}. 
For these two-point statistics, the Limber approximation provides a relation between the redshift and Fourier mode $k$ contributing to a given angular scale. Hence the mass and scale sensitivity $S_{\hat{\mathbf{d}}}(\log M_{200}, k)$ is computed using the relation
\begin{equation}
    \frac{\partial C(\ell)}{\partial k} = \frac{\partial C(\ell)}{\partial z}|_{z(\chi = \ell/k)} \frac{\partial z}{\partial k} = -\frac{H(z)}{c}\frac{\ell}{k^2}\frac{\partial C(\ell)}{\partial z}|_{z(\chi = \ell/k)}
    \label{eqn:limber_mapping}
\end{equation}

To illustrate the sensitivity of FRB DM variance, we consider the log-normal distribution of DM$_\mathrm{cosmic}$. For a single FRB at redshift $z_\mathrm{s}$, the likelihood of obtaining the measurement $\hat{X} = \log(\mathrm{DM}_\mathrm{cosmic}(z_{\mathrm{s}}))$ is
\begin{equation}
    \mathcal{L}_\mathrm{DM}(\hat{X}|\mu(z_{\mathrm{s}}), \sigma(z_{\mathrm{s}})) = \frac{1}{\sqrt{2\pi}\sigma(z_{\mathrm{s}})\hat{X}}
    \exp\left[
    -\frac{(\hat{X}-\mu(z_{\mathrm{s}}))^2}{2\sigma^2(z_{\mathrm{s}})}
    \right]\, 
\end{equation}
where the mean and variance of the log-normal distribution are related to the mean and variance of DM$_\mathrm{cosmic}$ as following:
\begin{equation}
\mu(z_\mathrm{s}) = \log\left(\frac{\langle \mathrm{DM}_\mathrm{cosmic}(z_\mathrm{s})\rangle^2}{\sqrt{\langle \mathrm{DM}_\mathrm{cosmic}(z_\mathrm{s})\rangle^2+\sigma^2[\mathrm{DM}_\mathrm{cosmic}(z_\mathrm{s}})]}\right)\,,
\end{equation}
\begin{equation}
\sigma^2(z_\mathrm{s}) =\log\left(1+\frac{\sigma^2[\mathrm{DM}_\mathrm{cosmic}(z_\mathrm{s})]}{\langle \mathrm{DM}_\mathrm{cosmic}(z_\mathrm{s})\rangle^2}\right)\,.
\end{equation}
For a sample of FRBs with host redshifts $z_i$, the sensitivity of observed FRB sample is then computed by combining the individual sensitivities,
\begin{equation}
    S_{\mathrm{DM}_\mathrm{cosmic}}(\log M_{200}, z) \propto \sum_{z_{\mathrm{s}_i}}\frac{1}{\sigma^2(z_{\mathrm{s}_i})}\left|\frac{\partial \mu(z_{\mathrm{s}_i})}{\partial z}\frac{\partial \mu(z_{\mathrm{s}_i})}{\partial \log M_{200}}\right|\,.
    \label{FRBsensitivity:Mz}
\end{equation}
The computation of $\sigma[\mathrm{DM}_{\mathrm{cosmic}}(z_{\rm s})]$ (Eqn.~\ref{eqn:variance}) contains integrals over both redshift and Fourier modes. Hence the mass and scale sensitivity is calculated as 
\begin{equation}
    S_{\mathrm{DM}_\mathrm{cosmic}}(\log M_{200}, k) \propto \sum_{z_{\mathrm{s}_i}}\frac{1}{\sigma^2(z_{\mathrm{s}_i})}\left|\frac{\partial \mu(z_{\mathrm{s}_i})}{\partial k}\frac{\partial \mu(z_{\mathrm{s}_i})}{\partial \log M_{200}}\right|\,,
    \label{FRBsensitivity:Mk}
\end{equation}
without the Limber approximation mapping between scale and redshift (Eqn.~\ref{eqn:limber_mapping}). We include the noise from variance in host DMs in the observer frame at FRB redshift by adding it in quadrature to $\sigma^2[\mathrm{DM}_\mathrm{cosmic}(z_\mathrm{s})]$ in our calculations, where we assume rest-frame host DM variance of 100~pc\,cm$^{-3}$. Note that in the above sensitivity calculations, the derivative of $\langle \mathrm{DM}_\mathrm{cosmic} (z_{\mathrm{s}}) \rangle$ w.r.t $M_\mathrm{200}$ and $k$ is zero (see Eqn.~\ref{eqn:meanDM}).

For all sensitivity calculations, we use $M_\mathrm{200}$ halo mass definition, which is the total mass enclosed within a sphere whose average density is 200-times the critical density of the Universe. We use the Tinker (2008) halo mass function~\protect\citemethods{2008ApJ...688..709T}, Tinker (2010) halo bias~\protect\citemethods{2010ApJ...724..878T} and Diemer \& Kravtsov (2015) halo mass-concentration relation~\protect\citemethods{2015ApJ...799..108D} in our calculations.

\textbf{Sensitivity synopsis of baryon probes:} The tSZ effect traces hot gas in massive galaxy clusters, making the tSZ Y-M relation~\protect\citemethods{2021MNRAS.504.4312G, 2021ApJS..253....3H}$^,$\citep{2025arXiv250704476D} sensitive to $M_\mathrm{200} \sim 10^{14.5}\,M_\odot$ halos at redshifts $z \sim 0.15-0.5$~\protect\citemethods{2021ApJS..253....3H}. The effective sensitivity of tSZ is extended to $M_\mathrm{200} \sim 10^{14}-10^{15}\,M_\odot$ halos at redshifts $z \sim 0.1-0.6$ and scales $k \sim 1-15~h\,\mathrm{Mpc}^{-1}$ by cross-correlating Compton $y$ maps from Planck~\protect\citemethods{2020A&A...643A..42P} and ACT DR4–6~\protect\citemethods{2020JCAP...12..047A, 2024PhRvD.109f3530C, 2025JCAP...11..061N} with DES-Y3 cosmic shear measurements~\citep{2025arXiv250607432P}. Cosmic shear correlations, for example from DES-Y3~\citep{2022PhRvD.105b3514A}$^,$\protect\citemethods{2022PhRvD.105b3515S} are sensitive to halo masses $M_\mathrm{200} \sim 10^{13}-10^{15}\,M_\odot$, redshifts $z \sim 0.05-0.45$, and scales $k \sim 0.5-10~h\,\mathrm{Mpc}^{-1}$. The ACT kSZ stacks~\protect\citemethods{2021PhRvD.103f3513S, 2025PhRvD.112j3512R} on DESI BGS~\protect\citemethods{2023AJ....165..253H} and LRG~\protect\citemethods{2023AJ....165...58Z} samples~\citep{2025arXiv250910455S} span halo mass ranges of $M_\mathrm{200} \sim 10^{13}-10^{14}\,M_\odot$ and $\sim 10^{13}-10^{13.5}\,M_\odot$, at redshifts $z \sim 0.4-1$ and $\sim 0.1-0.5$, respectively. X-ray observations, for example from eROSITA eRASS1 clusters~\protect\citemethods{2024A&A...685A.106B, 2024A&A...688A.210K}$^,$\citep{2025arXiv250910455S}, on the other hand, are sensitive to hot gas in the inner regions of $M_\mathrm{200} \sim 10^{13}-10^{15}\,M_\odot$ galaxy clusters at $z < 0.25$.

\textbf{Halo mass sensitivity of matter power spectrum suppression:} We assess the range of halo masses contributing to the suppression of matter power spectrum at scales of interest by computing the ratio of a modified power spectrum to its gravity-only analog, where the modified power spectrum is defined such that the halos below mass scale, denoted $M_0$, are in a gravity-only regime, whereas the halos above this mass scale include the effects of baryonic feedback. The results from this exercise are shown in Fig.~\ref{fig:sensitivity_analysis}, where we show that at intermediate scales of $0.1-5~h\,$Mpc$^{-1}$, halos with mass $> 10^{13}\,M_\odot$ dominantly contribute to the matter power spectrum suppression. This overlaps with the scales and halo mass sensitivity of most baryon probes.

\newpage

\setlength{\parskip}{12pt}
\noindent{\bfseries \large Inference Framework}
\setlength{\parskip}{3pt}

The conditional probability of observing an FRB with extragalactic DM contribution, DM$_\mathrm{exgal}$, at redshift $z_\mathrm{s}$, $p(\mathrm{DM}_\mathrm{exgal} | z_\mathrm{s})$, is sensitive to the influence of feedback on gas power spectrum~\citep{2014ApJ...780L..33M, 2024ApJ...973..151K, 2025NatAs...9.1226C, 2025ApJ...989...81S}\protect\citemethods{2022ApJ...928....9L, 2023arXiv230909766R, 2024ApJ...965...57B, 2025ApJ...983...46M}. The DM$_\mathrm{exgal}$, obtained by removing the Galactic component~\protect\citemethods{2002astro.ph..7156C, 2003astro.ph..1598C, 2019ascl.soft08022Y, 2019MNRAS.485..648P, 2020ApJ...888..105Y, 2025AJ....169..330R, 2023ApJ...946...58C}, includes contribution from the IGM and intervening halos, $\mathrm{DM}_\mathrm{cosmic}$, and the FRB host galaxy, $\mathrm{DM}_\mathrm{host}/(1+z_\mathrm{s})$, where DM$_\mathrm{host}$ is the rest-frame host galaxy DM. Assuming a log-normal DM$_\mathrm{host}$ distribution~\citep{2020ApJ...900..170Z} with parameters $\mu_\mathrm{host}$ and $\sigma_\mathrm{host}$, 
the distribution $p(\mathrm{DM}_\mathrm{exgal} | z_\mathrm{s})$ is written as
\begin{equation}
        p(\mathrm{DM}_\mathrm{exgal} | z_\mathrm{s}) = \int\limits_0^{\mathrm{DM}_\mathrm{exgal}} 
        p(\mathrm{DM}_\mathrm{cosmic}|z_\mathrm{s}) \times p\left( \mathrm{DM}_\mathrm{host} | z_\mathrm{s}, \mu_\mathrm{host}, \sigma_\mathrm{host} \right) 
        \mathrm{d} \mathrm{DM}_\mathrm{cosmic}.
    \label{eqn:DMexgal}
\end{equation}

The \textit{mean} of the $p(\mathrm{DM}_\mathrm{cosmic}|z_\mathrm{s})$ distribution explicitly depends on the diffuse baryon fraction, $f_\mathrm{diffuse}(z)$ (defined as the baryon fraction not locked in stellar matter ($f_\mathrm{star}(z)$) and atomic or molecular hydrogen gas ($f_\mathrm{ISM}(z)$)~\protect\citemethods{2025A&A...695A.163B, 2020ARA&A..58..363P}), and cosmology\footnote{Throughout this work, we use Planck18 cosmology~\protect\citemethods{2020A&A...641A...6P} with Hubble constant $H_0=67.66$\,km\,s$^{-1}$Mpc$^{-1}$, baryon density $\Omega_\mathrm{b}=0.049$, matter density $\Omega_\mathrm{m}=0.31$, dark energy density $\Omega_{\Lambda}=0.69$ and the root-mean-square amplitude of the linear matter density fluctuations smoothed with a top-hat filter of comoving radius 8\,$h^{-1}$Mpc, $\sigma_8=0.811$.}~\citep{2020Natur.581..391M}:

\begin{equation}
    \langle \mathrm{DM}_\mathrm{cosmic} (z_\mathrm{s}) \rangle = \int\limits_0^{z_\mathrm{s}} \frac{3c \chi_\mathrm{e} \Omega_\mathrm{b} H_0}{8 \pi G m_\mathrm{p}} \frac{f_\mathrm{diffuse}(z) (1+z)~\mathrm{d}z}{\sqrt{\Omega_\mathrm{m}(1+z)^3 + \Omega_{\Lambda}}}
\label{eqn:meanDM}
\end{equation}
where the integrand of an equivalent integral using comoving distance ($\chi$) is the radial kernel $W_\mathrm{DM}(\chi)$. The free electron fraction depends on the primordial helium abundance, $\chi_\mathrm{e} = Y_\mathrm{H} + Y_\mathrm{He}/2 \approx 1 - Y_\mathrm{He}/2$~\protect\citemethods{2020A&A...641A...6P}.

The \textit{variance} of the $p(\mathrm{DM}_\mathrm{cosmic}|z_{\mathrm{s}})$ distribution arises from intervening collapsed structures along the line of sight~\citep{2014ApJ...780L..33M} and can be written under the flat-sky and Limber approximation as~\protect\citemethods{2023MNRAS.524.2237R}$^,$\citep{2025ApJ...989...81S}:
\vspace{-0.3cm}
\begin{equation}
    \sigma^2 [\mathrm{DM}_\mathrm{cosmic}(z_\mathrm{s})] = \int\limits_0^{\chi_\mathrm{s}} \mathrm{d}\chi W_\mathrm{DM}^2(\chi) \int\limits_0^\infty \frac{k \mathrm{d}k}{2\pi} P_\mathrm{ee} (k, z(\chi)|\theta),
\label{eqn:variance}
\end{equation}
where $P_\mathrm{ee}(k,z)$ is the electron power spectrum, which depends on the set of feedback parameters, $\theta$. Therefore, the spread in DM$_\mathrm{exgal} - z$ relation is sensitive to the degree of clustering of baryons through $P_\mathrm{ee}(k,z)$ and correlates with the suppression in matter power spectrum upto scales of $k \lesssim 10~h$~Mpc$^{-1}$ (see Fig.~7 of reference\citep{2025ApJ...989...81S}). Equations~\ref{eqn:meanDM} and \ref{eqn:variance} establish that both, the mean DM$_\mathrm{exgal} - z$ relation and the scatter in DM$_\mathrm{exgal}$ at a given $z$ carry cosmological and astrophysically valuable information.

During Bayesian inference, assuming that FRB observations are independent, the posterior for the parameters $\Gamma = \{\theta, \mu_\mathrm{host}, \sigma_\mathrm{host}\}$ is written as
\begin{equation}
    p(\Gamma | \mathrm{obs}) \propto \prod_{i=1}^{N_\mathrm{FRB}} \mathcal{L}(\mathrm{DM}_\mathrm{exgal,i} | z_i, \Gamma) \times \pi(\Gamma),
\end{equation}
where $N_\mathrm{FRB}$ is the size of our FRB sample, $\theta$ is the set of feedback parameters and $\pi(\Gamma)$ is the prior distribution on free parameters, as summarized in Extended Data Table~\ref{table:parameters}. The log-likelihood for each observed FRB is constructed as $\mathcal{L}(\mathrm{DM}_\mathrm{exgal,i} | z_i, \Gamma) = p(\mathrm{DM}_{\mathrm{exgal}, i} | z_{\mathrm{s},i}, \Gamma)$, as defined using Eqn.~\ref{eqn:DMexgal}. The $\mathrm{DM}_\mathrm{cosmic}$ distribution, $p(\mathrm{DM}_\mathrm{cosmic} | z_{\mathrm{s}})$, has been demonstrated to be well-approximated by a log-normal distribution in hydrodynamical simulations~\citep{2025ApJ...989...81S}\protect\citemethods{2025arXiv250707090K}, with mean and variance defined in equations~\ref{eqn:meanDM} and \ref{eqn:variance}, respectively. The dependence on feedback parameters $\theta$ is explicit through the mean and variance of DM$_\mathrm{cosmic}$ distribution (see Eqn.~\ref{eqn:meanDM} and \ref{eqn:variance}). While covariance between sightlines arising from large-scale structure is not included here, its inclusion will be critical for unbiased inference with larger FRB samples~\protect\citemethods{2023MNRAS.524.2237R}. We sample the posteriors using the affine-invariant ensemble sampler, \textsc{emcee}~\protect\citemethods{2013PASP..125..306F}.

\setlength{\parskip}{12pt}
\noindent{\bfseries \large Halo Model Approach}
\setlength{\parskip}{3pt}

The variance in FRB DMs explicitly depend on the degree of clustering of baryons through the electron power spectrum, $P_\mathrm{ee}$, which is approximately equal to the gas power spectrum $P_\mathrm{gas}$ upto $k \sim 5~h$\,Mpc$^{-1}$ at $\lesssim$1\%-level accuracy (the electron bias, $b_\mathrm{e} \approx 1$)~\citep{2025arXiv250919514L}. We model the gas power spectrum using two halo model prescriptions where the gas profiles are: (i) calibrated to variations of hydrodynamical simulations \textsc{HMcode}~\protect\citemethods[][]{2020A&A...641A.130M, 2021MNRAS.502.1401M}, and (ii) defined analytically with flexible parameters \textsc{BCEmu}~\protect\citemethods[][]{2015JCAP...12..049S}$^,$\citep{2019JCAP...03..020S, 2021JCAP...12..046G}. These gas profiles are implemented in \textsc{BaryonForge}\footnote{\url{https://github.com/DhayaaAnbajagane/BaryonForge}}~\protect\citemethods{2024OJAp....7E.108A} and the power spectra are computed using \textsc{pyccl}\footnote{\url{https://github.com/LSSTDESC/CCL}} software package. We summarize the free parameters in both, \textsc{HMcode} and \textsc{BCEmu} in Extended Data Table~\ref{table:parameters}. Below, we summarize the key equations, and refer the reader to references there-in for more detailed discussion/implementation details.

In \textsc{HMcode}~\protect\citemethods{2020A&A...641A.130M, 2021MNRAS.502.1401M}, the gas profile parameters were calibrated to the variations of AGN heating temperature, $T_\mathrm{AGN}$, in the BAHAMAS hydrodynamical simulations~\protect/citemethods{2017MNRAS.465.2936M}. Formally, $T_\mathrm{AGN}$ is related to the BAHAMAS subgrid heating parameter $\Delta T_\mathrm{heat}$, which controls the occurrence of AGN feedback such that it activates only after assembling sufficient energy to heat a fixed number of gas particles by $\Delta T_\mathrm{heat}$~\protect/citemethods{2017MNRAS.465.2936M}. Therefore, varying $T_\mathrm{AGN}$ alters the strength of baryonic feedback, thus varying the halo stellar fraction, bound gas fraction, concentration and shape of the gas profiles. Calibrated to scales $k < 10~h$\,Mpc$^{-1}$ at redshifts $z < 2$, fitting for matter, cold dark matter, stars and gas profiles, the performance of this model on BAHAMAS simulations is accurate at $< 2$\%-level, with a prior range $T_\mathrm{AGN} \in [7.6, 8]$. However, we caution that a single parameter is generally not sufficient to fully capture the scale-dependent complexity of baryonic feedback~\citep{2019JCAP...03..020S} and the accuracy of this model is specific to the feedback implementation within BAHAMAS simulations, which may or may not be close to the true scenario. Due to its inherent limitations, we use this model primarily as a cross-check against our baseline results obtained with \textsc{BCEmu} model. Its simplicity also enables controlled tests of systematic uncertainties associated with the host galaxy DM contribution (see Extended Data Fig.~\ref{fig:HMcode_cornerplot}).

Analytical \textsc{BCEmu} is a more flexible approach which allows modeling baryonic feedback effects on the matter distribution without explicitly assuming a specific subgrid feedback physics. The radial gas density profile, $\rho_\mathrm{gas}$, of a halo of mass $M$ is modeled as~\protect\citemethods{2015JCAP...12..049S}$^,$\citep{2019JCAP...03..020S, 2021JCAP...12..046G}
\begin{equation}
    \rho_\mathrm{gas}(r|M) = \dfrac{\rho_\mathrm{gas, 0}}{\left[ 1+\left( \dfrac{r}{\theta_\mathrm{co} r_\mathrm{200c}} \right)^\beta \right] \left[ 1+\left( \dfrac{r}{\theta_\mathrm{ej} r_\mathrm{200c}} \right)^\gamma \right]^{\frac{\delta-\beta}{\gamma}}},
    \label{eqn:gas_profile}
\end{equation}
where $\rho_\mathrm{gas, 0}$ is a normalization constant. Essentially, this includes a cored radial profile (with the core of the profile fixed to $\theta_\mathrm{co} = 0.1$~\protect\citemethods{2024OJAp....7E.108A}$^,$\citep{2024MNRAS.534..655B}) truncated at the ejection radius, $\theta_\mathrm{ej} r_\mathrm{200c}$. The halo mass dependent slope of the gas profile ($\beta$) is written as
\begin{equation}
    \beta = \dfrac{3(M/M_c)^{\mu_\beta}}{1+(M/M_c)^{\mu_\beta}},
    \label{eqn:gas_profile_slope}
\end{equation}
where $M_c$ is the halo mass below which the gas profile becomes shallower than the NFW profile and $\mu_\beta$ controls the halo mass scaling of the slope, thus accounting for more efficient gas retention in galaxy clusters as opposed to galaxy groups~\protect\citemethods{2015JCAP...12..049S}$^,$\citep{2019JCAP...03..020S, 2021JCAP...12..046G}. The remaining slope parameters $\{\delta, \gamma\}$ control the slope of the profile after truncation. The total gas fraction in this model is written as 
\begin{equation}
    f_\mathrm{gas}(M) = \frac{\Omega_\mathrm{b}}{\Omega_\mathrm{m}} - f_\mathrm{star}(M),
    \label{eqn:f_gas}
\end{equation}
where the total stellar fraction (including central and satellite galaxies) is
\begin{equation}
    f_\mathrm{star}(M) = \dfrac{2A}{\left(\dfrac{M}{M_1}\right)^\tau + \left(\dfrac{M}{M_1}\right)^\eta}.
    \label{eqn:f_star}
\end{equation} 

Here, $\tau$ and $\eta$ are the low and high mass-end slopes of stellar-to-halo mass relation, respectively, and $A$ is a normalization. This double power law is essential to model the astrophysical down-turn of stellar fraction in low mass halos. We fix $\tau = -1.5$~\protect\citemethods{2019MNRAS.488.3143B} and pivot mass $M_1=2.5\times10^{11}\,M_\odot h^{-1}$~\protect\citemethods{2013MNRAS.428.3121M}, following standard practices in baryonification literature~\protect\citemethods{2015JCAP...12..049S}$^,$\citep{2019JCAP...03..020S, 2021JCAP...12..046G}. The stellar content locked in central galaxy indirectly affect the amount of gas that can be pushed out by feedback, and is parameterized as 
\begin{equation}
    f_\mathrm{cga}(M) = \dfrac{2A}{\left(\dfrac{M}{M_1}\right)^{\tau+\tau_\delta} + \left(\dfrac{M}{M_1}\right)^{\eta+\eta_\delta}},
    \label{eqn:f_cga}
\end{equation} 
where $\tau_\delta$ (fixed to zero~\protect\citemethods{2025arXiv250713317P, 2024OJAp....7E.108A}) and $\eta_\delta$ are the power law indices.

We use state-of-the-art software and procedures to compute and emulate power spectra under \textsc{HMcode} and \textsc{BCEmu} frameworks during inference~\protect\citemethods{2024OJAp....7E.108A, 2021MNRAS.506.4070A}$^,$\citep{2026ApJ...999..202S}. These neural network emulators, built on 50,000 samples, exhibit $<4$\% accuracy in predicting $\sigma [\mathrm{DM}_\mathrm{cosmic}(z)]$ upto redshift $z \sim 5$ and $<2$\% accuracy in predicting $P_\mathrm{hydro}(k)/P_\mathrm{gravity}(k)$ at $z=0$ upto scales $k \leq 100~h$\,Mpc$^{-1}$. For \textsc{HMcode}, the only feedback parameter is $\theta = \{\log T_\mathrm{AGN}\}$, whereas, for the \textsc{BCEmu7} model, the feedback parameters are $\theta = \{\log M_c, \theta_\mathrm{ej}, \mu_\beta, \delta, \eta, \gamma, \eta_\delta\}$.

\setlength{\parskip}{12pt}
\noindent{\bfseries \large Plausible DM$_\mathrm{host}$ redshift evolution}
\setlength{\parskip}{3pt}

We show the posterior distributions for inference conducted using \textsc{HMcode} halo model in Extended Data Fig.~\ref{fig:HMcode_cornerplot}. At the feedback parameter level, the extended FRB sample and the fiducial subset yield consistent constraints within 
$1\sigma$, indicating that current measurements are not strongly sensitive to sample selection. However, the apparent shift in posteriors between the fiducial and extended FRB sample may be indicative of a potential DM$_\mathrm{host}$ redshift evolution. 

The scenario of extreme $\mathrm{DM}_{\mathrm{host}}$ redshift evolution is unlikely. Excess $\mathrm{DM}_{\mathrm{cosmic}}$ around intervening halos has been robustly detected through stacking analyses of FRBs on foreground galaxies~\protect\citemethods{2025ApJ...993L..27H} and through cross-correlations with galaxies~\protect\citemethods{2025arXiv250608932W} at $\sim 5\sigma$ significance. Such detections would not be expected if the variance in $\mathrm{DM}_{\mathrm{exgal}}$ were dominated by redshift evolution of $\mathrm{DM}_{\mathrm{host}}$ alone. The observations of local FRB samples indicate that the typical magnitude of $\mathrm{DM}_{\mathrm{host}}$ in rest-frame is $\sim 150$~pc\,cm$^{-3}$~\citep{2025ApJ...991L..25L}\protect\citemethods{2025AJ....169..330R}. In addition, no strong redshift evolution is observed in FRB rotation measures~\protect\citemethods{2023ApJ...957L...8S} and scattering delays~\protect\citemethods{2022ApJ...934...71O}. If $\mathrm{DM}_{\mathrm{host}}$ exhibited strong redshift evolution, corresponding trends in these observables would be anticipated, but are not seen. 

To assess the impact of a plausible redshift evolution, expected in the scenario where FRBs trace star-formation~\protect\citemethods{2023ApJ...954...80G}$^,$\citep{2024Natur.635...61S}, we examine a model in which the median $\mathrm{DM}_{\mathrm{host}}$ follows the cosmic star-formation rate history \protect\citemethods{2014ARA&A..52..415M, 2023ApJ...954...80G}$^,$\citep{2024Natur.635...61S}. While the $\log T_{\mathrm{AGN}}$ posterior remains stable within the $1\sigma$ uncertainties, both the mean and variance of $\mathrm{DM}_{\mathrm{host}}$ shift to lower values (see Extended Data Fig.~\ref{fig:HMcode_cornerplot}). This indicates that our current constraints on baryonic feedback are robust to plausible redshift evolution of $\mathrm{DM}_{\mathrm{host}}$. However, these results highlight the need to develop the redshift lever arm required to directly constrain $\mathrm{DM}_{\mathrm{host}}$ evolution, which may become an important systematic for next-generation FRB experiments.

\setlength{\parskip}{12pt}
\noindent{\bfseries \large Constrained Baryonic Feedback Parameters}
\setlength{\parskip}{3pt}

We show the posterior distributions for inference conducted using \textsc{BCEmu7} halo model in Extended Data Fig.~\ref{fig:BCEmu_cornerplot}\footnote{While we impose an informed prior on $\eta$ using independent measurements of stellar mass fractions in groups and clusters throughout this work in \textsc{BCEmu5} and \textsc{BCEmu7} models, for the purpose of identifying directions/parameters constrained by FRBs, we remove this prior and conduct CPC analysis with posterior samples from an analysis assuming wide flat $\eta$ prior.}. We summarize the corresponding parameter constraints in Extended Data Table~\ref{table:HMcode_BCEmu_constraints}. While the 7 parameter model offers more flexibility, from the posterior distribution, it appears that many of the parameters are unconstrained/prior-dominated.

To quantify the effective degrees of freedom constrained by the data, identify parameter combinations that are actually constrained, and the ones that remain prior-dominated, we analyze the derived 7-dimensional posterior and compare it against the corresponding prior. We treat the prior and posterior as Gaussian approximations~\protect\citemethods{2019PhRvD..99d3506R} defined by the covariances $C_\Pi$ (prior) and 
$C_p$ (posterior). The covariances satisfy $C_p^{-1} = C_\Pi^{-1} + C^{-1}$, where $C^{-1}$ is the empirical Fisher matrix representing the information from data alone. The effective number of degrees of freedom constrained by the data is defined as
\begin{equation}
N_{\rm eff} \equiv 
N - \mathrm{Tr}\!\left(C_\Pi^{-1} C_p\right),
\label{eq:Neff}
\end{equation}
with $N=7$ in our case. Applying Eqn.~(\ref{eq:Neff}), we obtain $N_{\rm eff} \simeq 1.24$, indicating that the data effectively constrain only one linear combination of the seven parameters; the remaining directions are prior-dominated. To identify which directions in parameter space are constrained, we perform a Karhunen-Loève (KL) decomposition of the posterior relative to the prior.  
We solve the generalized eigenvalue problem:
\begin{equation}
C_p \, \mathbf{v}_i = \lambda_i \, C_\Pi \, \mathbf{v}_i,
\end{equation}
where $\lambda_i$ quantifies the improvement of the posterior variance over the prior and $\mathbf{v}_i$ is the corresponding KL eigenvector. Modes with $\lambda_i \gg 1$ are constrained by the data and modes with $\lambda_i \approx 1$ (or $<1$) are unconstrained/prior-dominated.

The KL decomposition yields the following improvement factors of the
posterior variance over the prior: $\lambda = 
(6.00, 1.21, 1.11, 1.09, 1.05, 0.97, 0.60)$, corresponding to fractional error improvements $\sqrt{\lambda - 1} =
(2.24, 0.46, 0.34, 0.31, 0.22, 0, 0, 0)$. The largest eigenvalue corresponds to an improvement factor of $\lambda_1 \simeq 6$ for the first mode, indicating a strongly constrained parameter combination, while all remaining modes show no meaningful improvement ($\lambda \lesssim 1$) relative to the prior. Together with $N_{\rm eff} \simeq 1.24$, these values confirm that the FRB likelihood constrains $\sim$ one direction in the 7-dimensional parameter space.

The Covariant Principal Components (CPC) analysis using the \textsc{tensiometer} framework\footnote{\url{https://github.com/mraveri/tensiometer}} yields the following variance improvements per mode: $(5.00,\ 0.21,\ 0.11,\ 0.09,\ -0.05,\ -0.03,\ -0.40)$. Thus, CPC mode 1 is improved by a factor $5$ (i.e.\ $224\%$ reduction in variance), the rest of the modes are noise-dominated and unconstrained. The best constrained CPC combination:
\begin{equation}
\begin{aligned}
    \mathrm{CPC}_1: \quad & +1.00\,(\log M_c - 13.05) +0.41\,(\theta_{\rm ej}-4.68) -0.93(\mu_\beta -0.97)
    \,-0.35(\delta - 7.28),
\end{aligned}
    \label{eqn:CPCmode1}
\end{equation} 
strongly projects onto $\log M_c$, with subdominant contributions from $\theta_{\rm ej}$, $\mu_\beta$ and $\delta$. For reference, from the fractional Fisher matrix decomposition, we get the following KL mode 1 contributions: $(\log M_c: 0.554, \theta_{\rm ej}: 0.107, \delta: 0.116, \gamma: 0.064, \mu_\beta: 0.157, \eta: 0.000, \eta_\delta: 0.003)$. Thus, the FRB data constrain a single dominant combination primarily aligned with $\log M_c$ ($\sim 55$\% of KL mode 1). Weak projections on $\theta_{\rm ej}$, $\mu_\beta$ and $\delta$ indicate mild partial sensitivity through parameter degeneracies. Parameters $\eta$, $\gamma$ and $\eta_\delta$ project almost entirely onto unconstrained KL modes and are therefore prior-dominated. 

We visualize the contribution of this constrained degree of freedom to the measured power spectrum suppression by using posterior samples for these 4 parameters and randomly sampling the prior for the rest of the three parameters. We show the results of this exercise in Extended Data Fig.~\ref{fig:spk_BCEmu7_CPC}. The first mode of CPC demonstrates a significant improvement over the prior volume of \textsc{BCEmu7}, thus strengthening the robustness of our constraining power.

We tested the data-constrained structure of the posterior by projecting the posterior samples onto the KL basis and zeroing out the unconstrained modes ($i > N_{\rm eff}=1$):
$\mathbf{s}_{\rm filtered} 
= V \, 
\begin{pmatrix}
\alpha_1, 0, 0, \hdots, 0
\end{pmatrix}
+ \boldsymbol{\mu}_{\Pi},$
where $V$ is the matrix of KL eigenvectors, $\alpha_i$ are coordinates of each sample in the KL basis, and $\boldsymbol{\mu}_{\Pi}$ is the prior mean. Rotating back to the original parameter basis yields a KL-filtered posterior. This posterior includes only the directions that the data genuinely constrain. A parameter whose contour remains unchanged under this procedure is one whose variation lies primarily in the constrained subspace. A parameter whose contour changes significantly must have substantial projection onto unconstrained KL modes and is therefore prior-driven. In our analysis, the only parameter whose contour remain seemingly unchanged is $\log M_c$ - this is genuinely constrained by the data. The rest are all prior-dominated or only connected to constrained directions through parameter degeneracies.

Overall, this quantitative analysis confirms that the FRB DM-$z$ likelihood effectively constrains $\sim$1.24 degrees of freedom (1 strong mode), dominated by $\log M_c$, with weak contributions from $\theta_\mathrm{ej}$, $\mu_\beta$ and $\delta$. Parameters such as $\eta$, $\gamma$ and $\eta_\delta$ project strongly onto unconstrained KL modes, so their apparent posterior structure is largely prior-driven. Based on this analysis, we define our reduced complexity model, \textsc{BCEmu4}, where we vary $\log M_c$, $\theta_\mathrm{ej}$, $\mu_\beta$ and $\delta$, while fixing other parameters to their fiducial value listed in Extended Data Table~\ref{table:parameters}.

\setlength{\parskip}{12pt}
\noindent{\bfseries \large Fixed Parameters and Cosmology Dependence}
\setlength{\parskip}{3pt}

Throughout this work, we have fixed the cosmology to Planck18 TT,TE,EE+lowE+lensing cosmology~\protect\citemethods{2020A&A...641A...6P}. Furthermore, motivated by our CPC analysis, we fixed the \textsc{BCEmu} model parameters $\eta$, $\eta_\delta$ and $\gamma$. We test the sensitivity of the measured matter power spectrum suppression to different cosmology and aforementioned feedback parameters. We test changing from Planck18 to DES-Y3 cosmology with $\Omega_\mathrm{m}=0.293$ and $\sigma_8=0.764$. The observed change in matter power spectrum suppression in Extended Data Fig.~\ref{fig:spk_vary_feedback_cosmology} can be attributed to the change in Universe baryon fraction $\Omega_\mathrm{b}/\Omega_\mathrm{m}$. We also vary $\eta$, $\eta_\delta$ and $\gamma$ from their fiducial value one-by-one and conduct inference using the \textsc{BCEmu4} model. We show the results from this exercise in Extended Data Fig.~\ref{fig:spk_vary_feedback_cosmology}. While our measurement with \textsc{BCEmu4} model is fairly robust to variations of $\eta_\delta$ and $\gamma$ at scales that FRBs are sensitive to, $\eta$ on the other hand significantly impact the measurement across all scales. While the impact of $\eta$ on power spectrum suppression might seem counterintuitive at first, one can understand this better by looking at corresponding variations in fraction of baryons in gas, $f_\mathrm{gas} = M_\mathrm{gas}/M_\mathrm{halo}$ as a function of $R/R_{200}$ (see Supplementary Fig.~\ref{fig:f_gasR_f_starR_f_baryonsR_SPk}). The change in $\eta$ impacts the total gas fraction, $f_\mathrm{gas} (R \rightarrow \infty)$, in $10^{13}-10^{14}~M_\odot$ halos by 40-45\%. This stems from the boundary condition that $f_\mathrm{gas} (M_\mathrm{halo}, R \rightarrow \infty)+f_\mathrm{star} (M_\mathrm{halo}, R \rightarrow \infty) = \Omega_\mathrm{b}/\Omega_\mathrm{m}$ (see Eqn.~\ref{eqn:f_gas}). Since matter power spectrum suppression is specifically sensitive to the fraction of baryons in gas phase (the baryons impacted by feedback physics), the imprint of stellar-to-halo mass relation variations ($\eta$) is reflected at intermediate scales. This suggests we need to extend the model to vary $\eta$ (\textsc{BCEmu5}). However, the inability of FRBs (and indeed, any probe of baryons) to constrain $\eta$ due to a lack of constraining power necessitates a better-motivated prior on the SHM relation.

\setlength{\parskip}{12pt}
\noindent{\bfseries \large Stellar-to-Halo Mass Relation Prior}
\setlength{\parskip}{3pt}

\textbf{Slope at high-mass end:} Since we find that the choice of fiducial $\eta$ impacts the final measurement of interest (see Fig.~\ref{fig:spk_vary_feedback_cosmology}, left panel), we extend the model to fit for stellar-to-halo mass (SHM) relation, in addition to FRB DM variance, while imposing informed priors on SHM model parameters. Within the \textsc{BCEmu} modeling framework, we fit for the SHM relation measurements from two datasets: (i) stellar mass measurements from SDSS optical data and halo mass measurements from Chandra X-ray observations of galaxy clusters with median halo mass $M_\mathrm{500} = 14.48$~\citep{2018AstL...44....8K}\protect\citemethods{2006ApJ...640..691V, 2013ApJ...778...14G, 2013MNRAS.429.3288S} and (ii) stellar mass measurements from DES optical, WISE/Spitzer near-infrared data and halo mass measurements from SPT SZ effect scaling relation of galaxy clusters with median halo mass $\log M_\mathrm{500} = 14.68$~\protect\citemethods{2018MNRAS.478.3072C}. We show these two measurement sets in Extended Data Fig.~\ref{fig:eta_prior}. We conduct joint fits to FRBs with these two datasets, assuming a fixed normalization of $A=0.055/2$ and a fixed slope of SHM relation at low-mass end $\tau=-1.5$ (following standard practices~\protect\citemethods{2019MNRAS.488.3143B, 2024OJAp....7E.108A, 2025arXiv250713317P}). We use a broad $\eta$ prior, uniform in the range 0.05-0.4. The posteriors on the power law index are $\eta = 0.22_{-0.05}^{+0.06}$ and $\eta = 0.28_{-0.02}^{+0.02}$ for the two datasets, respectively. We show the best-fit SHM relation at the high-mass end for the two datasets in Extended Data Fig.~\ref{fig:eta_prior}. We also show the corresponding constraints on power spectrum suppression from \textsc{BCEmu5}, which are consistent between the two datasets.

\textbf{Slope at low-mass end:} The fiducial $\tau=-1.5$ is based on literature measurement~\protect\citemethods{2019MNRAS.488.3143B}, which is inferred by forward-modeling galaxy growth in N-body halo merger trees and fitting to observed stellar mass functions, star-formation rates, and clustering data (stellar masses come from SED analysis-based galaxy survey measurements, while halo masses come entirely from dark-matter simulations). Fixing $\tau$ is justified by the fact that the primary physical quantity of interest in this work - the matter power spectrum suppression on intermediate scales - is only weakly sensitive to the baryon content of low-mass halos. Indeed, varying $\tau$ over an extreme range, from $\tau = 0$ to $\tau = -4$, changes the matter power spectrum suppression at $\lesssim 2\%$-level, which is well below the current constraining power.

\textbf{Normalization of SHM relation and the impact of Stellar Initial Mass Function:} The fixed normalization of SHM relation in this work ($A=0.028$) assumes a Chabrier IMF~\protect\citemethods{2003PASP..115..763C}. Since the variations at the low-mass end of the IMF\footnote{We note that testing modifications to the high-mass end of the IMF is considerably more challenging. Such changes alter the colors of stellar population models and therefore require a full re-fitting of galaxy spectral energy distributions to robustly assess their impact on stellar mass estimates. Moreover, the resulting effects are non-monotonic: while a higher fraction of massive stars initially lowers the mass-to-light ratio in star-forming systems, the subsequent build up of stellar remnants can increase the mass-to-light ratio in older, quiescent populations~\protect\citemethods{2024ApJ...973L..32V}. Given the absence of a consistent framework for modeling the time evolution of the high-mass end of the IMF, these variations cannot be systematically explored within the scope of this study.} primarily rescale stellar masses without significantly altering galaxy luminosities, we can apply standard conversion factors to convert this normalization to different IMFs. We apply the standard +0.24~dex correction to convert from Chabrier to Salpeter~\protect\citemethods{1955ApJ...121..161S} IMF, and +0.05~dex correction for the Kroupa~\protect\citemethods{2001MNRAS.322..231K} IMF. For clarity, we focus on the more extreme Chabrier-to-Salpeter conversion (resulting in $A=0.047$) in what follows. We show the variations in $f_\mathrm{gas} = M_\mathrm{gas}/M_\mathrm{halo}$ as a function of $R/R_{200}$ and power spectrum suppression for this range of normalizations in Supplementary Fig.~\ref{fig:f_gasR_f_starR_f_baryonsR_SPk}. The variations in $A$ impact power spectrum suppression on intermediate scales at $\lesssim 5$\% level, which is below the current constraining power. Consequently, we conclude that the choice of stellar IMF is not a dominant systematic for power spectrum suppression measurement with current FRB samples (and any other probe of baryons).

We validate this conclusion by perturbing the stellar fraction measurements, adopting alternative IMF prescriptions. Supplementary Fig.~\ref{fig:impact_of_IMF} shows the resulting stellar mass fractions~\protect\citemethods{2018MNRAS.478.3072C} under alternative IMF assumptions. We fit the SHM relation within our \textsc{BCEmu5} framework, where the normalization is fixed to $A=0.028$ for measurements with Chabrier IMF and $A=0.047$ for Salpeter IMF. The corresponding posterior distributions are shown in Supplementary Fig.~\ref{fig:impact_of_IMF}. Clearly, the power spectrum suppression constraints are currently unbiased by the choice of stellar IMF.

\textbf{$f_\mathrm{diffuse}$ as a metric of feedback strength:} When imposing a prior on the SHM relation derived from SED-based stellar mass measurements, the assumed stellar IMF has a non-negligible impact on the inferred diffuse baryon fraction. This effect is illustrated in Supplementary Fig.~\ref{fig:impact_of_IMF}, where different IMF choices produce a systematic shift in the posterior of $f_\mathrm{diffuse}$. This behavior is expected by construction. The quantity $f_\mathrm{diffuse}$ is defined by subtracting the stellar mass from the total baryon budget, and is therefore directly sensitive to the normalization of the SHM relation. As shown in Supplementary Fig.~\ref{fig:parameter_variations-f_gasR200M_f_starR200M_SPk}, variations in the IMF (and hence, $A$) primarily rescale $f_\mathrm{star}$, leading to a corresponding shift in $f_\mathrm{diffuse}$.

At first glance, Supplementary Fig.~\ref{fig:parameter_variations-f_gasR200M_f_starR200M_SPk} may suggest that changes in the SHM normalization affect $f_\mathrm{diffuse}$ more strongly than the suppression of the matter power spectrum, despite both quantities being fit simultaneously. This apparent discrepancy arises from the limited halo-mass sensitivity of the matter power spectrum at the relevant scales. In the intermediate-$k$ regime, our measurements primarily probe $f_\mathrm{star}$ in massive halos with $M_{200} \gtrsim 10^{13}\,M_\odot$. For these systems, changing the IMF alters $f_\mathrm{star}$ at the $\lesssim 20\%$ level, which is subdominant given the current statistical precision. As a result, the induced change in power spectrum suppression is limited to $\lesssim 5\%$, well below our present constraining power. In contrast, the inferred shift in $f_\mathrm{diffuse}$ is of order $7\%$, reflecting its direct dependence on the SHM normalization.

Looking ahead, with larger FRB samples and correspondingly improved constraining power, a more conservative and self-consistent approach will be to jointly fit the parameters governing the high-mass end of the SHM relation, namely the normalization $A$ and the high-mass slope $\eta$, directly using FRB observations, without relying on external priors. As a proof of concept, we perform this analysis on our current FRB sample adopting broad, uninformative priors of $A \in [0.01, 0.1]$ and $\eta \in [0.05, 0.4]$. Under these assumptions, we obtain a constraint of $f_\mathrm{diffuse} = 0.95^{+0.03}_{-0.04}$, which directly excludes $f_\mathrm{diffuse} < 80\%$ at $3\sigma$ confidence level. The corresponding constraints on the matter power spectrum suppression are consistent with those shown in Fig.~\ref{fig:spk_constraints}, with a mild extension toward lower suppression amplitudes driven by the reduced constraints on $\eta$. As FRB samples grow, increasingly precise measurements of $f_\mathrm{diffuse}$ will provide a novel avenue for constraining the stellar IMF.

\textbf{Our feedback measurement in the landscape of existing studies:} Given the current constraining power, there are two robust and unbiased probes of feedback strength in the Universe: (1) halo gas mass fractions within $R_{200}$, and (2) the matter power spectrum suppression at intermediate scales. As illustrated in Supplementary Fig.~\ref{fig:parameter_variations-f_gasR200M_f_starR200M_SPk}, both of these measurements are essentially insensitive to stellar physics uncertainties. With this in mind, we compare our measurements of gas mass fractions within $R_{200}$ as a fraction of the cosmic baryon budget ($\Omega_\mathrm{b}/\Omega_\mathrm{m}$) with previous works in literature. 

For clusters ($M_{200} \geq 10^{14}\,M_\odot$), we infer an intracluster medium baryon fraction of $f_\mathrm{ICM} = 0.035^{+0.005}_{-0.007}$, in good agreement with $f_\mathrm{ICM} = 0.0375^{+0.005}_{-0.005}$~\citep{2025NatAs...9.1226C}, derived from Chandra and eROSITA cluster mass functions \protect\citemethods{2024A&A...685A.106B} combined with halo baryon fraction relations \protect\citemethods{2009ApJ...693.1142S, 2010ApJ...719..119D}. At group scales ($5\times10^{12} \leq M < 10^{14}\,M_\odot$), we find $f_\mathrm{IGrM} = 0.048^{+0.019}_{-0.024}$, consistent with the Chandra-based estimate $f_\mathrm{IGrM} = 0.054^{+0.010}_{-0.010}$ \citep{2025NatAs...9.1226C}. For lower-mass halos ($10^9 \leq M < 5\times10^{12}\,M_\odot$), we measure a CGM fraction of $f_\mathrm{CGM} = 0.013^{+0.031}_{-0.010}$, in agreement with reference~\citep{2025NatAs...9.1226C}. The combined fraction of diffuse gas in halos, $f_\mathrm{CGM}+f_\mathrm{IGrM} = 0.06^{+0.03}_{-0.04}$, is statistically consistent with reference~\citep{2024ApJ...973..151K}, who reconstruct the foreground density field of FRBs \protect\citemethods{2022ApJ...928....9L}. Collectively, these results demonstrate coherence between diverse FRB-based measurement techniques.

Adopting our conservative approach to $f_\mathrm{diffuse}$, where we impose no external priors on the SHM relation, we obtained
$f_\mathrm{diffuse} = 0.95^{+0.03}_{-0.04}$. Using baryon budgeting,
\begin{equation}
f_\mathrm{diffuse} = 1 - f_\mathrm{stars} - f_\mathrm{ISM} = f_\mathrm{IGM} + f_\mathrm{ICM} + f_\mathrm{IGrM} + f_\mathrm{CGM},
\end{equation}
we infer the fraction of baryons beyond $R_{200}$ of halos, $f_\mathrm{IGM} = 0.85^{+0.05}_{-0.05}$. This is in excellent agreement with the hydrodynamical simulation-calibrated inference of reference~\citep{2025NatAs...9.1226C}, despite employing an entirely independent methodology. The modest apparent tension ($\sim 2.3\sigma$) with Khrykin et al.~\citep{2024ApJ...973..151K} arises primarily from differences in the treatment of $f_\mathrm{diffuse}$. In our conservative analysis, we explicitly model $f_\mathrm{diffuse}(z)$ without invoking external priors on the SHM relation. In contrast, reference~\citep{2024ApJ...973..151K} adopt a flat prior $f_\mathrm{diffuse} \in [0.75, 0.95]$, meaning the lower bound effectively truncates the posterior and can bias the inferred $f_\mathrm{IGM}$ downward. This underscores that when $f_\mathrm{diffuse}$ is not fitted directly, the choice of prior, particularly its edge, needs to be physically motivated. At the level of halo gas fraction measurements across FRB studies, all results remain consistent; discrepancies in $f_\mathrm{IGM}$ stem from $f_\mathrm{diffuse}$ prior assumptions (when $f_\mathrm{diffuse}$ is not explicitly modeled or fit for during the analyses) rather than the underlying data. These results are encouraging and highlight that future FRB samples with improved statistics will allow us to measure $f_\mathrm{diffuse}$ directly and precisely, further advancing our understanding of baryon partition in the Universe.

\setlength{\parskip}{12pt}
\noindent{\bfseries \large Jackknife Resampling Test}
\setlength{\parskip}{3pt}

We performed jackknife (leave-one-out) resampling to assess whether our constraints are disproportionately driven by a single sightline intersecting a massive galaxy cluster. This test is particularly relevant given that FRBs are expected to be sensitive to halos with masses $\gtrsim 10^{13.5}~M_\odot$. We find that the posterior constraints remain stable across all jackknife realizations (see Supplementary Fig.~\ref{fig:drop_one_out_validation}), indicating that our results are not biased by any individual sightline.

\setlength{\parskip}{12pt}
\noindent{\bfseries \large Constraining Power across Scales}
\setlength{\parskip}{3pt}

We test the ability of current FRB sample to constrain the suppression of matter power spectrum through our analysis framework. We conduct two mock tests, where for each observed redshift of FRBs, we assign synthetic extragalactic DMs assuming (i) no variance in cosmic DMs (the scatter is only from the host DMs), and (ii) randomly high variance in cosmic DMs (the scatter is very large). For both these mock scenarios, we show the DM$_\mathrm{exgal}-z$ relations in Supplementary Fig.~\ref{fig:extreme_scenario_test}. We run these synthetic FRB samples through our analysis framework comprising of 5 models: \textsc{HMcode}, \textsc{BCEmu1}, \textsc{BCEmu4}, \textsc{BCEmu5} and \textsc{BCEmu7}. We show the corresponding predictions of matter power spectrum suppression and halo gas mass fractions in Supplementary Fig.~\ref{fig:extreme_scenario_test}. Compared to the prior volume of each of these models, the extreme feedback scenario (zero scatter in DM$_\mathrm{exgal}-z$ relation) pushes against the prior, favoring large suppression in the matter power spectrum and low gas fractions. On the other hand, the no feedback scenario (large scatter in DM$_\mathrm{exgal}-z$ relation) favors a close to zero suppression in the matter power spectrum with higher gas mass fractions. These extreme scenario tests indicate that the current FRB sample has the constraining power to distinguish between extreme feedback cases.

We quantify the constraining power of the data using the Kullback-Leibler (KL) divergence between the posterior and prior distributions~\protect\citemethods{2014PhRvD..89h3501A, 2017JCAP...10..045N, 2017JCAP...10..045N}. The KL divergence measures the information gain obtained when updating the prior belief with the data, and is defined as
\begin{equation}
D_{\mathrm{KL}} \left(p_{\mathrm{post}} \,\|\, p_{\mathrm{prior}}\right)
= \int p_{\mathrm{post}}(\theta)\,
\log \left[\frac{p_{\mathrm{post}}(\theta)}{p_{\mathrm{prior}}(\theta)}\right]
\,\mathrm{d}\theta .
\end{equation}
By construction, $D_{\mathrm{KL}} = 0$ when the posterior is identical to the prior, indicating that the data provide no additional information about the parameter. As the posterior becomes narrower or shifts relative to the prior, the KL divergence increases, reflecting a corresponding increase in the information content of the data. The KL divergence therefore provides a natural, prior-referenced measure of how strongly the data constrain the parameter of interest. KL divergence is measured in units of bits; so, for example, in the context of data compression in information theory, this would represent how many bits of data is lost. In our context, it represents the number of bits of information gained relative to the prior.  When evaluated as a function of scale, $D_{\mathrm{KL}}(k)$ directly identifies the scales on which the data are informative, enabling a scale-dependent assessment of constraining power that is insensitive to arbitrary choices of parameter normalization. Using the matter power spectrum suppression prior and posterior samples from observed FRBs, we find that the FRB data contribute an information gain of $\sim 15$~bits at scale $k \sim 1~h\,\mathrm{Mpc}^{-1}$. The overlap of constrained scale range with the sensitivity range of FRBs shown in Fig.~\ref{fig:sensitivity_analysis} is reassuring.

\setlength{\parskip}{12pt}
\noindent{\bfseries \large Sensitivity and Fisher Forecasts}
\setlength{\parskip}{3pt}

\textbf{Theoretical set-up:} We forecast the halo mass and redshift sensitivity of next-generation large-scale structure and CMB experiments -- including the Vera Rubin Observatory, DSA and the Simons Observatory -- using a suite of two-point observables. Our primary observable is the angular cross-power spectrum between projected fields. Under the Limber approximation, the angular cross-power spectrum $C_{ij}^{AB}(\ell)$ between redshift bin $i$ of projected field $A(\hat{x})$ and redshift bin $j$ of projected field $B(\hat{x})$ is
\begin{equation}
C^{AB}_{ij} (\ell) = \int\limits_0^\infty \frac{\mathrm{d}\chi}{\chi^2} \, W_A^i(\chi) \, W_B^j(\chi) \, P_\mathrm{AB}\Big(k = \frac{\ell}{\chi}, z(\chi)\Big),
\end{equation}
where $P_{AB}(k,z)$ is the three-dimensional cross-power spectrum of fields $A$ and $B$, and $W^i_A(\chi)$ and $W^j_B(\chi)$ are their respective line-of-sight weight functions. For projected galaxy density in redshift bin $i$, the weighting function is given by the normalized redshift distribution,
\begin{equation}
    W_g^i(\chi) = \frac{n_g^i(z(\chi))}{\bar{n}_g^i} \frac{\mathrm{d}z}{\mathrm{d}\chi},
\end{equation}
where $n_g^i(z)$ is the redshift distribution of galaxies in tomography bin $i$ and $\bar{n}_g^i$ is the corresponding angular number density. For the weak lensing convergence field, the weight function is the lensing efficiency,
\begin{equation}
    W_\gamma(\chi) = \frac{3H_0^2\Omega_\mathrm{m}}{2c^2} \chi (1+z(\chi)) \int\limits_\chi^{\infty} d\chi^\prime \frac{n_\mathrm{s}^i(z(\chi^\prime)) \mathrm{d}z/\mathrm{d}\chi^\prime}{\bar{n}_\mathrm{s}^i} \frac{\chi^\prime - \chi}{\chi^\prime},
\end{equation}
where $n_\mathrm{s}^i(z)$ and $\bar{n}_\mathrm{s}^i$ denote the redshift distribution and angular number density of source galaxies in bin $i$. For the angular DM field sourced by FRBs, the DM perturbation weighting function is
\begin{equation}
W_{\mathcal{D}}(\chi) = W_\mathrm{DM}(\chi) 
\int\limits_\chi^{\infty} \mathrm{d}\chi^\prime \, \frac{n_\mathrm{f}(z(\chi^\prime))}{\bar{n}_\mathrm{f}(z(\chi^\prime))},
\end{equation}
where $n_\mathrm{f}(z)$ and $\bar{n}_\mathrm{f}(z)$ are the redshift distribution and angular number density of FRBs, respectively, and $W_\mathrm{DM}(\chi)$ encodes the line-of-sight response of the DM to free electrons (see Eqn.~\ref{eqn:meanDM}).

We model the three-dimensional power spectra using \textsc{BaryonForge} and \textsc{pyccl}. For lensing convergence ($A=\gamma$) or DM ($A=\mathcal{D}$), we compute cross-spectra with matter or gas as appropriate, i.e.,
\begin{equation}
P_{AB}(k,z) = P_{mB}(k,z) \quad \mathrm{or} \quad P_{gB}(k,z),
\end{equation}
where subscripts $m$ and $g$ denote matter and gas, respectively. For galaxy density ($A=g$), we assume a linear bias model such that
\begin{equation}
P_{AB}(k,z) = b(z)\, P_{mB}(k,z),
\end{equation}
with $b(z)$ the effective galaxy bias.

The Gaussian covariance, derived from the definition of the covariance and Wick’s theorem for Gaussian fields,~\protect\citemethods{2017MNRAS.470.2100K} between two generic multi-probe angular power spectra can be expressed as
\begin{equation}
\mathrm{Cov} [C^{AB}(\ell_i),C^{CD}(\ell_j)] = \frac{\delta^\mathrm{K}_{ij}}{(2\ell_i+1)\Delta\ell\, f_\mathrm{sky}} \times \left[\hat{C}^{AC}(\ell_i)\hat{C}^{BD}(\ell_i) + \hat{C}^{AD}(\ell_i)\hat{C}^{BC}(\ell_i)\right],
\label{eqn:generalized_covariance}
\end{equation}
where $\Delta \ell$ denotes the multipole bin width and $f_\mathrm{sky}$ is the observed sky fraction. The quantity $\hat{C}^{AB}$ represents the measured (i.e., signal-plus-noise) angular power spectrum,
\begin{equation}
\hat{C}^{AB}(\ell) = C^{AB}(\ell) + \delta^\mathrm{K}_{AB} N_A,
\end{equation}
with $N_A$ the noise contribution associated with field $A$. For galaxies, the noise is Poisson shot noise,
\begin{equation}
N_\mathrm{g} = \frac{1}{\bar{n}_{\mathrm{g}}}.
\end{equation}
For FRB DMs, the noise receives two contributions: the variance of the DM field itself and the variance arising from the host galaxy contribution,
\begin{equation}
\label{eq:noise_level_dm}
N_\mathcal{D} = \frac{\sigma_{\mathcal{D}}^2}{\bar{n}_{\mathrm{f}}} + \frac{\sigma^2[\mathrm{DM}_\mathrm{host}]}{(1+\bar{z}_\mathrm{f})^2 \bar{n}_{\mathrm{f}}},
\end{equation}
where $\bar{z}_\mathrm{f}$ is the mean redshift of the FRB sample (which converts the rest-frame host variance to the observed frame). The variance of the DM field is given by
\begin{equation}
\sigma_{\mathcal{D}}^2 = \frac{1}{2\pi} \int \mathrm{d}\ell\, \ell\, C^{\mathcal{DD}}(\ell),
\end{equation}
and represents the intrinsic variance of the cosmological DM field, an irreducible component arising from the stochasticity of large-scale structure, reflecting the sample-averaged dispersion of DM values for the observed FRBs. For other tracers, such as cosmic shear, this field-variance term is typically negligible because the intrinsic ellipticity dispersion dominates over the variance of the underlying cosmological field. Finally, we note that cross-power spectra do not receive a noise contribution. This is because galaxy shot noise, shape noise and stochastic DM fluctuations arise from independent, uncorrelated processes, and therefore their cross-correlation vanishes.

Under the assumption of a Gaussian likelihood, we construct the Fisher information matrix as
\begin{equation}
    F_{ij} = \frac{\partial C^{AB} (\ell)}{\partial p_i} \mathrm{Cov}^{-1}_{AB}[\ell] \left(\frac{\partial C^{AB} (\ell)}{\partial p_j}\right)^T,
\end{equation}
and compute covariance on the estimate of parameters $p_i$ and $p_j$ as
\begin{equation}
    \langle \Delta p_i \Delta p_j \rangle = (F^{-1})_{ij}.
\end{equation}

\textbf{Survey specifications:} For our forecasts, we adopt survey specifications consistent with the Year-1 configuration of the Vera Rubin Observatory\footnote{\url{[https://github.com/CosmoLike/cocoa_baryons_lssty1}} and projected performance for one year of operations of DSA. We assume five source and lens tomographic bins with median redshifts $\{0.32, 0.53, 0.74, 1.00, 1.62\}$, shape noise $\sigma_e = 0.37$, source number density $\bar{n}_s = 2.25~\mathrm{arcmin}^{-2}$, lens galaxy number density $\bar{n}_g = 3.59~\mathrm{arcmin}^{-2}$, and galaxy biases $b = \{1.24, 1.36, 1.47, 1.60, 1.77\}$. The FRB redshift distribution is modeled as $n_\mathrm{f}(z) \propto z^2 \exp(-\alpha z)$ with $\alpha = 3.5$, representative of next-generation FRB surveys. We assume $10^4$ well-localized FRBs overlapping with the Rubin footprint, corresponding to a sky coverage of $\sim 40\%$, and adopt a rest-frame host-galaxy dispersion measure uncertainty of $100~\mathrm{pc\,cm}^{-3}$. Following LSST DESC forecasts configuration~\protect\citemethods{2018arXiv180901669T}, we restrict our analysis to multipoles $\ell \leq 3000$.

\textbf{Sensitivity Forecast:} We present our halo mass ($M_{200}$) and redshift ($z$) sensitivity forecasts in Extended Data Fig.~\ref{fig:sensitivity_forecast} and summarize the techniques used to compute sensitivity of several observables here. We adopt the aforementioned survey settings to forecast the sensitivity of cosmic shear analysis ($C^{\gamma\gamma}$, 15 two-point functions) using five tomography bins. For the angular correlation functions, we adopt the sensitivity estimator defined in Eqn.~\ref{eqn:sensitivity_estimator_correlation_functions}. For the DM-$z$ analysis using a DSA-like FRB sample, we instead use the sensitivity estimator introduced in Eqn.~\ref{FRBsensitivity:Mz}. When stacking FRB DMs on foreground galaxies, we approximate $\mathrm{DM} \propto M^{2/3}$ and assume the number of sightlines scales as $N \propto n(M) M^{2/3}$, where $n(M)$ is the halo mass function. The resulting signal-to-noise ratio scales as $\propto M^{2/3} \sqrt{N}$. Motivated by this scaling, we estimate the sensitivity as
\begin{equation}
    S(\log M_{200},z) \propto \sqrt{n(M_{200})M_{200}^2} \times \int\limits_z^\infty n_\mathrm{f}(z),
\end{equation}
where the integral over the FRB redshift distribution beyond $z$ gives the expected fraction of background FRBs that can be stacked on galaxies at redshift $z$. We define the sensitivity of kSZ stacks from the Simons Observatory as
\begin{equation}
    S(\log M_{200},z) \propto \sqrt{n(M_{200})M_{200}^2} \times \Theta\left(\dfrac{R_\mathrm{vir}(M_{200}, z)}{D_\mathrm{A}(z)} > \theta_\mathrm{res}\right),
\end{equation}
where $\Theta$ denotes the Heaviside step function enforcing the requirement that only halos whose angular extent on the sky, given by the ratio of the virial radius $R_\mathrm{vir}$ to the angular diameter distance $D_\mathrm{A}$, exceeds the instrument beam resolution $\theta_\mathrm{res}$ (assumed to be 1~arcminute) contribute to the stacked signal. For tSZ effect galaxy clusters, we use the signal-to-noise ratio forecasts for Simons Observatory in reference\protect\citemethods{2024JCAP...11..018Z}.

\textbf{Fisher Forecast:} We conduct Fisher parameter forecasts for two analyses: (i) Vera Rubin Observatory-only 3x2-point analysis with five tomography bins for source and lens galaxies, including galaxy clustering ($C^{gg}$, 15 two-point functions), galaxy-galaxy lensing ($C^{g\gamma}$, 25 two-point functions) and cosmic shear ($C^{\gamma\gamma}$, 15 two-point functions), and (ii) joint Vera Rubin Observatory and DSA FRBs 6x2-point analysis, adding in galaxy-DM cross-correlation ($C^{g\mathcal{D}}$, 5 two-point functions), shear-DM cross-correlation ($C^{\gamma\mathcal{D}}$, 5 two-point functions) and DM-DM auto-correlation ($C^{\mathcal{DD}}$, 1 two-point function) to the standard 3x2-point analysis. The parameter constraint forecasts are presented in Extended Data Fig.~\ref{fig:fisher_forecast}. Expanding the analysis from 3x2-point to 6x2-point statistic leads to substantial improvements in constraining power across both, cosmological and astrophysical feedback parameters. In particular, the uncertainty on $\log M_c$ improves from $0.072$ to $0.017$ (a factor of $\sim 4.2$), thus breaking several degeneracies with cosmological parameters: factor of $\sim 3.1,~2.1,~1.5,~2.5$ gains in $\Omega_\mathrm{m}$, $\Omega_\mathrm{b}$, $h$ and $\sigma_8$. The constraints on neutrino masses improve modestly ($\sim 1.2$), whereas dark energy equation of state parameters show notable gains: $w_0$ by $\sim 1.83$ and $w_a$ by $\sim 1.40$. This 6x2-point analysis incorporating FRBs helps mitigate baryonic uncertainties in cosmological inference, approaching the optimal 3x2-point limits under the assumption of perfectly know feedback model, consistent with the findings of reference~\protect\citemethods{2026arXiv260212174W}. These results highlight the additional information content and degeneracy-breaking power provided by including FRBs in the data vector. The resulting precision will enable detailed studies of feedback as a function of halo mass and redshift, offering a first step toward refining models and improving our understanding of baryonic processes in the Universe. Although these Fisher forecasts assume a simplified model of baryonic feedback, they illustrate the statistical potential achievable with FRB cross-correlations.

\newpage

\noindent{\bfseries \LARGE Extended Data}\setlength{\parskip}{12pt}

\newcolumntype{C}{>{\footnotesize\raggedright\arraybackslash}c}

\newcolumntype{L}{>{\footnotesize\raggedright\arraybackslash}l}

\newcolumntype{R}{>{\footnotesize\raggedright\arraybackslash}r}

\begin{table*}[ht!]
    \setlength{\tabcolsep}{4pt}
    \centering
    \captionsetup{labelformat=tablelabel}
    \caption{FRB sample.}
    \begin{tabular}{lrrlrrrrllll}
        \toprule
        FRB&R.A.&Decl.&$P_\mathrm{host}$&DM$_\mathrm{obs}$&DM$_\mathrm{MW}$&$z$&DM$_\mathrm{exgal}$&Survey&Reference \\
        \midrule
        \multicolumn{10}{l}{\textbf{Sub-arcsecond to arcsecond-scale localizations with confident host associations}} \\
        20240304B&182.997&11.813&0.9750&2458.20&28.11&2.1480&2430.09&MeerKAT&\protect\citemethods{2025arXiv250801648C} \\
        20230521B&351.036&71.138&0.9680&1342.90&138.75&1.3540&1204.15&DSA-110&\citep{2025NatAs...9.1226C} \\
        20240104A&348.874&72.821&0.9900&1348.90&120.67&1.3300&1228.23&DSA-110&Verdi+2025 \\
        20220610A&351.073&-33.514&1.0000&1457.62&30.94&1.0160&1426.68&ASKAP&\protect\citemethods{2023Sci...382..294R, 2024ApJ...963L..34G, 2025PASA...42...36S} \\
        20221029A&141.966&72.452&0.9200&1391.05&43.91&0.9750&1347.14&DSA-110&\protect\citemethods{2024ApJ...964..131S}$^,$\citep{2025NatAs...9.1226C, 2024Natur.635...61S} \\
        20240123A&68.263&71.945&0.9980&1462.00&90.28&0.9680&1371.72&DSA-110&\citep{2025NatAs...9.1226C, 2024Natur.635...61S} \\
        20220222C&203.904&-28.027&0.944&1071.20&56.01&0.853&1015.19&MeerKAT&\protect\citemethods{2025MNRAS.tmp.2025P} \\
        20190523A&207.064&72.471&0.8407&760.80&37.23&0.6600&723.57&DSA-10&\protect\citemethods{2019Natur.572..352R} \\
        20250518&207.992&71.282&0.9900&920.60&36.61&0.6392&883.99&DSA-110&Verdi+2025 \\
        20220224C&166.678&-22.940&0.938&1140.20&52.46&0.6271&1087.74&MeerKAT&\protect\citemethods{2025MNRAS.tmp.2025P} \\
        20220418A&219.106&70.095&0.9740&623.25&36.69&0.6220&586.56&DSA-110&\protect\citemethods{2024ApJ...964..131S, 2024ApJ...967...29L}$^,$\citep{2025NatAs...9.1226C, 2024Natur.635...61S} \\
        20190614D&65.076&73.707&0.9900&959.20&87.90&0.6000&871.30&RealFast&\protect\citemethods{2020ApJ...899..161L} \\
        20221219A&257.630&71.627&0.9960&706.70&44.39&0.5540&662.31&DSA-110&\citep{2025NatAs...9.1226C, 2024Natur.635...61S}\protect\citemethods{2024arXiv240514182F} \\
        20230814B&335.976&73.026&0.9960&696.40&104.88&0.5535&591.52&DSA-110&\citep{2025NatAs...9.1226C} \\
        20190711A&329.419&-80.358&1.0000&593.10&56.42&0.5220&536.68&ASKAP&\protect\citemethods{2020ApJ...903..152H, 2021ApJ...917...75M, 2023ApJ...954...80G, 2025PASA...42...36S} \\
        20220918A&17.592&-70.811&0.9965&656.80&40.75&0.4910&616.05&ASKAP&\protect\citemethods{2025PASA...42...36S, 2025ApJ...993..119G} \\
        20220310F&134.721&73.491&0.9700&462.24&46.28&0.4780&415.96&DSA-110&\protect\citemethods{2024ApJ...964..131S, 2024ApJ...967...29L}$^,$\citep{2025NatAs...9.1226C, 2024Natur.635...61S} \\
        20231020B&57.278&-37.770&0.998&952.20&34.44&0.4775&917.76&MeerKAT&\protect\citemethods{2025MNRAS.tmp.2025P} \\
        20230907D&187.142&8.658&0.942&1030.79&28.68&0.4638&1002.11&MeerKAT&\protect\citemethods{2025MNRAS.tmp.2025P} \\
        20230712A&167.360&72.558&0.9950&586.96&39.20&0.4525&547.76&DSA-110&\citep{2025NatAs...9.1226C, 2024Natur.635...61S} \\
        20220204A&274.228&69.722&0.9940&612.20&50.74&0.4000&561.46&DSA-110&\citep{2025NatAs...9.1226C, 2024Natur.635...61S} \\
        20230613A&356.853&-27.053&0.999&483.51&30.04&0.3923&453.47&MeerKAT&\protect\citemethods{2025MNRAS.tmp.2025P} \\
        20220501C&352.379&-32.491&1.0000&449.50&30.61&0.3810&418.89&ASKAP&\protect\citemethods{2025PASA...42...36S} \\
        20190611B&320.746&-79.398&1.0000&321.40&57.83&0.3780&263.57&ASKAP&\citep{2020Natur.581..391M} \\
        20240119A&224.467&71.612&0.9840&483.10&37.91&0.3700&445.19&DSA-110&\citep{2025NatAs...9.1226C} \\
        20200906A&53.496&-14.083&1.0000&577.80&35.82&0.3688&541.98&ASKAP&\protect\citemethods{2022AJ....163...69B, 2023ApJ...954...80G, 2025PASA...42...36S} \\
        20220717A&293.304&-19.288&0.9700&637.34&118.41&0.3630&518.93&MeerKAT&\protect\citemethods{2024MNRAS.532.3881R} \\
        20230902A&52.140&-47.333&1.0000&440.10&34.11&0.3619&405.99&ASKAP&\protect\citemethods{2025PASA...42...36S} \\

        20220726A&73.946&69.930&0.9940&686.55&89.52&0.3610&597.03&DSA-110&\citep{2025NatAs...9.1226C, 2024Natur.635...61S} \\
        20211203C&204.562&-31.380&1.0000&635.00&63.63&0.3439&571.37&ASKAP&\protect\citemethods{2023ApJ...954...80G, 2025PASA...42...36S} \\
        20231220A&123.909&73.660&0.9960&491.20&49.88&0.3355&441.32&DSA-110&\citep{2025NatAs...9.1226C, 2024Natur.635...61S} \\
        20180301A&93.227&4.671&1.0000&536.00&151.62&0.3304&384.38&RealFAST&\protect\citemethods{2022AJ....163...69B, 2023ApJ...954...80G} \\
        20230626A&235.630&71.134&0.9990&451.20&39.17&0.3270&412.03&DSA-110&\citep{2025NatAs...9.1226C, 2024Natur.635...61S} \\

        20230125D&150.205&-31.545&0.9996&640.08&88.13&0.3265&551.95&MeerKAT&\protect\citemethods{2025MNRAS.tmp.2025P} \\
        20180924B&326.105&-40.900&1.0000&361.42&40.45&0.3214&320.97&ASKAP&\protect\citemethods{2019Sci...365..565B, 2020ApJ...903..152H, 2021ApJ...917...75M, 2023ApJ...954...80G, 2025PASA...42...36S} \\
        20230913&305.037&70.793&0.9400&518.60&74.50&0.3024&444.10&DSA-110&Verdi+2025 \\
        20230501A&340.027&70.922&0.9930&532.50&125.57&0.3010&406.93&DSA-110&\citep{2025NatAs...9.1226C, 2024Natur.635...61S} \\
        20220506D&318.045&72.827&0.9910&396.97&84.49&0.3004&312.48&DSA-110&\protect\citemethods{2024ApJ...964..131S, 2024ApJ...967...29L}$^,$\citep{2025NatAs...9.1226C, 2024Natur.635...61S} \\
        20190102C&322.416&-79.476&1.0000&364.50&57.43&0.2913&307.07&ASKAP&\protect\citemethods{2020ApJ...903..152H, 2021ApJ...917...75M, 2022AJ....163...69B, 2023ApJ...954...80G, 2025PASA...42...36S} \\

        20240229A&169.984&70.676&0.9970&491.15&37.94&0.2870&453.21&DSA-110&\citep{2025NatAs...9.1226C} \\
        20221012A&280.799&70.524&0.9960&441.08&54.36&0.2847&386.72&DSA-110&\protect\citemethods{2024ApJ...964..131S, 2024ApJ...967...29L}$^,$\citep{2025NatAs...9.1226C, 2024Natur.635...61S} \\
        
        \bottomrule
    \end{tabular}
    \label{table:FRBsample}
\end{table*}

\begin{table*}
    \setlength{\tabcolsep}{4pt}
    \ContinuedFloat
    \centering
    \captionsetup{labelformat=tablelabel}
    \caption{FRB sample {\textit{(cont.)}}.}
    \begin{tabular}{lrrlrrrrllll}
        \toprule
        FRB&R.A.&Decl.&$P_\mathrm{host}$&DM$_\mathrm{obs}$&DM$_\mathrm{MW}$&$z$&DM$_\mathrm{exgal}$&Survey&Reference \\
        \midrule
        20210320C&204.459&-16.123&1.0000&384.80&39.18&0.2797&345.62&ASKAP&\protect\citemethods{2023ApJ...954...80G, 2025PASA...42...36S} \\
        20220105A&208.804&22.467&1.0000&580.00&21.94&0.2785&558.06&ASKAP&\protect\citemethods{2023ApJ...954...80G, 2025PASA...42...36S} \\
        20221116A&21.211&72.654&0.9430&640.60&132.34&0.2764&508.26&DSA-110&\citep{2025NatAs...9.1226C, 2024Natur.635...61S} \\
        20230307A&177.782&71.695&0.9680&608.90&37.60&0.2710&571.30&DSA-110&\citep{2025NatAs...9.1226C, 2024Natur.635...61S} \\
        20231123B&242.538&70.785&0.9080&396.70&40.25&0.2625&356.45&DSA-110&\citep{2025NatAs...9.1226C, 2024Natur.635...61S} \\
        20221113A&71.411&70.307&0.9950&411.40&91.67&0.2505&319.73&DSA-110&\citep{2025NatAs...9.1226C, 2024Natur.635...61S} \\
        20220307B&350.875&72.192&0.9830&499.27&128.19&0.2481&371.08&DSA-110&\citep{2025NatAs...9.1226C, 2024Natur.635...61S}\protect\citemethods{2024ApJ...964..131S, 2024ApJ...967...29L} \\
        20191228A&344.431&-29.594&1.0000&297.50&32.85&0.2432&264.65&ASKAP&\protect\citemethods{2022AJ....163...69B, 2025PASA...42...36S} \\
        20220825A&311.982&72.585&0.9930&651.24&78.49&0.2414&572.75&DSA-110&\citep{2025NatAs...9.1226C, 2024Natur.635...61S}\protect\citemethods{2024ApJ...964..131S, 2024ApJ...967...29L} \\
        20221101B&342.216&70.682&0.9870&490.70&131.20&0.2395&359.50&DSA-110&\citep{2025NatAs...9.1226C, 2024Natur.635...61S} \\
        20190714A&183.980&-13.021&1.0000&504.13&38.39&0.2365&465.74&ASKAP&\protect\citemethods{2020ApJ...903..152H, 2021ApJ...917...75M, 2022AJ....163...69B, 2023ApJ...954...80G, 2025PASA...42...36S} \\
        20191001A&323.351&-54.748&0.9995&507.90&44.16&0.2340&463.74&ASKAP&\protect\citemethods{2020ApJ...903..152H, 2021ApJ...917...75M, 2023ApJ...954...80G, 2024ApJ...973...64W, 2025PASA...42...36S} \\
        20240215A&268.441&70.232&0.9980&549.50&48.01&0.2100&501.49&DSA-110&\citep{2025NatAs...9.1226C} \\
        20221106A&56.705&-25.570&0.9446&343.80&34.73&0.2044&309.07&ASKAP&\protect\citemethods{2025PASA...42...36S} \\
        20121102A&82.995&33.148&1.0000&557.00&188.40&0.1927&368.60&VLBI&\protect\citemethods{2017ApJ...843L...8B, 2021ApJ...917...75M, 2023ApJ...954...80G} \\
        20220725A&353.315&-35.990&1.0000&290.40&30.77&0.1926&259.63&ASKAP&\protect\citemethods{2025PASA...42...36S} \\
        20220529A&19.104&20.632&0.9870&246.00&39.99&0.1839&206.01&CHIME&\protect\citemethods{2026Sci...391..280L} \\
        20210603A&10.274&21.226&0.9990&500.15&39.41&0.1770&460.74&CHIME&\protect\citemethods{2024NatAs...8.1429C} \\
        20200430A&229.706&12.377&1.0000&380.25&27.09&0.1600&353.16&ASKAP&\protect\citemethods{2020ApJ...903..152H, 2022AJ....163...69B, 2023ApJ...954...80G, 2025PASA...42...36S} \\
        20220920A&240.257&70.919&0.9850&314.99&39.89&0.1582&275.10&DSA-110&\citep{2025NatAs...9.1226C, 2024Natur.635...61S}\protect\citemethods{2024ApJ...964..131S, 2024ApJ...967...29L} \\
        20230526A&22.233&-52.717&1.0000&361.40&31.88&0.1570&329.52&ASKAP&\protect\citemethods{2025PASA...42...36S} \\
        20231226A&155.364&6.110&1.0000&329.90&38.07&0.1569&291.83&ASKAP&\protect\citemethods{2025PASA...42...36S} \\
        20210410D&326.086&-79.318&0.9957&578.78&56.23&0.1415&522.55&MeerKAT&\protect\citemethods{2023MNRAS.524.2064C} \\
        20240209A&289.887&86.064&0.9900&176.52&55.50&0.1384&121.02&CHIME&\protect\citemethods{2025ApJ...979L..22E, 2025ApJ...979L..21S} \\
        20210807D&299.220&-0.762&0.9990&251.30&121.28&0.1293&130.02&ASKAP&\protect\citemethods{2025PASA...42...36S, 2024ApJ...973...64W, 2023ApJ...954...80G} \\
        20240310A&166.786&72.281&0.9884&601.80&39.11&0.1270&562.69&ASKAP&\protect\citemethods{2025PASA...42...36S} \\
        20230628A&166.787&72.282&0.9530&345.15&39.11&0.1265&306.04&DSA-110&\citep{2025NatAs...9.1226C, 2024Natur.635...61S} \\
        20240213A&166.168&74.075&0.9950&357.40&40.11&0.1185&317.29&DSA-110&\citep{2025NatAs...9.1226C} \\
        20190608B&334.020&-7.899&1.0000&339.50&37.19&0.1178&302.31&ASKAP&\protect\citemethods{2021ApJ...922..173C, 2022AJ....163...69B, 2020ApJ...903..152H, 2021ApJ...917...75M, 2023ApJ...954...80G, 2025PASA...42...36S} \\
        20220914A&282.058&73.336&0.9740&631.28&54.70&0.1139&576.58&DSA-110&\citep{2025NatAs...9.1226C, 2024Natur.635...61S}\protect\citemethods{2024ApJ...964..131S,  2024ApJ...967...29L} \\
        20201124A&77.015&26.061&0.9500&413.52&139.94&0.0980&273.58&DSA-110&\protect\citemethods{2022MNRAS.513..982R, 2022ApJ...927L...3N} \\
        20230124A&231.917&70.968&0.9990&590.57&38.53&0.0939&552.05&DSA-110&\citep{2025NatAs...9.1226C, 2024Natur.635...61S} \\
        20230930A&10.507&41.417&0.9994&456.00&68.09&0.0925&387.91&RealFAST&\protect\citemethods{2025ApJ...993..221A} \\
        20220509G&282.670&70.244&0.9700&269.50&55.58&0.0894&213.92&DSA-110&\citep{2025NatAs...9.1226C, 2024Natur.635...61S}\protect\citemethods{2024ApJ...967...29L, 2023ApJ...950..175S, 2023ApJ...949L..26C} \\
        20220912A&347.270&48.707&0.9500&219.46&125.24&0.0771&94.22&DSA-110&\protect\citemethods{2023ApJ...949L...3R} \\
        20240203&312.621&73.900&0.9960&272.40&76.22&0.0740&196.18&DSA-110&Verdi+2025 \\
        20211212A&157.351&1.361&1.0000&206.00&38.84&0.0715&167.16&ASKAP&\protect\citemethods{2023ApJ...954...80G, 2025PASA...42...36S} \\
        20231120A&143.985&73.285&0.9870&438.90&43.77&0.0700&395.13&DSA-110&\citep{2025NatAs...9.1226C, 2024Natur.635...61S} \\
        20231204A&207.999&48.116&0.9760&221.00&29.71&0.0644&191.29&CHIME&\protect\citemethods{2025ApJS..280....6C, 2023ApJ...950..134M} \\
        20201123A&263.660&-50.764&0.9150&433.55&252.56&0.0507&180.99&MeerKAT&\protect\citemethods{2022MNRAS.514.1961R} \\
        20211127I&199.808&-18.838&0.9998&234.83&42.46&0.0469&192.37&ASKAP&\protect\citemethods{2023ApJ...954...80G} \\
        20220207C&310.200&72.882&0.9700&262.30&76.01&0.0430&186.29&DSA-110&\citep{2025NatAs...9.1226C, 2024Natur.635...61S}\protect\citemethods{2024ApJ...967...29L} \\
        20240201A&149.906&14.088&1.0000&374.50&38.57&0.0427&335.93&ASKAP&\protect\citemethods{2025PASA...42...36S} \\
        20180916B&29.503&65.717&1.0000&348.76&198.84&0.0337&149.92&CHIME&\protect\citemethods{2021ApJ...908L..12T, 2022evlb.confE..35M, 2021ApJ...917...75M, 2023ApJ...954...80G} \\

        \bottomrule
    \end{tabular}
\end{table*}

\begin{table*}
    \setlength{\tabcolsep}{4pt}
    \ContinuedFloat
    \centering
    \captionsetup{labelformat=tablelabel}
    \caption{FRB sample {\textit{(cont.)}}.}
    \begin{tabular}{lrrlrrrrllll}
        \toprule
        FRB&R.A.&Decl.&$P_\mathrm{host}$&DM$_\mathrm{obs}$&DM$_\mathrm{MW}$&$z$&DM$_\mathrm{exgal}$&Survey&Reference \\

        \midrule

        20240210A&8.780&-28.271&1.0000&283.73&28.63&0.0237&255.10&ASKAP&\protect\citemethods{2025PASA...42...36S} \\
        20171020A&333.827&-19.670&0.9800&114.10&36.55&0.0087&77.55&ASKAP&\protect\citemethods{2023PASA...40...29L, 2018ApJ...867L..10M} \\

        \midrule

        \multicolumn{10}{l}{\textbf{Sub-arcminute to arcminute-scale localizations with confident host associations}} \\
        20231025B&270.788&63.989&0.9250&368.70&48.64&0.3238&320.06&CHIME&\protect\citemethods{2025ApJS..280....6C} \\
        20231017A&346.754&36.653&0.9180&344.20&64.55&0.2450&279.65&CHIME&\protect\citemethods{2025ApJS..280....6C} \\
        20230730A&54.665&33.159&0.9520&312.50&85.24&0.2115&227.26&CHIME&\protect\citemethods{2025ApJS..280....6C} \\
        20230203A&151.662&35.694&0.9380&420.10&36.30&0.1464&383.80&CHIME&\protect\citemethods{2025ApJS..280....6C} \\
        20230222A&106.960&11.225&0.9850&706.10&134.12&0.1223&571.98&CHIME&\protect\citemethods{2025ApJS..280....6C} \\
        20230703A&184.624&48.730&0.9730&291.30&26.96&0.1184&264.34&CHIME&\protect\citemethods{2025ApJS..280....6C} \\
        20231201A&54.589&26.818&0.9110&169.40&70.04&0.1119&99.36&CHIME&\protect\citemethods{2025ApJS..280....6C} \\
        20230222B&238.739&30.899&0.9810&187.80&27.70&0.1100&160.10&CHIME&\protect\citemethods{2025ApJS..280....6C} \\
        20231128A&199.578&42.993&0.9850&331.60&24.97&0.1079&306.63&CHIME&\protect\citemethods{2025ApJS..280....6C} \\
        20231223C&259.545&29.498&0.9530&165.80&47.90&0.1059&117.90&CHIME&\protect\citemethods{2025ApJS..280....6C} \\
        20231011A&18.242&41.749&0.9913&186.30&70.34&0.0783&115.96&CHIME&\protect\citemethods{2025ApJS..280....6C} \\
        20231123A&82.623&4.474&0.9559&302.10&89.74&0.0729&212.36&CHIME&\protect\citemethods{2025ApJS..280....6C} \\
        20190418A&65.817&16.067&0.9200&182.78&70.14&0.0713&112.64&CHIME&\protect\citemethods{2024ApJ...971L..51B} \\
        20231005A&246.028&35.449&0.9850&189.40&33.37&0.0713&156.03&CHIME&\protect\citemethods{2025ApJS..280....6C} \\
        20231206A&112.443&56.256&0.9890&457.70&59.13&0.0659&398.57&CHIME&\protect\citemethods{2025ApJS..280....6C} \\
        20200223B&8.270&28.831&0.9940&201.80&45.60&0.0602&156.20&CHIME&\protect\citemethods{2024ApJ...961...99I} \\
        20230926A&269.125&41.814&0.9960&222.80&52.71&0.0553&170.09&CHIME&\protect\citemethods{2025ApJS..280....6C} \\
        20181223C&180.929&27.553&0.9600&111.61&19.96&0.0302&91.65&CHIME&\protect\citemethods{2024ApJ...971L..51B} \\
        20231230A&72.789&2.368&0.9180&131.40&61.49&0.0298&69.91&CHIME&\protect\citemethods{2025ApJS..280....6C} \\
        20181220A&348.717&48.3403&0.9900&209.40&118.48&0.0275&90.92&CHIME&\protect\citemethods{2024ApJ...971L..51B} \\
        20231229A&26.468&35.108&0.9980&198.50&58.16&0.0190&140.34&CHIME&\protect\citemethods{2025ApJS..280....6C} \\
        20181030A&158.583&73.751&0.9975&103.50&41.05&0.0039&62.45&CHIME&\protect\citemethods{2021ApJ...919L..24B} \\

        \midrule

        \multicolumn{10}{l}{\textbf{Insecure host galaxy association}} \\
        20230216A&156.473&3.437&0.4200&828.00&38.48&0.5310&789.52&DSA-110&\citep{2025NatAs...9.1226C, 2024Natur.635...61S} \\
        20181112A&327.349&-52.971&0.8862&589.27&41.73&0.4755&547.54&ASKAP&\protect\citemethods{2019Sci...366..231P, 2020ApJ...903..152H, 2023ApJ...954...80G, 2025PASA...42...36S} \\
        20220330D&163.751&70.351&0.6300&468.10&38.65&0.3714&429.45&DSA-110&\citep{2025NatAs...9.1226C, 2024Natur.635...61S} \\
        20220208A&322.575&70.041&0.5600&437.00&101.64&0.3510&335.36&DSA-110&\citep{2025NatAs...9.1226C, 2024Natur.635...61S} \\
        20230808F&53.304&-51.935&0.5900&653.20&35.38&0.3472&617.82&MeerKAT&\protect\citemethods{2025MNRAS.538.1800H} \\
        20221027A&130.872&72.101&0.6160&452.50&47.16&0.2290&405.34&DSA-110&\citep{2025NatAs...9.1226C, 2024Natur.635...61S} \\
        20230311A&91.110&55.946&0.7740&364.30&92.39&0.1918&271.91&CHIME&\protect\citemethods{2025ApJS..280....6C} \\
        20190110C&249.318&41.443&0.7790&221.60&36.97&0.1224&184.63&CHIME&\protect\citemethods{2024ApJ...961...99I} \\
        20190425A&255.662&21.577&0.7870&128.20&48.72&0.0312&79.48&CHIME&\protect\citemethods{2023MNRAS.519.2235P, 2025ApJ...979...95Q} \\

        \midrule

        \multicolumn{10}{l}{\textbf{Evidence for excess host galaxy dispersion measure}} \\
        20220831A&338.696&70.538&0.9980&1146.25&126.72&0.2620&1019.53&DSA-110&\citep{2025NatAs...9.1226C} \\
        20190520B&240.518&-11.288&0.9984&1210.30&60.09&0.2410&1150.21&FAST&\protect\citemethods{2023ApJ...954...80G, 2022Natur.606..873N} \\
        20210117A&339.979&-16.151&1.0000&728.95&34.26&0.2140&694.69&ASKAP& \protect\citemethods{2023ApJ...948...67B, 2023ApJ...954...80G, 2025PASA...42...36S} \\
        20240114A&321.916&4.329&0.9974&527.70&49.58&0.1300&478.12&CHIME&\protect\citemethods{2025ApJ...980L..24C, 2024arXiv241201478B} \\
        20230708A&303.116&-55.356&1.0000&411.51&60.33&0.1050&351.18&ASKAP&\protect\citemethods{2025PASA...42...36S, 2025arXiv250620774M} \\

        \midrule

        \multicolumn{10}{l}{\textbf{Negative extragalactic dispersion measure or less than Milky Way halo dispersion measure}} \\

        20210405I&255.340&-49.545&0.9975&565.17&516.42&0.0660&48.75&MeerKAT&\protect\citemethods{2024MNRAS.527.3659D} \\

        \bottomrule
    \end{tabular}
\end{table*}

\begin{table*}[ht!]
    \setlength{\tabcolsep}{4pt}
    \ContinuedFloat
    \centering
    \captionsetup{labelformat=tablelabel}
    \caption{FRB sample {\textit{(cont.)}}.}
    \begin{tabular}{lrrlrrrrllll}
        \toprule
        FRB&R.A.&Decl.&$P_\mathrm{host}$&DM$_\mathrm{obs}$&DM$_\mathrm{MW}$&$z$&DM$_\mathrm{exgal}$&Survey&Reference \\

        \midrule
        20230718A&128.162&-40.452&0.9400&476.67&423.71&0.0350&52.96&ASKAP&\protect\citemethods{2024ApJ...962L..13G} \\
        20220319D&32.178&71.035&0.9900&110.95&139.78&0.0111&-28.83&DSA-110&\citep{2025NatAs...9.1226C, 2024Natur.635...61S}\protect\citemethods{2025AJ....169..330R, 2024ApJ...967...29L} \\
        20200120E&149.486&68.826&1.0000&87.82&40.77&0.0008&47.05&CHIME&\protect\citemethods{2021ApJ...910L..18B, 2022Natur.602..585K} \\

        \bottomrule
    \end{tabular}
    
    \footnotesize{{\raggedright N{\scriptsize{OTE}}: The complete catalog of FRBs used in this work, including sub-arcsecond, arcsecond, sub-arcminute and arcminute-scale localizations with confident host associations. For completeness, we also list the FRBs excluded from our sample, including the ones with insecure host galaxy association, evidence for excess host galaxy DM, negative extragalactic DM and extragalactic DM less than the assumed Milky Way halo contribution (50~pc\,cm$^{-3}$).
    \par}}
\end{table*}

\begin{table*}
    \setlength{\tabcolsep}{3.2pt}
    \centering
    \captionsetup{labelformat=tablelabel}
    \caption{Priors on the parameters of two halo model prescriptions.}
    \begin{tabular}{lllll}
        \toprule
        Parameter & Equation & Description & Prior & Fixed \\
        \midrule

        \multicolumn{3}{l}{\textbf{Hydrosimulations-calibrated halo model: \textsc{HMcode}}} \\

        $\log T_\mathrm{AGN}$ & - & AGN heating temperature calibrated to BAHAMAS simulations & U$[7.6, 8.0]$ & 7.8 \\
        
        \midrule

        \multicolumn{3}{l}{\textbf{Flexible analytical gas profiles halo model : \textsc{BCEmu}}} \\
        $\log M_c$ & \ref{eqn:gas_profile_slope} & Mass scale below which gas profile becomes shallower than & U[11.0, 15.0] & 14 \\
        & & the NFW profile & & \\
        \cdashline{1-5}
        $\mu_\beta$ & \ref{eqn:gas_profile_slope} & Mass dependence of inner gas profile slope $\beta$ & U[0.0, 2.0] & 1 \\
        $\delta$ & \ref{eqn:gas_profile} & Outer slope of bound gas profile & U[3.0, 11.0] & 7 \\
        $\theta_\mathrm{ej}$ & \ref{eqn:gas_profile} & Maximum radius of gas ejection with respect to virial radius & U[2.0, 8.0] & 3.5 \\
        \cdashline{1-5}
        $\eta$ & \ref{eqn:f_gas}, \ref{eqn:f_star} & Governs the total stellar fraction in high-mass halos & U[0.05, 0.4] & 0.2 \\
        \cdashline{1-5}
        $\gamma$ & \ref{eqn:gas_profile} & Outer slope of bound gas profile & U[1.0, 4.0] & 2.5 \\
        $\eta_\delta$ & \ref{eqn:f_cga} & Governs the partition of stellar content between central and & U[0.05, 0.4] & 0.2 \\
        & & satellite galaxies in high-mass halos & & \\

        \midrule

        \multicolumn{3}{l}{\textbf{FRB host galaxy DM distribution}} \\
        $\mu_\mathrm{host}$ & - & Lognormal DM$_\mathrm{host}$ distribution parameter & U[4.0, 6.0] & - \\
        $\sigma_\mathrm{host}$ & - & Lognormal DM$_\mathrm{host}$ distribution parameter & U[0.2, 1.0] & - \\
        
        \midrule
    \end{tabular}
    \footnotesize{{\raggedright N{\scriptsize{OTE}}: Summary of priors and fiducial values of baryonic feedback parameters in the hydrodynamical simulations-calibrated (\textsc{HMcode}) and the flexible analytical gas profiles (\textsc{BCEmu}) halo model prescriptions (including reduced complexity models), and the host galaxy DM distribution parameters. The \textsc{BCEmu1} model only varies $\log M_c$, \textsc{BCEmu4} varies $\{\log M_c, \mu_\beta, \delta, \theta_\mathrm{ej}\}$, \textsc{BCEmu5} varies $\{\log M_c, \mu_\beta, \delta, \theta_\mathrm{ej}, \eta\}$ and \textsc{BCEmu7} varies all the seven parameters.\par}}
    \label{table:parameters}
\end{table*}

\newpage
\clearpage
\newpage
\newpage

\begin{table*}[ht!]
    \centering
    \captionsetup{labelformat=tablelabel}
    \caption{Constraints on the parameters of two halo model prescriptions.}
    \begin{tabular}{lrrrrr}
        \toprule
        Parameter & \textsc{HMcode} & \textsc{BCEmu1} & \textsc{BCEmu4} & \textsc{BCEmu5} & \textsc{BCEmu7} \\
        \midrule
        $\log T_\mathrm{AGN}$ & $7.72_{-0.14}^{+0.12}$ & $\times$ & $\times$ & $\times$ & $\times$ \\
        $\log M_c$ & $\times$ & $13.44_{-0.30}^{+0.26}$ & $12.87_{-1.18}^{+1.18}$ & $12.81_{-1.16}^{+1.16}$ & $12.84_{-1.05}^{+1.14}$ \\
        $\mu_\beta$ & $\times$ & - & $0.91_{-0.57}^{+0.71}$ & $0.89_{-0.54}^{+0.71}$ & $0.94_{-0.52}^{+0.67}$ \\
        $\delta$ & $\times$ & - & $7.59_{-3.00}^{+2.44}$ & $7.28_{-2.63}^{+2.50}$ & $7.33_{-2.59}^{+2.56}$ \\
        $\theta_\mathrm{ej}$ & $\times$ & - & $4.55_{-1.97}^{+2.37}$ & $4.72_{-2.14}^{+2.35}$ & $4.93_{-2.24}^{+2.27}$ \\
        $\eta$ & $\times$ & - & - & $0.24_{-0.05}^{+0.06}$ & $0.22_{-0.04}^{+0.04}$ \\
        $\gamma$ & $\times$ & - & - & - & $2.79_{-1.14}^{+0.88}$ \\
        $\eta_\delta$ & $\times$ & - & - & - & $0.21_{-0.10}^{+0.12}$ \\
        \midrule
        $\langle \mathrm{DM}_\mathrm{host} \rangle$ & $131.11_{-18.49}^{+24.15}$ & $131.43_{-16.62}^{+18.77}$ & $133.61_{-17.38}^{+17.79}$ & $128.60_{-17.50}^{+18.33}$ & $129.25_{-16.89}^{+15.95}$ \\
        $\sigma[\mathrm{DM}_\mathrm{host}]$ & $179.40_{-44.39}^{+71.66}$ & $154.96_{-25.70}^{+27.72}$ & $157.92_{-26.26}^{+27.19}$ & $150.25_{-25.66}^{+28.88}$ & $149.49_{-22.77}^{+26.54}$ \\
        \bottomrule
    \end{tabular}
    
    \footnotesize{{\raggedright N{\scriptsize{OTE}}: The constraints on parameters of the hydrodynamical simulations-calibrated (\textsc{HMcode}) and the flexible analytical gas profiles (\textsc{BCEmu1}, \textsc{BCEmu4}, \textsc{BCEmu5}, \textsc{BCEmu7}) halo models, as inferred using FRBs. The tabulated values are the median and 68\% confidence intervals.\par}}
    \label{table:HMcode_BCEmu_constraints}
\end{table*}

\begin{figure}[ht!]
    \centering
    \includegraphics[width=0.375\columnwidth]{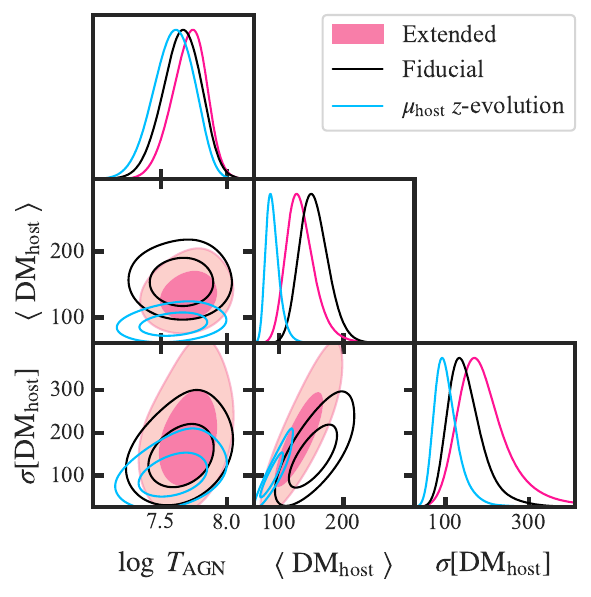}
    \caption{{\edfigurelabel{fig:HMcode_cornerplot}} \textbf{An evaluation of the susceptibility of feedback inference conducted with observed sample of localized FRBs to plausible host galaxy DM redshift evolution.} The marginalized posteriors for the feedback parameter ($T_\mathrm{AGN}$~\protect/citemethods{2017MNRAS.465.2936M}) in hydrodynamical simulations-calibrated halo model framework (\textsc{HMcode}), mean FRB host galaxy DM contribution ($\langle \mathrm{DM}_\mathrm{host} \rangle$) and its variance ($\sigma[\mathrm{DM}_\mathrm{host}]$) are shown. The inner and outer contours show the 68\% and 95\% confidence intervals, respectively. We show the posteriors with the extended sample and the fiducial sample that excludes local-universe FRBs with arcminute-scale localizations~\protect\citemethods{2021ApJ...919L..24B, 2024ApJ...961...99I, 2024ApJ...971L..51B, 2025ApJS..280....6C}. While the marginalized posterior of $T_\mathrm{AGN}$ remains stable, we observe a shift in $\langle \mathrm{DM}_\mathrm{host} \rangle$ and $\sigma[\mathrm{DM}_\mathrm{host}]$, which may be an indication of the redshift evolution of DM$_\mathrm{host}$. We test the scenario where the median DM$_\mathrm{host}$ evolves with redshift, proportional to the cosmic star-formation history of the universe~\protect\citemethods{2014ARA&A..52..415M, 2023ApJ...954...80G}$^,$\citep{2024Natur.635...61S}. While we observe a drop in both, $\langle \mathrm{DM}_\mathrm{host} \rangle$ and $\sigma[\mathrm{DM}_\mathrm{host}]$, the marginalized posterior of $T_\mathrm{AGN}$ does not shift, indicating that our feedback measurements with the current FRB sample are robust to potential redshift evolution of DM$_\mathrm{host}$.}
\end{figure}

\newpage
\clearpage
\newpage
\newpage

\begin{figure}
    \includegraphics[width=\columnwidth]{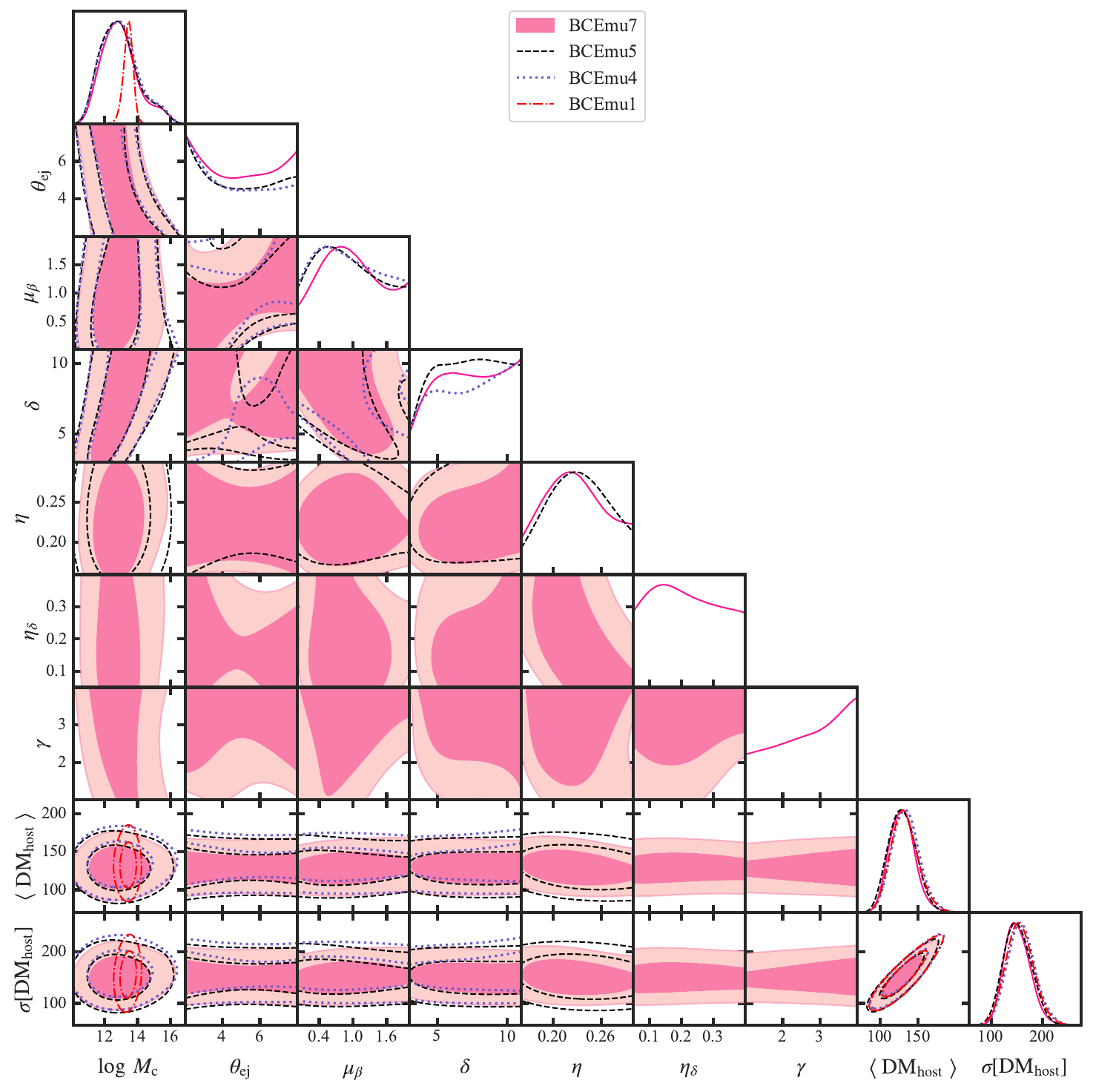}
    \caption{{\edfigurelabel{fig:BCEmu_cornerplot}} \textbf{An MCMC fit to the observed sample of localized FRBs using a halo model prescription with flexible analytical gas profiles.} The marginalized posteriors for parameters defining the gas profiles (\textsc{BCEmu}), mean FRB host galaxy DM contribution ($\langle \mathrm{DM}_\mathrm{host} \rangle$) and its variance ($\sigma[\mathrm{DM}_\mathrm{host}]$) are presented. The inner and outer contours show the 68\% and 95\% confidence intervals, respectively. The constraints on all parameters are relatively stable across all reduced complexity \textsc{BCEmu} variants.}
\end{figure}

\newpage
\clearpage
\newpage
\newpage

\begin{figure}[h]
    \centering
    \includegraphics[width=0.5\columnwidth]{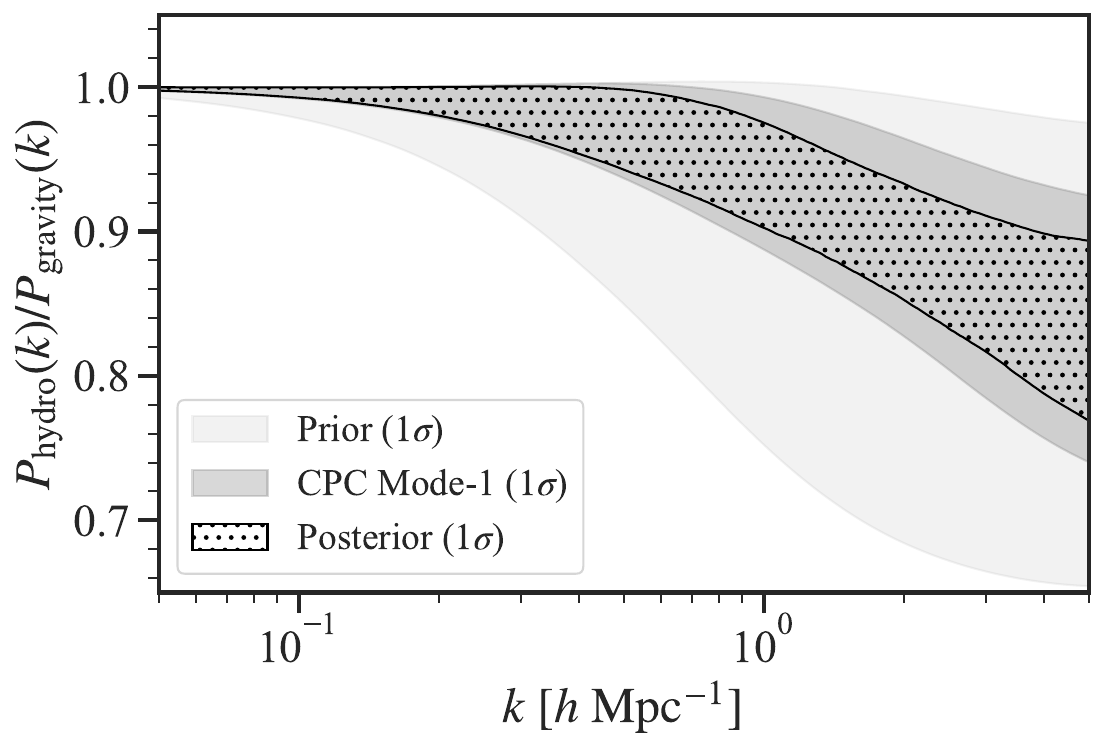}
    \caption{{\edfigurelabel{fig:spk_BCEmu7_CPC}} \textbf{Contribution of the constrained degree of freedom to the measured matter power spectrum suppression}. In the halo model prescription with flexible analytical gas profiles (\textsc{BCEmu7}), we identify $\sim 1$ constrained degree of freedom through covariant principal component (CPC) analysis of posterior samples. We show the $1\sigma$ range of the \textsc{BCEmu7} prior volume, measurement from the first constrained CPC-mode (a combination of $\log M_c, \theta_\mathrm{ej}, \mu_\beta, \delta$) and the \textsc{BCEmu7} posterior. This constrained CPC mode leads to $\sim 224$\% reduction in the variance of prior relative to posterior.}
\end{figure}

\begin{figure}[ht!]
    \includegraphics[width=\columnwidth]{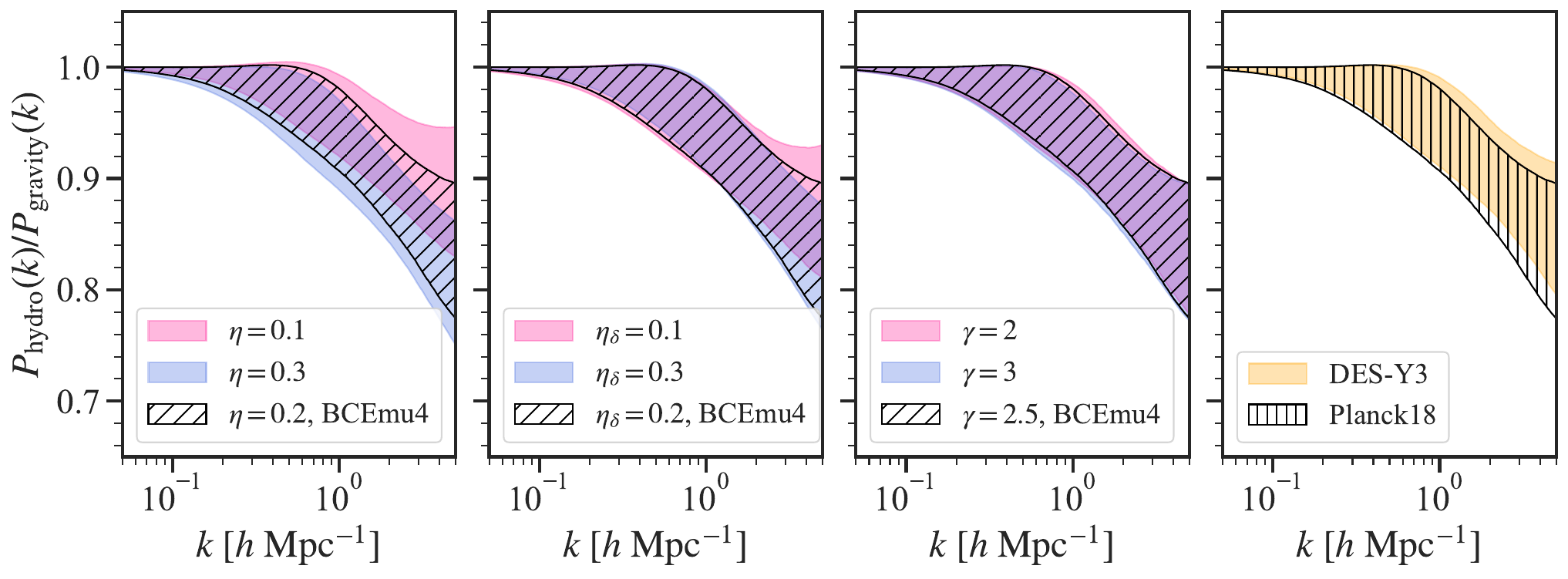}
    \caption{{\edfigurelabel{fig:spk_vary_feedback_cosmology}} \textbf{Robustness of matter power spectrum suppression measurement to variations of fixed halo model parameters and cosmology.} We test the susceptibility of our constraints from the \textsc{BCEmu4} model to changing $\eta$: the power-law index of stellar mass fraction at high-mass end, $\eta_\delta$: the power-law index for halo central galaxy star fraction at high-mass end ($\eta_\mathrm{cga} = \eta + \eta_\delta$) and $\gamma$: the outer slope of bound gas profile. The choice of fixed $\eta$ impacts constraints, necessitating a better physically motivated prior on the stellar-to-halo mass relation. The observed shift in our constraints when changing from Planck18~\protect\citemethods{2020A&A...641A...6P} to DES-Y3~\citep{2022PhRvD.105b3514A}\protect\citemethods{2022PhRvD.105b3515S} cosmology in our baseline \textsc{BCEmu5} model is due to the change in universe baryon fraction ($\Omega_{\mathrm{b}}/\Omega_{\mathrm{m}}$) under two cosmologies.}
\end{figure}

\newpage
\clearpage
\newpage
\newpage

\begin{figure}[h]
    \includegraphics[width=\columnwidth]{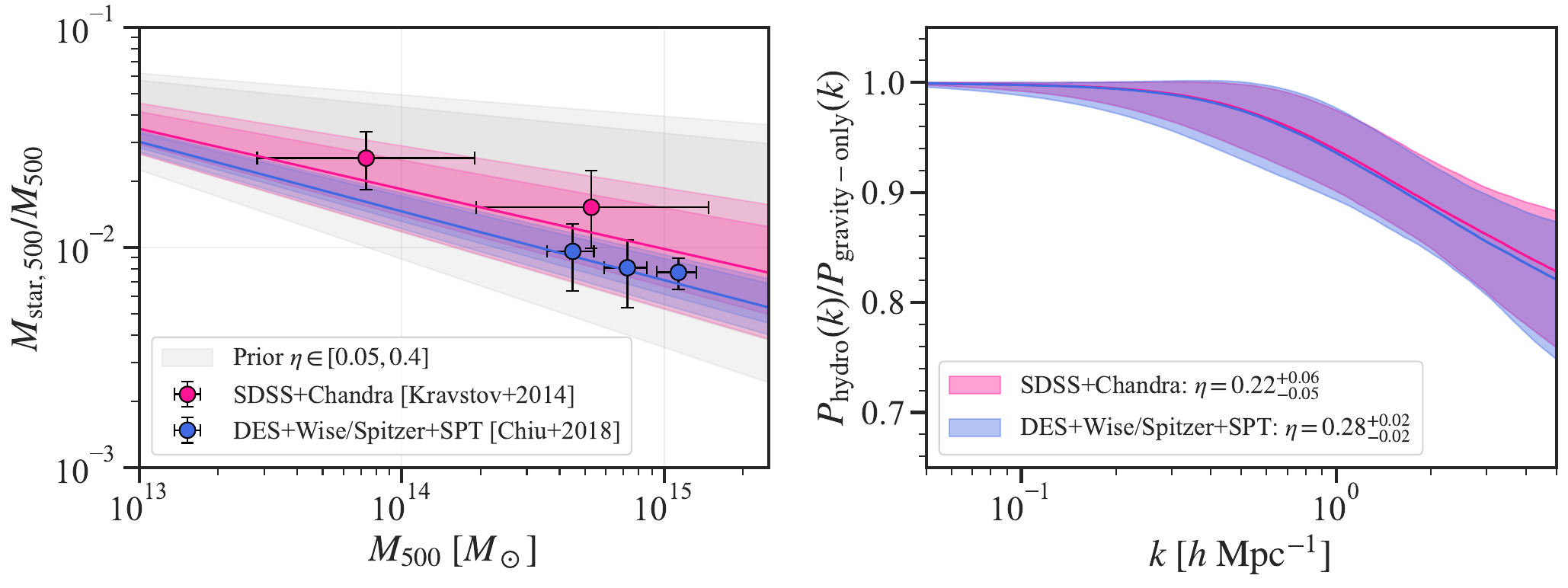}
    \caption{{\edfigurelabel{fig:eta_prior}} \textbf{The stellar-to-halo mass relation prior motivated by observations.} In panel \textbf{a}, we present MCMC fits to the power-law index of stellar-to-halo mass (SHM) relation (including central and satellite galaxies) at the high-mass end ($\eta$) using two datasets: (i) SDSS optical data and Chandra X-ray observations of galaxy clusters (median $\log M_\mathrm{500} = 14.48$)~\citep{2018AstL...44....8K}\protect\citemethods{2006ApJ...640..691V, 2013ApJ...778...14G, 2013MNRAS.429.3288S}, and (ii) DES optical data, WISE/Spitzer near-infrared data and SPT SZ effect scaling relation of galaxy clusters (median $\log M_\mathrm{500} = 14.68$)~\protect\citemethods{2018MNRAS.478.3072C}. Assuming a wide prior on $\eta \in [0.05, 0.40]$, with fixed normalization of the SHM relation, $A = 0.055/2$, the posteriors on the power law index are $\eta = 0.22_{-0.05}^{+0.06}$ and $\eta = 0.28_{-0.02}^{+0.02}$ for the two datasets, respectively. The matter power spectrum suppression constraints from an analysis conducted using either of the two aforementioned datasets as additional priors, as shown in panel \textbf{b}, attest robustness of current measurements to stellar astrophysics assumptions.}
\end{figure}

\newpage
\clearpage
\newpage
\newpage

\begin{figure}[ht!]
    \includegraphics[width=\columnwidth]{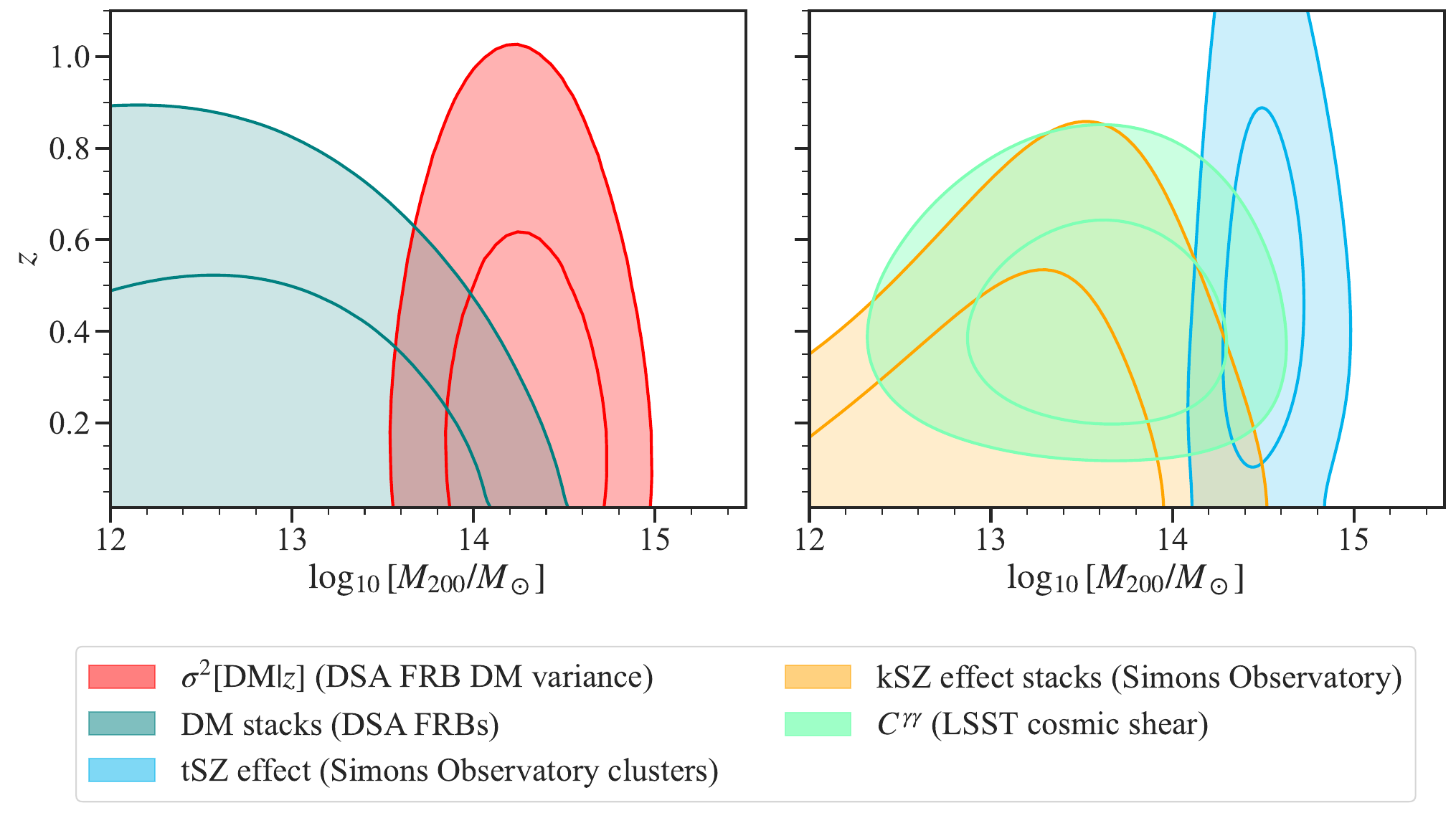}
    \caption{{\edfigurelabel{fig:sensitivity_forecast}} \textbf{The next decade of baryon mapping: sensitivity forecasts for FRBs and their complementarity with traditional large-scale structure and CMB probes.} Contours illustrate the halo mass ($M_{200}$) and redshift ($z$) regimes accessible through a range of observables: FRB DM variance and stacked DM measurements from forthcoming FRB facilities, such as DSA, SKA, and CHORD; Vera C. Rubin Observatory (LSST) cosmic shear; and thermal and kinetic Sunyaev-Zel'dovich (tSZ/kSZ) measurements from Simons Observatory. FRBs provide a direct probe of ionized baryons, enabling sensitivity to diffuse low density gas that is difficult to access with lensing or CMB observables alone. While FRB DM variance primarily probes cluster-scale halos ($M_{200} \gtrsim 10^{13.5}~M_\odot$) at redshifts $z \lesssim 1$, stacking FRB sightlines on LSST galaxies extends sensitivity to lower-mass, galaxy-scale halos ($10^{11} \lesssim M_{200} \lesssim 10^{14.5}~M_\odot$). In addition, stacking of FRB DMs on LSST galaxies enable tomographic studies of astrophysical feedback. In combination with tSZ and kSZ measurements from next-generation CMB experiments and weak lensing measurements from LSST, FRBs enhance the mass-redshift coverage and break degeneracies between cosmology and feedback. Together, these complementary tracers will, over the coming decade, bridge the longstanding divide between precision cosmology and the physics of galaxy formation.}
\end{figure}

\newpage
\clearpage
\newpage
\newpage

\begin{figure}[ht!]
    \includegraphics[width=0.96\columnwidth]{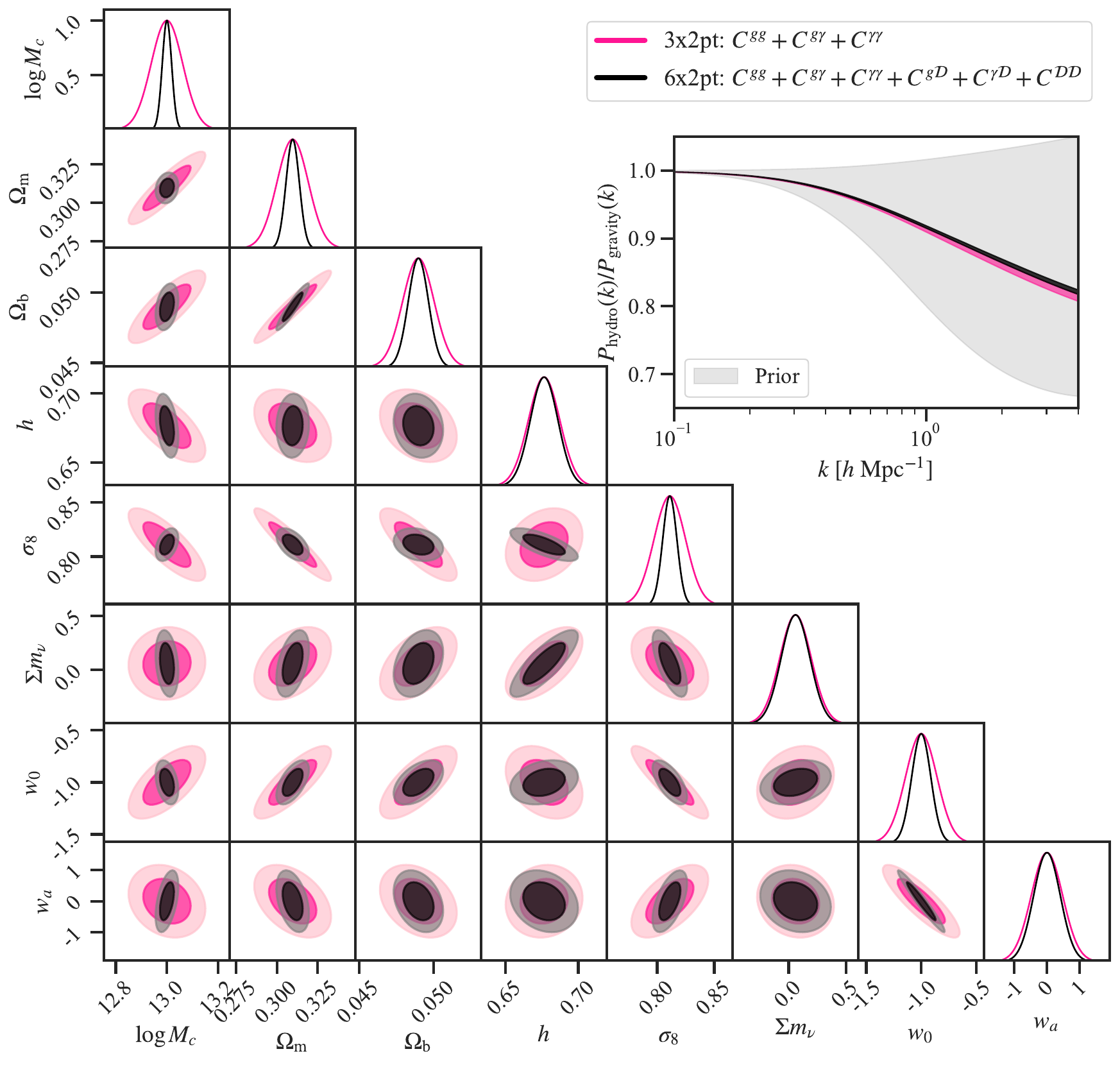}
    \caption{{\edfigurelabel{fig:fisher_forecast}} \textbf{Fisher parameter forecast for a multi-probe cosmological analysis.} The triangle plot compares marginalized constraints on cosmological and baryonic parameters from an LSST-only 3x2-point analysis -- galaxy clustering ($C_\ell^{gg}$), galaxy-galaxy lensing ($C_\ell^{g\gamma}$) and cosmic shear ($C_\ell^{\gamma\gamma}$) -- with those from an extended 6x2-point analysis that additionally incorporates FRB-based observables from DSA. The 6x2-point combination includes cross-correlations between FRB DMs and large-scale structure tracers -- DM-galaxy ($C_\ell^{g\mathcal{D}}$), DM-shear ($C_\ell^{\gamma\mathcal{D}}$) -- as well as the DM auto-correlation ($C_\ell^{\mathcal{DD}}$). Parameters shown include the baryonic feedback characteristic mass scale ($\log M_c$), standard cosmological parameters ($\Omega_\mathrm{m}$, $\Omega_\mathrm{b}$, $h$, $\sigma_8$), the sum of neutrino masses ($\Sigma m_\nu$), and dark energy equation-of-state parameters ($w_0, w_a$). The inset panel displays the corresponding constraint on the suppression of the matter power spectrum, comparing the 3x2-point and 6x2-point analyses. Including FRB DM statistics substantially tightens constraints on baryonic physics ($\Delta \log M_c \sim 0.02$), thereby  breaking key degeneracies with cosmological parameters, such as $\Omega_\mathrm{m}$, $\Omega_\mathrm{b}$, $h$, and $\sigma_8$. The improved control over baryonic uncertainties propagates into sharper constraints on extensions to the standard $\Lambda$CDM cosmological model, including neutrino masses ($\Delta~\Sigma m_\nu \sim 0.12$~eV) and dynamical dark energy ($\Delta w_0 \sim 0.08, \Delta w_a \sim 0.4$), highlighting the power of FRBs in enhancing both, astrophysical and cosmological inference in next-generation surveys. }
\end{figure}

\newpage
\clearpage
\newpage
\newpage

\vspace{0.5cm} 
\setlength{\parskip}{17pt}%
\bibliographystylemethods{naturemag}
\bibliographymethods{manuscript}
\vspace{-0.5cm}  

\newpage
\clearpage
\newpage
\newpage

\noindent{\bfseries \LARGE Supplementary Information}

\begin{figure}[h]
    \centering
    \includegraphics[width=0.5\columnwidth]{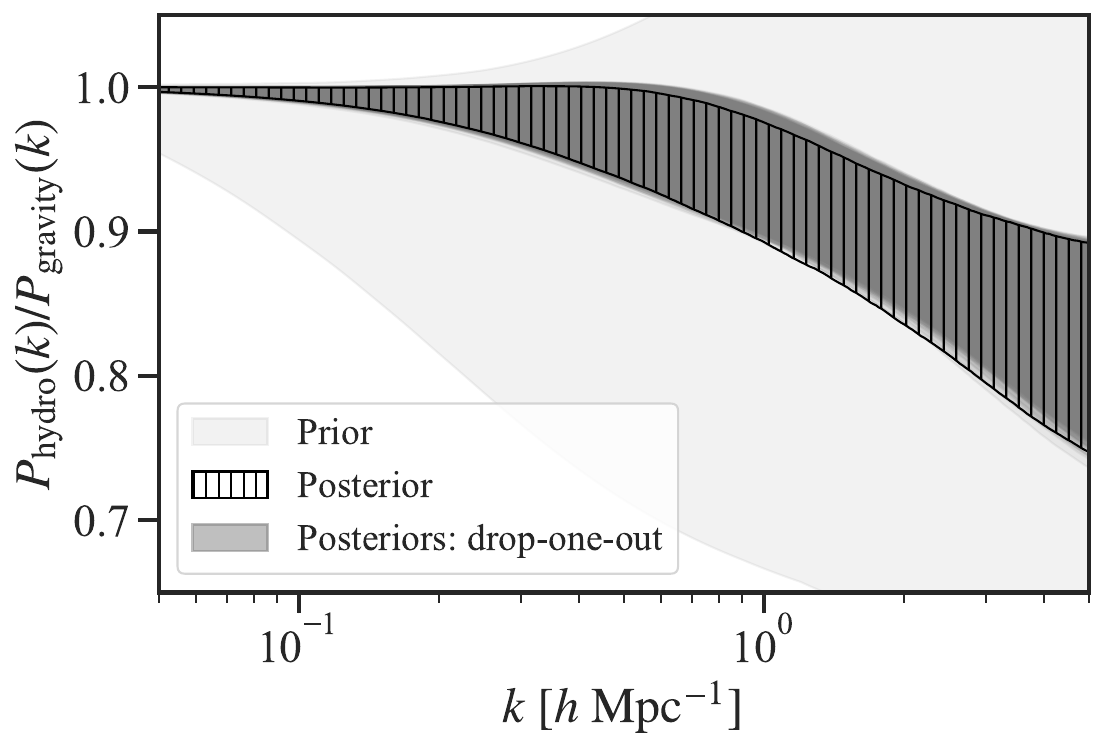}
    \caption{{\supfigurelabel{fig:drop_one_out_validation}} \textbf{Jackknife resampling validation of matter power spectrum suppression constraints.} We perform jackknife resampling, sequentially excluding individual FRBs from the analysis to assess potential bias. This is essential to test whether the current constraining power is driven by a single FRB sightline intersecting a massive galaxy cluster. The resulting posterior constraints remain stable across all jackknifed realizations, demonstrating that our results are not driven by any particular sightline.}
\end{figure}

\newpage
\clearpage
\newpage
\newpage

\begin{figure}[ht!]
    \centering
    \includegraphics[width=\columnwidth]{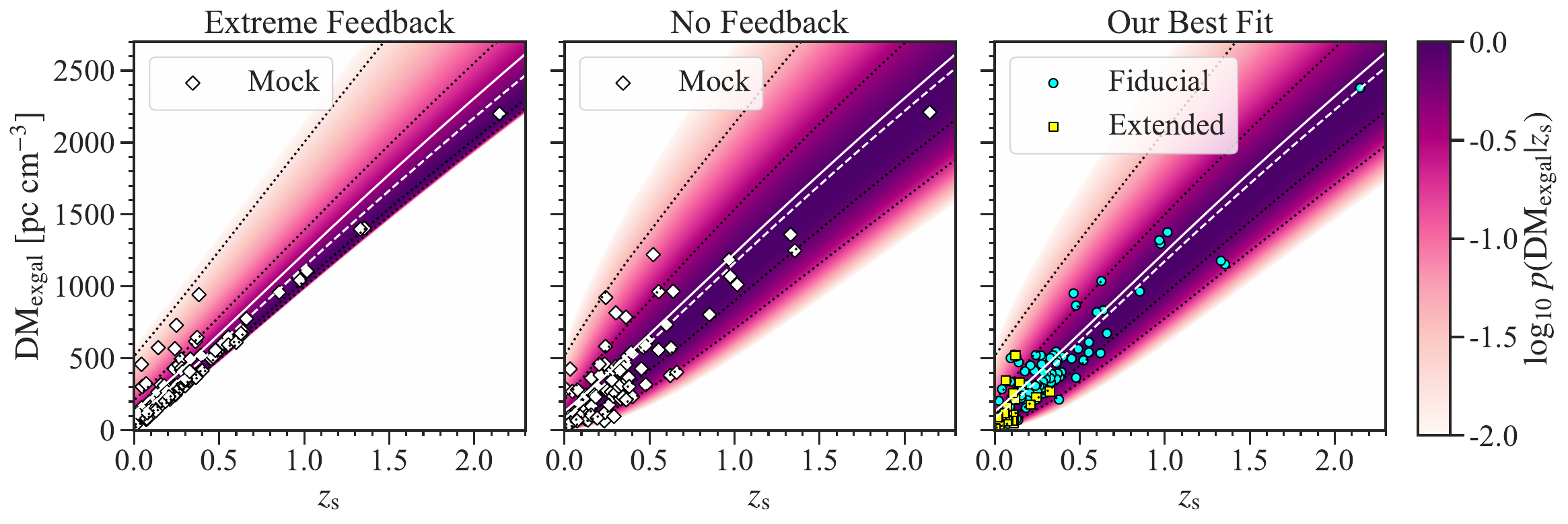}
    \includegraphics[width=\columnwidth]{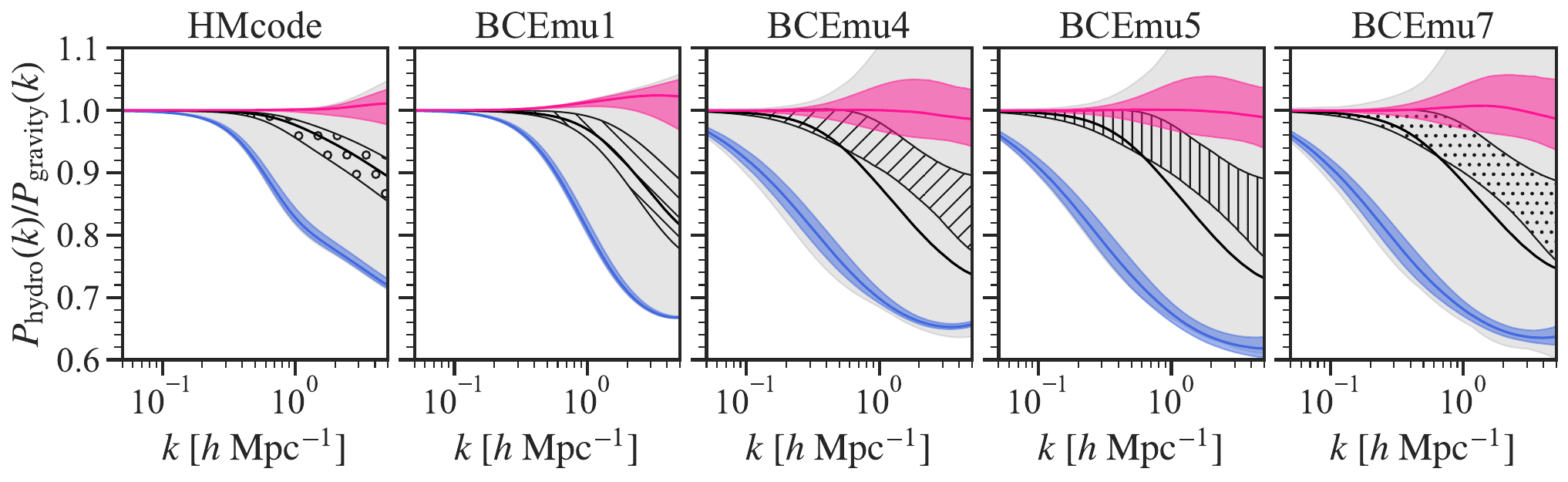}
    \includegraphics[width=\columnwidth]{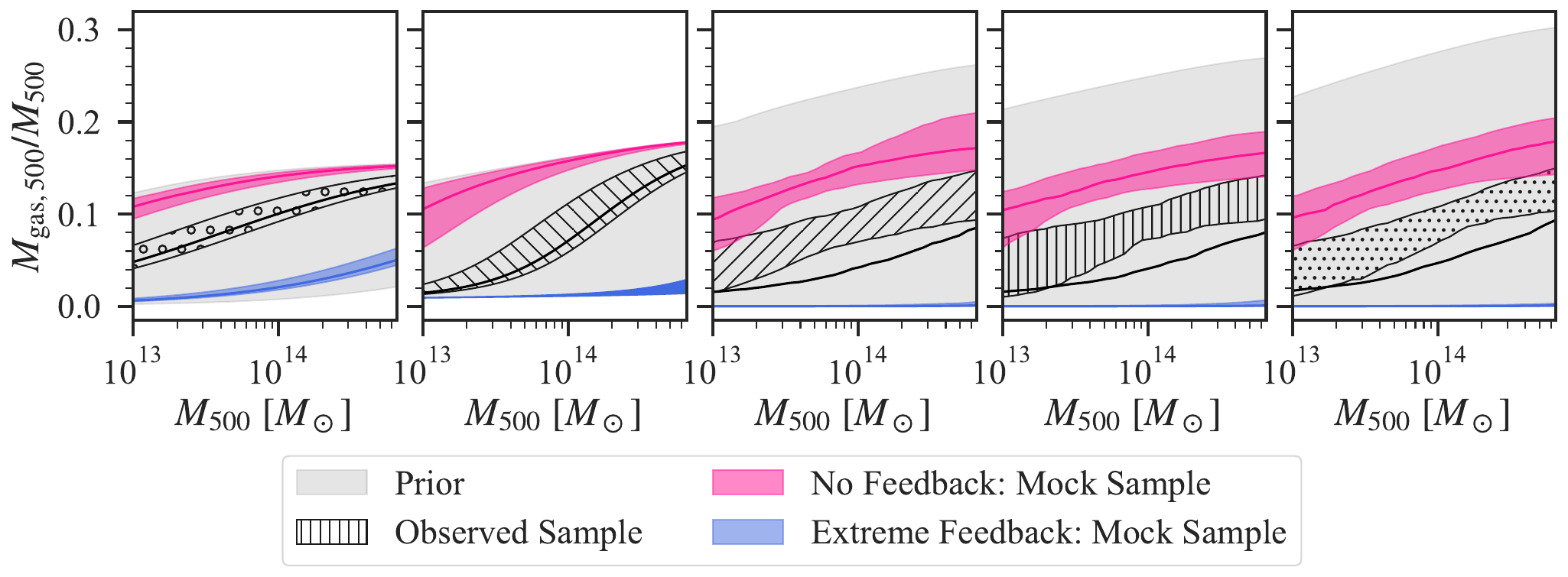}
    \caption{{\supfigurelabel{fig:extreme_scenario_test}} \textbf{Validation of the ability of current FRB dataset to distinguish between extreme feedback scenarios.} The DM$_\mathrm{exgal}-z$ relation of FRBs in universes with extreme feedback (zero variance in DM$_\mathrm{cosmic}$), no feedback (large variance in DM$_\mathrm{cosmic}$) and feedback strength indicated by current dataset is illustrated in panel \textbf{a}. In the two mock universes (extreme and no feedback), for redshifts of observed FRB sample, we generate synthetic DM$_\mathrm{exgal}$ and fit for the DM$_\mathrm{exgal}-z$ relation to measure the suppression in matter power spectrum ($P_\mathrm{hydro}/P_\mathrm{gravity}$) and gas mass fractions ($M_\mathrm{gas,500}/M_{500}$). The posteriors for the two mock scenarios for models, including the hydrodynamical simulations-calibrated halo model (\textsc{HMcode}) and variants of flexible analytical model (\textsc{BCEmu1}, \textsc{BCEmu4}, \textsc{BCEmu5}, \textsc{BCEmu7}), are shown in panels \textbf{b} and \textbf{c}. The mock FRBs from extreme feedback scenario predict a strong suppression/low gas fractions, whereas the mock FRBs from no feedback scenario predict a close to zero suppression/high gas fractions -- both consistent with our expectations. This validates the constraining power of current FRB sample.}
\end{figure}

\newpage
\clearpage
\newpage
\newpage

\begin{figure}[ht!]
    \centering
    \includegraphics[width=\columnwidth]{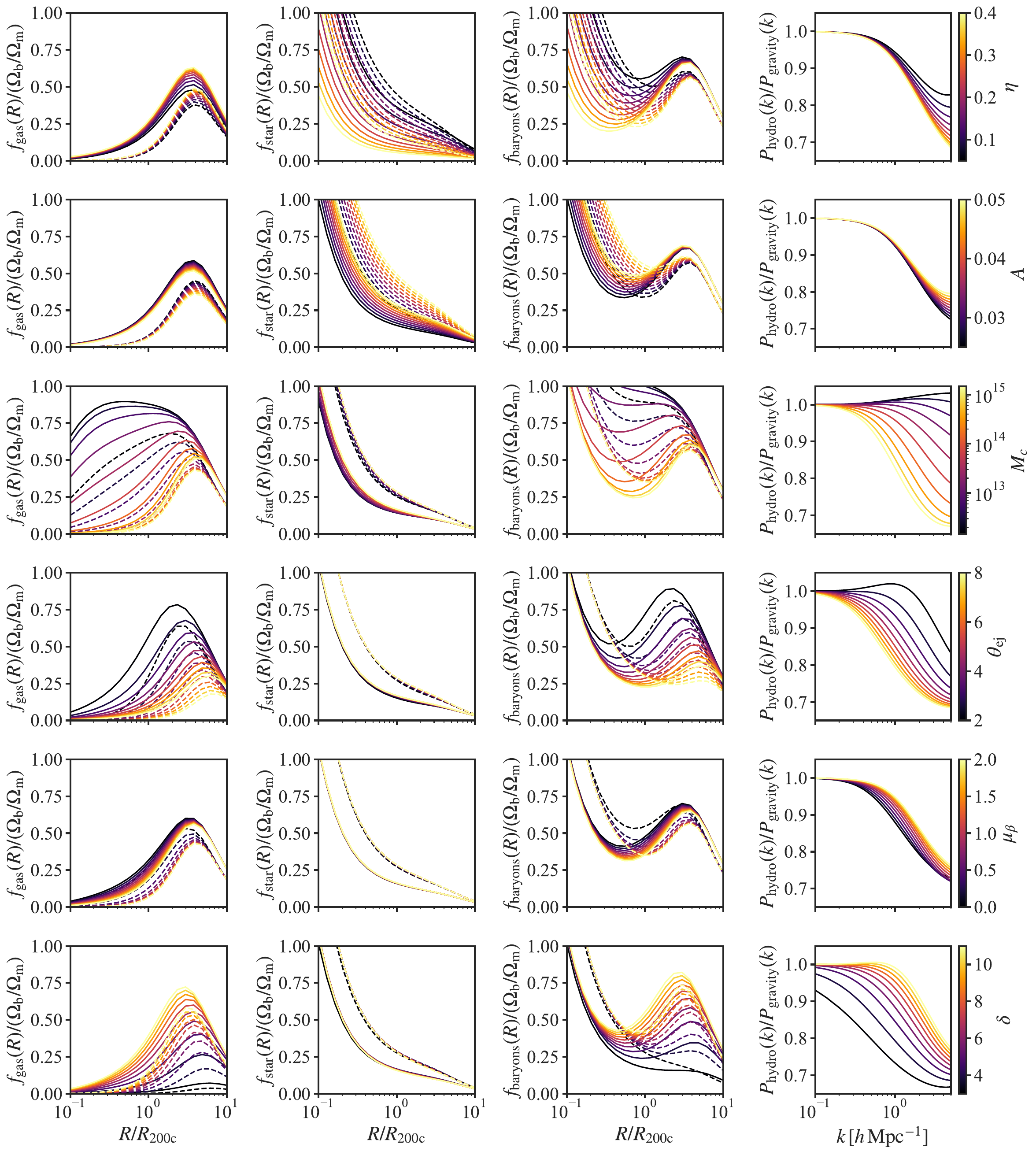}
    \caption{{\supfigurelabel{fig:f_gasR_f_starR_f_baryonsR_SPk}} \textbf{Sensitivity of halo gas, stellar, and baryon mass fractions and matter power spectrum suppression to stellar and gas physics.} The parameters governing the stellar-to-halo mass relation ($A$ and $\eta$) primarily affect the gas fractions integrated out to infinity, but within $R_{200}$, gas fractions are largely insensitive to stellar physics. A similar trend is seen for matter power spectrum suppression: the influence of stellar physics at intermediate-scales is below the current sensitivity of baryon probes. In contrast, the parameters controlling the halo gas profile ($\log M_c$, $\theta_\mathrm{ej}$, $\mu_\beta$, and $\delta$) have a strong impact on both, the matter power spectrum suppression and the gas fractions within $R_{200}$. This demonstrates that both, the matter power spectrum suppression and gas fractions within $R_{200}$ are robust metrics of gas physics.}
\end{figure}

\newpage
\clearpage
\newpage
\newpage

\begin{figure}[ht!]
    \includegraphics[width=\columnwidth]{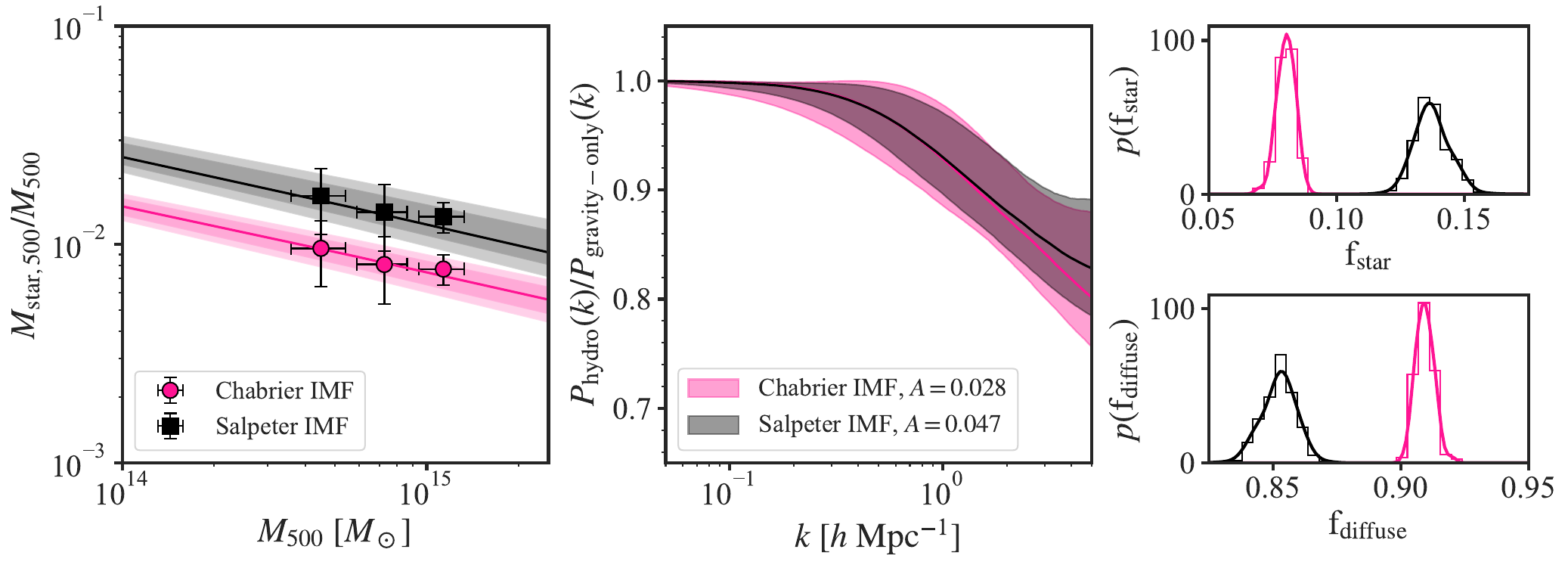}
    \caption{{\supfigurelabel{fig:impact_of_IMF}} \textbf{An evaluation of the impact of stellar initial mass function on the matter power spectrum suppression measurement when using observed stellar-to-halo mass relation data.} The stellar-to-halo mass (SHM) relation from DES, WISE/Spitzer and SPT SZ scaling relation of galaxy clusters~\protect\citemethods{2018MNRAS.478.3072C}, assuming Chabrier~\protect\citemethods{2003PASP..115..763C} and Salpeter~\protect\citemethods{1955ApJ...121..161S} initial mass functions (IMFs), is shown in panel \textbf{a}. The corresponding constraints on the matter power spectrum suppression, as illustrated in panel \textbf{b}, are robust and unbiased by a change in the normalization (consistent with the two IMFs) of SHM relation. However, we show in panel \textbf{c} that the choice of stellar IMF introduces bias in the inferred global stellar (f$_\mathrm{star}$) and diffuse (f$_\mathrm{diffuse}$) baryon fractions. This motivates the need to directly fit for diffuse gas fraction, without imposing external priors, to avoid biased inference, as larger FRB samples become accessible in near future. Such an inference will also provide direct constraints on the stellar IMF.}
\end{figure}

\begin{figure}[ht!]
    \includegraphics[width=\columnwidth]{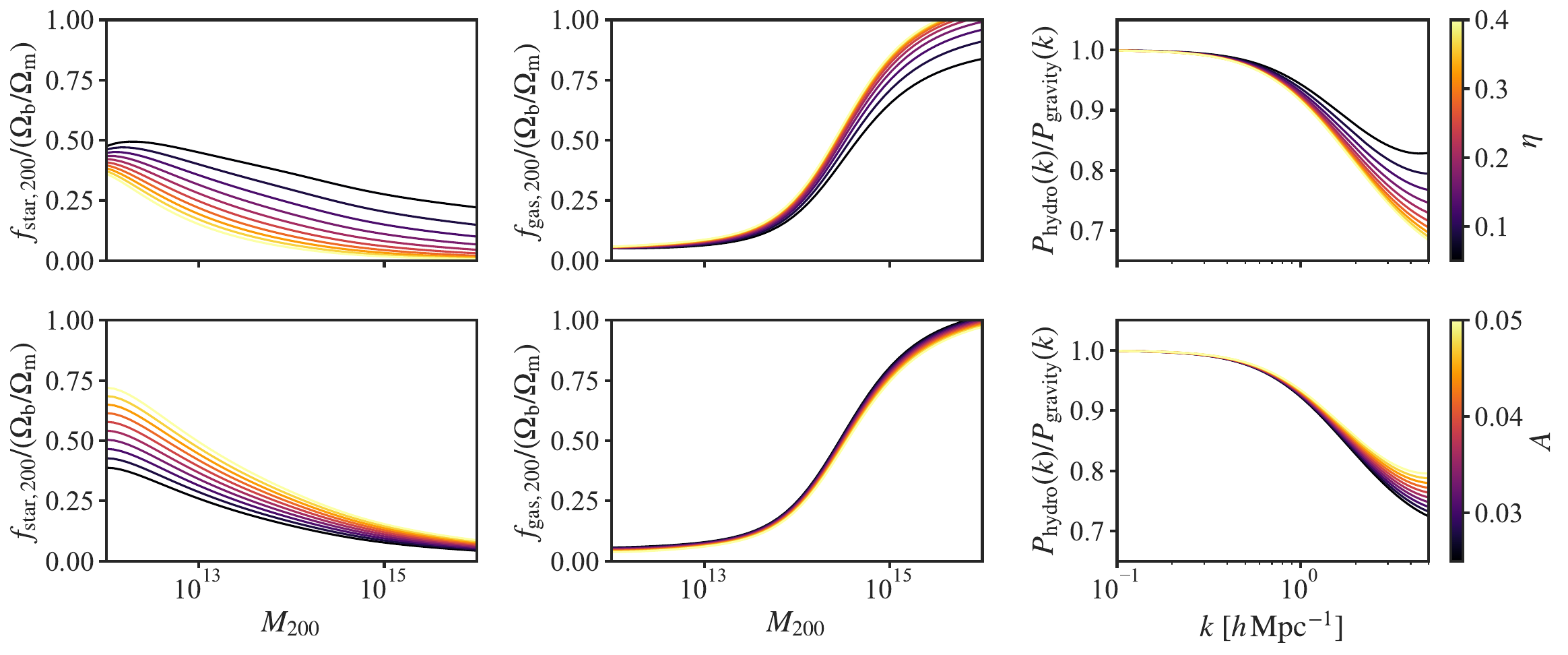}
    \caption{{\supfigurelabel{fig:parameter_variations-f_gasR200M_f_starR200M_SPk}} \textbf{Sensitivity to stellar physics of gas and stellar mass fractions enclosed within $R_{200}$ as a function of halo mass.} Varying the stellar-to-halo mass relation normalization parameter, $A$ is equivalent to changing stellar initial mass function (IMF). The impact of stellar IMF on gas mass fractions within $R_{200}$ across a broad halo mass range and matter power spectrum suppression at intermediate scales is below the current constraining power of baryon probes.}
\end{figure}

\end{document}